\documentclass[a4paper,11pt]{article}
\usepackage[dvipsnames]{xcolor}

\usepackage[T1]{fontenc} 
\usepackage{amsmath}
\usepackage{amssymb}
\usepackage{amsfonts}
\usepackage{mathtools}
\usepackage{microtype}
\usepackage{tikz}
\usepackage{bm}
\usepackage[utf8]{inputenc}
\usepackage{blkarray}
\usepackage{bigstrut}
\usepackage{nccmath}
\usepackage{enumitem}
\usepackage{braket}
\usepackage{dsfont}
\usepackage{lipsum}
\usepackage{leftindex}
\usepackage{array,multirow}
\usepackage{physics}
\usepackage{float}
\usepackage{subfig}
\usepackage[export]{adjustbox}
\usepackage{graphicx}
\usepackage{caption}
\usepackage{comment}
\usepackage{lscape}
\usepackage{lmodern}
\usepackage[normalem]{ulem}
\usetikzlibrary{decorations.pathmorphing,decorations.pathreplacing,calligraphy,calc,snakes,cd,external,arrows.meta}
\usepackage[compat=1.1.0]{tikz-feynman}

\usepackage{jheppub}

\def\>{\rangle}
\def\<{\langle}

\DeclareFontFamily{U}{mathx}{}
\DeclareFontShape{U}{mathx}{m}{n}{<-> mathx10}{}
\DeclareSymbolFont{mathx}{U}{mathx}{m}{n}
\DeclareMathAccent{\widehat}{0}{mathx}{"70}
\DeclareMathAccent{\widecheck}{0}{mathx}{"71}
\usepackage{cleveref}

\newcommand{\agl}[2]{\langle#1\, #2 \rangle}
\newcommand{\sqr}[2]{\lbrack #1\, #2 \rbrack}
\newcommand{\txt}[1]{\textup{#1}}

\newcommand{\hol}[0]{\mathsf{H}}
\newcommand{\ahol}[0]{\mathsf{A}}

\newcommand{\zb}[0]{\bar{z}}

\makeatletter\g@addto@macro\bfseries{\boldmath}

\title{Renormalization of effective field theories via on-shell methods: the case of axion-like particles 
}

\author[a,b]{Luigi~C.~Bresciani,}
\author[a,b,c]{Giacomo~Brunello,}
\author[a,b,d]{Gabriele~Levati,}
\author[a,b]{Pierpaolo~Mastrolia}
\author[a,b]{and Paride~Paradisi}

\affiliation[a]{Dipartimento di Fisica e Astronomia, Università di Padova, Via Marzolo 8, I-35131 Padova, Italy}
\affiliation[b]{INFN, Sezione di Padova,
Via Marzolo 8, I-35131 Padova, Italy}
\affiliation[c]{Institut de Physique Théorique, CEA, CNRS, Université Paris-Saclay, F-91191 Gif-sur-Yvette cedex, France}
\affiliation[d]{Albert Einstein Center for Fundamental Physics, Institute for
Theoretical Physics, University of Bern, Sidlerstrasse 5, CH-3012 Bern,
Switzerland}

\emailAdd{luigicarlo.bresciani@phd.unipd.it}
\emailAdd{giacomo.brunello@phd.unipd.it}
\emailAdd{gabriele.levati@unibe.ch}
\emailAdd{pierpaolo.mastrolia@unipd.it}
\emailAdd{paride.paradisi@unipd.it}

\abstract{
We consider the most general axion-like particle effective field theory, including both CP-odd and CP-even types of interactions, and evaluate the corresponding renormalization group equations, improving and extending previous results in the literature. Our calculations exploit on-shell and unitarity-based methods. The relevant phase-space cut-integrals are carried out using different integration methods, among which the double-cut integration via Stokes' theorem proves to be technically simpler. A close comparison between the standard Feynman diagrammatic approach and the unitarity-based method enables us to explicitly verify the reduction of complexity in the latter case, along with a more direct and elegant way to establish a connection among anomalous dimensions of operators that are dual under the CP symmetry.}

\begin{document}
\flushbottom
\pagestyle{myplain}

\maketitle

\allowdisplaybreaks

\section{\label{sec:intro}Introduction}

The Standard Model (SM) of particle physics, describing the fundamental interactions of Nature, is among the most successful theories of physics.
However, the SM alone is unable to provide a satisfactory answer to several open questions of both observational and theoretical nature. 
It should therefore be regarded as the low-energy remnant of a more complete ultraviolet theory entailing new dynamics emerging at some large 
--- yet unknown --- energy scale $\Lambda$.

SM extensions including light pseudoscalars, such as the so-called 
Axion-Like Particles (ALPs)~\cite{Jaeckel:2010ni,Marsh:2015xka,Irastorza:2018dyq,DiLuzio:2020wdo}, 
are among the most interesting and studied scenarios.
Their lightness, compared to the scale $\Lambda$, can be easily motivated if 
they are the pseudo-Nambu-Goldstone bosons of some spontaneously broken global symmetry.
ALPs can elegantly address several open questions in particle physics such as the strong CP~\cite{Peccei:1977hh,Peccei:1977ur,Weinberg:1977ma,Wilczek:1977pj} and flavor problems~\cite{Davidson:1981zd,Wilczek:1982rv,Berezhiani:1989fp,Calibbi:2016hwq,Greljo:2024evt}, as well as the stability of the electroweak scale~\cite{Graham:2015cka}. 
Moreover, they can be regarded as being natural dark matter candidates~\cite{Abbott:1982af,Preskill:1982cy,Dine:1982ah,Davis:1986xc}. 
From the experimental side, ALPs can be probed by cosmological and astrophysical  searches~\cite{ADMX:2003rdr,Barbieri:2016vwg,Caldwell:2016dcw,Zioutas:1998cc,Irastorza:2011gs,CAST:2017uph,Armengaud:2014gea,VanBibber:1987rq,Bahre:2013ywa,OSQAR:2015qdv,Arvanitaki:2014dfa}, beam-dump experiments~\cite{Alekhin:2015byh,Dobrich:2015jyk,Shan:2024pcc}, 
at colliders~\cite{Bauer:2017ris,Brivio:2017ije} and through a plethora of rare processes~\cite{Bauer:2021mvw,Gavela:2019wzg,Cornella:2019uxs}.
From the theoretical viewpoint, it is customary to describe the leading-order ALP interactions with SM particles via effective dimension-5 operators~\cite{Georgi:1986df}. Such an Effective Field Theory (EFT) approach allows to capture general features of broad classes of models without relying on specific ultraviolet completions. 
Physical observables are then obtained by computing matrix elements of the ALP EFT Lagrangian at those energy scales $E \ll \Lambda$ that are accessible by experiments. 
As a result, a crucial ingredient to make theoretical predictions is to run the ALP Lagrangian from the scale $\Lambda$ down to the experimental scale $E$.
This goal can be achieved by evaluating the anomalous dimension matrix of the ALP effective operators.

The renormalization group equations (RGEs) of the Standard Model EFT extended with a CP-odd ALP have been already computed at one-loop accuracy up to dimension-5 operators using diagrammatic methods~\cite{Bauer:2020jbp,Chala:2020wvs}. 
Instead, the case of ALPs with both CP even and odd components leads to CP-violating effects which have
been investigated in~\cite{Marciano:2016yhf,DiLuzio:2020oah}.
The corresponding RGEs of such a CP violating ALP framework
have been derived in~\cite{DasBakshi:2023lca}. 

Quite recently, anomalous dimensions have been evaluated  
through on-shell and unitarity-based techniques for scattering amplitudes~\cite{Caron-Huot:2016cwu,EliasMiro:2020tdv,Baratella:2020lzz,Jiang:2020mhe,Bern:2020ikv,Baratella:2020dvw,AccettulliHuber:2021uoa,EliasMiro:2021jgu,Baratella:2022nog,Machado:2022ozb}.
This approach, rooted in recent developments in the context of generalized unitarity \cite{Arkani-Hamed:2008owk, Huang:2012aq, Cheung:2015aba} and of the study of the dilatation operator in $\mathcal{N}=4$ super Yang-Mills theories \cite{Zwiebel:2011bx,Wilhelm:2014qua, Nandan:2014oga, Koster:2014fva, Brandhuber:2014pta, Brandhuber:2015boa, Loebbert:2015ova, Brandhuber:2016fni}, is particularly suited to unveil hidden structures with the emergence of zeros in the anomalous dimension matrix.
The origin of these vanishing elements is a direct consequence of selection rules~\cite{Elias-Miro:2014eia} based on operator lengths~\cite{Bern:2019wie}, helicity~\cite{Cheung:2015aba}, and angular momentum~\cite{Jiang:2020rwz}.\footnote{See also~\cite{Chala:2023jyx,Chala:2023xjy} for other zeros stemming from arguments based on the positivity of the $S$-matrix.}
Remarkably, anomalous dimensions can be 
related to the discontinuities of form factors of EFT operators,
therefore, they can be efficiently extracted from
unitarity cuts, evaluated via phase-space integrals ~\cite{Caron-Huot:2016cwu}.
This method has been proven to be
particularly promising for computing anomalous dimensions beyond one-loop order
\cite{Caron-Huot:2016cwu,Bern:2020ikv,EliasMiro:2021jgu}.

First studies concerned anomalous dimension matrices of non-renormalizable massless theories including mixing effects among operators of the same dimension~\cite{Caron-Huot:2016cwu}.
More recent studies have also systematically considered the inclusion of leading mass effects in this framework, which are extremely relevant in several EFT extensions of the SM~\cite{Bresciani:2023jsu}.
In particular, leading mass effects can be included in the massless limit~\cite{Bresciani:2023jsu} by exploiting the Higgs low-energy theorem~\cite{Ellis:1975ap,Shifman:1979eb,Shadmi:2018xan,Durieux:2019eor,Durieux:2020gip,Balkin:2021dko,Bertuzzo:2023slg}. 

The aim of this paper is to apply the above method~\cite{Caron-Huot:2016cwu,Bresciani:2023jsu} to the one-loop renormalization of CP violating 
ALP theories~\cite{Marciano:2016yhf,DiLuzio:2020oah} up to the phenomenologically most relevant dimension-6 operators, therefore reproducing and extending previous results~\cite{DasBakshi:2023lca}.
An extensive derivation of the relevant anomalous dimension matrix will be carried out at one-loop order both with standard techniques and through on-shell methods, aiming to show the strength of the latter approach, which drastically reduces the complexity of standard loop calculations. The relevant phase space cut-integrals will be evaluated by different parameterizations, both by angular integration~\cite{Zwiebel:2011bx,Caron-Huot:2016cwu,EliasMiro:2020tdv,Baratella:2020lzz,Bern:2020ikv,Baratella:2020dvw}, and using Stokes' theorem~\cite{Mastrolia:2009dr,Jiang:2020mhe,Shu:2021qlr,Baratella:2021guc}, 
for cross checks, as well as to highlight the strengths of the various approaches.

The paper is organized as follows. In Section~\ref{sec:method}, we summarize the 
on-shell method~\cite{Caron-Huot:2016cwu} and present different parameterizations to evaluate phase space integrals.
In Section~\ref{sec:ALPEFT}, we introduce the EFT for axion-like particles and in Section~\ref{sec:uvanomalousdim} we report a detailed derivation of the corresponding anomalous dimensions.
In Section~\ref{sec:comparison}, we compare our results as obtained with on-shell and standard methods.
Section~\ref{sec:conclusions} is dedicated to our conclusions.
In Appendix~\ref{app:notation} we report our notation and conventions and, finally, tree amplitudes and infrared anomalous dimensions of ALP operators are given in Appendix~\ref{app:Amplitudes} and \ref{app:IRanomalousdimensions}, respectively.

\section{\label{sec:method}
Renormalization of EFT via on-shell methods}

In this section, we first review the on-shell method for computing anomalous dimensions introduced in Ref.~\cite{Caron-Huot:2016cwu}. Then, we discuss two independent ways to perform 
phase-space integrals both via angular variables~\cite{Zwiebel:2011bx} and through the use of Stokes' 
Theorem~\cite{Mastrolia:2009dr}. 

\subsection{The on-shell method}
We consider an effective Lagrangian of the type
\begin{equation}
\mathcal L_{\text{EFT}} = \sum_i 
\frac{c_i}{\Lambda^{[\mathcal O_i]-4}} \,\mathcal O_i \, , 
\end{equation}
where $\mathcal{O}_i$ are local gauge-invariant operators, $c_i$ are the corresponding Wilson coefficients, and $\Lambda$ refers to the UV cut-off scale of our EFT.

{\it Form factors} of the operators $\mathcal{O}_i$ are generically defined as
\begin{equation}
    F_i(\vec n;q)= \frac{1}{\Lambda^{[\mathcal O_i]-4}}\mel*{\vec n}{\mathcal O_i(q)}{0}\,,
\end{equation}
namely as the matrix element between an outgoing on-shell state $\bra*{\vec n}=\bra*{1^{h_1},\dotsc,n^{h_n}}$ and an operator $\mathcal O_i$ that injects an additional off-shell momentum $q$.
Within the dimensional regularization scheme, form factors depend on the renormalization scale $\mu$, and satisfy the Callan-Symanzik equation
\begin{equation}
\label{eq:CSEq}
    \left(\delta_{ij} \mu\frac{\partial}{\partial \mu}+ \frac{\partial \beta_i}{\partial c_j}-\delta_{ij}\gamma_{i,\txt{IR}} +\delta_{ij}\beta_g\frac{\partial}{\partial g}\right)F_i=   0\,,
\end{equation}
where $g$ collectively denotes the couplings related to the renormalizable operators of our Lagrangian, while $\gamma_{i,\txt{IR}}$ is the infrared anomalous dimension. 
The renormalization of the operator $\mathcal{O}_i$ 
is described by
\begin{equation}
\beta_i(\{c_k\}) \equiv \mu \dv[]{c_i}{\mu} \,, 
\end{equation}
where $c_i$ are the Wilson coefficients of the effective Lagrangian $\mathcal{L}_{\text{EFT}}$.

Exploiting the analyticity of form factors, unitarity, and the CPT theorem, it can be shown that an elegant relation exists linking the action of the dilatation operator ($D$)
to the action of the $S$-matrix ($S$) on form factors~\cite{Caron-Huot:2016cwu}:
\begin{equation}
\label{eq:FundamentalRelation}
    e^{-i \pi D}F_i^* = SF_i^*
\end{equation}
where $S=\mathbf{1} + i\mathcal{M}$ while $D = \sum_{i}p_i\cdot \partial/\partial p_i$
(the sum runs over all particles $i$). 

It is precisely the combination of Eqs.~\eqref{eq:FundamentalRelation} and \eqref{eq:CSEq} that allows one to directly link the renormalization group coefficients to the $S$-matrix.
In particular, at one-loop order, it has been found that
\begin{equation}
\label{eq:masterformula}
\left(
\frac{\partial \beta_i^{(1)}}{\partial c_j}
-\delta_{ij}\gamma_{i,\txt{IR}}^{(1)} 
+\delta_{ij}\beta_g^{(1)}\frac{\partial}{\partial g}
\right)
F_i^{(0)} = -\frac{1}{\pi}(\mathcal M F_j)^{(1)}\,, 
\end{equation}
where the right-hand side of Eq.~\eqref{eq:masterformula} corresponds to a sum over all one-loop two-particle unitarity cuts,
\begin{align}
\label{eq:MF1Loop}
    (\mathcal M F_j)^{(1)}(1,\dots,n) = 
   &\sum_{k=2}^n \sum_{\{x,y\}}\int d\text{LIPS}_2
   \nonumber\\ 
   &\times\sum_{h_1,h_2}\!\! F_j^{(0)}\!(x^{h_1},y^{h_2},k \!+\! 1,\dots,n) \mathcal M^{(0)}\!(1,\dots,k;x^{h_1},y^{h_2}\!)\, ,
\end{align}
where $\mathcal M(\vec n; \vec m)=\mel*{\vec n}{\mathcal M}{\vec m}$, and $d\text{LIPS}_2$ is the (two-particle) Lorentz invariant phase-space measure. 
The corresponding cut-integral can be evaluated either by angular integration~\cite{Zwiebel:2011bx} or via Stokes' theorem~\cite{Mastrolia:2009dr},
which differ for the parameterizations of the on-shell loop momentum across the cut and for the employed integration techniques, 
as we will show in the following.

Two observations are in order. 
Let us first remark that, at one-loop level, the $\beta$ function does not contribute to minimal form factors, because 
the latter are expectation values of purely local products of fields, and, by definition, they are independent of the renormalizable couplings of the theory, which we have collectively denoted by $g$.
The contribution from the $\beta$ function of renormalizable couplings is however unavoidable at higher perturbative orders.
Secondly, we stress that, owing to the adopted dimensional regularization scheme, this method is sensitive only to the difference between UV and IR divergences. 
Therefore, in order to disentangle the renormalization group equations for the UV divergent part, the IR contribution must be computed independently (see Appendix~\ref{app:IRanomalousdimensions} for more details). 

The on-shell method, so far discussed, is well suited to compute anomalous dimensions in massless theories, where all 
EFT operators have the same mass-dimension, as shown in the case of Standard Model EFT~\cite{EliasMiro:2020tdv}.  

In particular, in the case of linear operator mixing, 
Eq.~\eqref{eq:masterformula}
becomes
\begin{equation}
\label{eq:masterformula1L}
\left(
\gamma_{i \leftarrow j}^{(1)}
-\delta_{ij}\gamma_{i,\txt{IR}}^{(1)} 
\right)
F_i|_{*}^{(0)} = -\frac{1}{\pi}(\mathcal M F_j)|_{*}^{(1)}
\end{equation}
which has been evaluated at the Gaussian fixed point $(*)$, where all the Wilson coefficients $c_i$ are vanishing.
Moreover, $\gamma_{i \leftarrow j}^{(1)}$ is obtained from
the Taylor expansion of the renormalization group equations for the Wilson coefficients $c_i$
\begin{equation}
    \mu\dv[]{c_i}{\mu} = 
    \gamma_{i\leftarrow j}c_j +
    \frac{1}{2}\gamma_{i\leftarrow j,k}c_jc_k + \dots\,,
\end{equation}
where 
\begin{equation}
    \gamma_{i\leftarrow j_1, \dotsc, j_n} = \left.\frac{\partial^n \beta_i}{\partial c_{j_1}\dotsm \partial c_{j_n}}\right |_{*}\,.
\end{equation}

The result of Eq.~\eqref{eq:masterformula1L} can be easily applied to the case of non-linear mixing among operators with different dimensions, which is of interest to our study. At one-loop order, one can find the following expression~\cite{Bresciani:2023jsu} 
\begin{equation}
\label{eq:mastergammaijk}
\gamma_{i\leftarrow j,k}^{(1)}F_i|_{*}^{(0)} = -\frac{1}{\pi}\left.\frac{\partial}{\partial c_k}\right|_{c_k=0}(\mathcal M F_j)|_{*,c_k\neq 0}^{(1)} \,,
\end{equation}
where $j,k\neq i$ and
\begin{align}
    \gamma_{i\leftarrow j,k}&=\left. \frac{\partial^2 \beta_i}{\partial c_j \partial c_k}\right|_{*} = \left.\frac{\partial}{\partial c_k}\right|_{c_k=0} \left.\frac{\partial \beta_i}{\partial c_j}\right|_{*,c_k\neq 0} \,.
    \label{eq:gamma_ijk}
\end{align}
Moreover, by making use of the Higgs low-energy theorem~\cite{Ellis:1975ap,Shifman:1979eb}, it is possible to include leading mass effects while still working in a massless formalism~\cite{Bresciani:2023jsu}. 
In practice, whenever an amplitude requires $N$ fermion mass insertions not to vanish, we consider an equivalent amplitude entailing $N$ extra massless Higgs fields, where 
\begin{equation}
N = 4-[\mathcal{O}_{i}] +\sum^n_{k=1} ([\mathcal{O}_{j_k}]-4)\geq 0
\label{eq:superficial_div}
\end{equation}
corresponds to the superficial degree of divergence associated with the loop diagram under consideration~\cite{Bresciani:2023jsu}.
Then, the anomalous dimension
$\gamma_{i\leftarrow j_1, \dotsc, j_n}$ 
is obtained by renormalizing the operator $(h/v)^N\mathcal{O}_i/N!$
instead of $\mathcal{O}_i$~\cite{Bresciani:2023jsu}.
Instead, for $N < 0$, $\gamma_{i\leftarrow j_1, \dotsc, j_n}$ does vanish.

The Lorentz-invariant phase space measure appearing in Eq.~\eqref{eq:MF1Loop} can be parameterized in different ways, depending on the employed method of integration. 
Here we will focus on two possible integration techniques: one based on an angular parameterization of the phase space and the other relying on the use of Stokes' theorem.
Whereas the former has the virtue of being quite intuitive, the latter turns out to be more suited to our purposes. Based on an elegant mathematical result, it offers a simpler and more direct way of carrying out phase-space integrations, as shown in several explicit examples.

\subsection{Phase-space integrals via angular variables}
\label{sec:phasespace_Angular}

The integration with angular variables relies on a parameterization of the virtual phase space
that is realized through the application of the following spinor rotation matrix~\cite{Zwiebel:2011bx}:
\begin{eqnarray}\label{eq:angular_param}
    \begin{pmatrix}
 \lambda_x \\
 \lambda_y
    \end{pmatrix}
     = 
    \begin{pmatrix}
 \cos\theta & -\sin\theta e^{i\phi} \\ 
 \sin\theta e^{-i\phi} & \cos\theta
\end{pmatrix}
\begin{pmatrix}
 \lambda_{a} \\
 \lambda_{b}
    \end{pmatrix}\,,
\end{eqnarray}
where $ \lambda_{a},\lambda_{b}$ correspond to the external momenta $p_a,p_b$,
and the integration measure is 
\begin{equation}
    \int d\text{LIPS}_2= \frac{1}{8 \pi}\int_0^{2 \pi} \frac{d\phi}{2\pi}\int_0^{\pi/2} 2 \cos\theta \sin\theta \, d\theta  \, . 
\end{equation}
An additional factor of $1/2$ has to be included in the measure if the two particles across the cut are indistinguishable.

Sometimes it is convenient to use the following parameterization
\begin{equation}
    \begin{pmatrix}
        \lambda_x \\ \lambda_y
    \end{pmatrix}
    =
    \frac{1}{\sqrt{1+t^2}}
    \begin{pmatrix}
        1 & -tz \\
        t/z & 1
    \end{pmatrix}
    \begin{pmatrix}
        \lambda_a \\ \lambda_b
    \end{pmatrix}\,,
\end{equation}
which rationalizes the integrand function and is derived from Eq.~\eqref{eq:angular_param} by setting $t=\tan\theta$ and $z=e^{i\phi}$.
The corresponding integration measure is given by
\begin{equation}
\label{eq:hybbrid}
    \int d\text{LIPS}_2 = \frac{1}{8\pi}\int_0^\infty \frac{2t \, dt}{(1+t^2)^2}\oint_{|z|=1}\frac{dz}{2\pi i z}\,.
\end{equation}

\subsection{Phase-space integrals via Stokes' Theorem}
\label{sec:phasespace_Stokes}
One-loop Feynman integrals, as well as scattering amplitudes in dimensional regularization, with $d= 4- \epsilon $ space-time dimensions,  can be decomposed in a finite basis of scalar integrals, known as master integrals.
Remarkably, up to order $\mathcal{O}(\epsilon^0)$, for any one-loop $n$-point amplitude, such master integrals are 4-point, 3-point, 2-point, and 1-point functions.
The latter do not contribute to processes with massless internal states; therefore, the UV singularities of massless 1-loop amplitudes,
together with collinear IR divergences,
are entirely contained in the 2-point function, and
they are proportional to the associated decomposition coefficient.
Singularities associated with 3- and 4-point functions are instead related only to IR divergences. 
Various techniques have been developed to evaluate the decomposition coefficients, including integration-by-parts identities \cite{Tkachov:1981wb,Chetyrkin:1981qh,Laporta:2000dsw} and generalized unitarity \cite{Bern:1994zx,Bern:1995db,Britto:2004nc,Britto:2006sj,Anastasiou:2006gt,Ossola:2006us}.
An efficient method to compute directly the 2-point function coefficients, projecting it out of a double-cut, relies on Stokes' theorem \cite{Mastrolia:2009dr}, and it is based on a reparameterization of the virtual spinors in terms of the external ones that is implemented via the following spinor rotation matrix:
\begin{eqnarray}
    \begin{pmatrix}
 \lambda_x \\
 \lambda_y
    \end{pmatrix}
     =  
   \frac{1}{\sqrt{1+ z \bar{z}}}\left(
\begin{array}{cc}
 1 & \bar{z} \\
 -z & 1 \\
\end{array}
\right)
\begin{pmatrix}
 \lambda_{a} \\
 \lambda_{b}
    \end{pmatrix}\,,
    \label{eq:stokes_par}
\end{eqnarray}
where $z$ and $\bar{z}$ are complex conjugate variables.
The integration measure is defined as:
\begin{eqnarray}
    \int d \text{LIPS}_2 = 
    \frac{-1}{8 \pi} 
    \oint \frac{dz}{2\pi i } 
    \int    
    \frac{d{\bar z}}{(1 + z \bar{z})^2} \, . 
\end{eqnarray}

The integrand is a generic rational function $g(z,\zb)$, which can be integrated, using complex analysis, by a two-step procedure: finding a primitive function in $\zb$, say $G(z,{\bar z})$ (considering $z$ as an independent variable), and applying Cauchy's residue theorem in $z$.
 The computation of the primitive generates two kinds of contributions: a rational part and a logarithmic one, i.e., 
 $G = G_{\rm rat} + G_{\log}$. 
 For the determination of the anomalous dimensions,
 it is sufficient to determine the UV singularities carried out by the 2-point integrals, neglecting the higher-point functions that contain only IR singularities.
 Since the cut of the 2-point scalar function is rational, the contribution of the 2-point integral within the double-cut of a generic one-loop integral appears exclusively in the rational contribution of the double-cut \cite{Mastrolia:2009dr}. 
 Therefore, in order to isolate the coefficient of the 2-point function, it is sufficient to retain only the rational part (Rat) of the phase-space integration, coming just from $G_{\rm rat}$, yielding:
    \begin{equation}
        {\rm Rat}\int d{\rm LIPS}_2 \ g(z,\zb)
        \equiv 
         \frac{-1}{8 \pi} 
         \oint \frac{dz }{ 2 \pi i} 
         G_{\rm rat}(z,{\bar z}) 
         =
         \frac{-1}{8 \pi} 
         \sum_{z_0\in {\cal P}_{G_{\rm rat}}} 
         \!\!\!
         \Res_{(z,\zb)=(z_0,z_0^*)} G_{\rm rat}(z,\zb)\, . 
    \end{equation}
  
The above formula considers the contributions of the residues at all the finite poles $z_0$ of $G_{\rm rat}$, collected in the set ${\cal P}_{G_{\rm rat}}$  
(when taking the residues, the variable ${\bar z}$  is evaluated at $z_0^* = (z_0)^{\rm c.c.}$). 
Stokes' integration method offers the advantage of projecting directly the contribution of the 2-point functions out of double-cuts, 
hence its application appears to be simpler than other techniques. 

Within the coefficient of the 2-point integrals, both UV and IR collinear divergences are combined.
Therefore, in order to isolate the genuine UV anomalous dimension,
one has to compute and add,
out of the complete IR anomalous dimension, only the collinear part.

\section{\label{sec:ALPEFT}
Effective Field Theory for Axion-Like Particles
}

The CP-violating interactions of an Axion-Like Particle (ALP) with SM fields below the electroweak (EW) scale can be conveniently described by the following $SU(3)_c \times U(1)_{\text{em}}$ invariant Lagrangian \cite{Marciano:2016yhf,DiLuzio:2020oah}: 
\begin{equation}
\label{eq:Lag1}
\mathcal{L}_{\text{EFT}} = \mathcal{L}_{\text{SM}} + 
\frac{\tilde{\mathcal C}_\gamma}{\Lambda} \mathcal{O}_{\tilde\gamma} + \frac{\tilde{\mathcal C}_g}{\Lambda} \mathcal{O}_{\tilde g} + \mathcal{Y}^{ij}_P \mathcal{O}_{P_{ij}} 
+\frac{\mathcal C_\gamma}{\Lambda} \mathcal{O}_\gamma + \frac{\mathcal C_g}{\Lambda} \mathcal{O}_g + \mathcal{Y}^{ij}_S \mathcal{O}_{S_{ij}} 
\end{equation}
where $\mathcal{L}_{\text{SM}}$ is the SM Lagrangian and
\begin{align}
\label{eq:LagA}
     \mathcal{O}_{\tilde \gamma} &= \phi\, F\tilde F\,, & \mathcal{O}_{\tilde g} &= \phi\, G\tilde G\,, & \mathcal O_{P_{ij}} &= \phi\, \bar{f}_i i \gamma_5 f_j\,,
     \\
\label{eq:LagB}     
 \mathcal O_\gamma &= \phi\, FF\,, & \mathcal O_g &= \phi\, GG\,, & \mathcal O_{S_{ij}} &= \phi\, \bar{f}_i f_j\,.
\end{align}
In the above expressions, $\phi$ is the ALP field and $\Lambda$ represents the new physics scale at which our effective description breaks down. $F_{\mu\nu}$ and $G_{\mu\nu}$ are the photonic and gluonic field-strength tensors, respectively, and $\tilde{F}_{\mu\nu} = \frac{1}{2} \varepsilon_{\mu\nu\alpha\beta}F^{\alpha \beta}$ and $\tilde{G}_{\mu\nu} = \frac{1}{2} \varepsilon_{\mu\nu\alpha\beta}G^{\alpha \beta}$ are their duals ($\varepsilon^{0123}=1$). 
$f \in \{ e, u, d \}$ represents a SM fermionic field and the indices $i$ and $j$ denote its generation.

The interactions in Eq.~(\ref{eq:LagA}) are manifestly invariant under the $\phi$ shift symmetry (up to non-perturbative effects) since $F\tilde{F}$ and $G\tilde{G}$ are total derivatives. Moreover, pseudoscalar interactions could be written in a shift-symmetric way through the dimension-5 operator 
$\frac{\partial_\mu \phi}{\Lambda}\bar f \gamma^\mu\gamma^5 f$ after applying the equations of motion and integrating by parts. This would justify the $v/\Lambda$ normalization factor~\cite{DiLuzio:2020wdo}. 
Instead, the interactions in Eq.~(\ref{eq:LagB}) break the shift symmetry explicitly. 
Since in the unbroken phase of the SM scalar interactions should be written through the dimension-5 operator $\phi H \bar f_L f_R + \mathrm{h.c.}$ being $H$ the SM Higgs doublet, it would be natural to introduce the normalization factor $v/\Lambda$ in the last term of Eq.~(\ref{eq:Lag1})~\cite{DiLuzio:2020wdo}. 
Moreover, in Eq.~(\ref{eq:Lag1}) we do not factor out the gauge couplings $e^2$ and $g^2_s$ from the coefficients $\tilde{\mathcal C}_{\gamma,g}$ and $\mathcal C_{\gamma,g}$ which would make them scale invariant at one-loop order.

Covariant derivatives are defined according to 
\begin{equation}
\label{eq:CovDev}
D_\mu f_i  = \left( \partial_\mu - i e Q_f A_\mu - i g_s c_f  G_\mu^a  T^a \right) f_i\,.
\end{equation}

The Lagrangian \eqref{eq:Lag1} necessarily violates the CP symmetry regardless of the scalar or pseudoscalar nature of the ALP field $\phi$, as the two pieces \eqref{eq:LagA} and \eqref{eq:LagB} possess opposite CP transformation properties.
The simultaneous presence of these groups of operators results in an extremely rich and interesting phenomenology, and contributions to the Electric Dipole Moments (EDMs) of particles, nucleons, atoms and molecules are generated either via tree- or loop-level exchanges of ALPs~\cite{Marciano:2016yhf,DiLuzio:2020oah}.
Besides such CP-violating effects one has then of course CP-preserving contributions to other low-energy observables, among which are, for instance, the Magnetic Dipole Moments (MDMs) of either elementary or composite particles.

The largest part of these effects are generated at loop-level and their leading contribution can be estimated by considering the running of the corresponding Wilson coefficient from the high-energy cutoff scale $\Lambda$ down to the energy scale at which experiments are performed. 

Running effects are encoded in a set of possibly coupled differential equations, the Renormalization Group Equations (RGEs), which can be schematically written as 
\begin{equation}
\label{eq:AnomalousDimension}
    \mu  \dv[]{c_i}{\mu} = \gamma_{i \leftarrow j} c_{j}\,,
\end{equation}
where the $c_i$ are the Wilson coefficients associated to local, 
gauge-invariant operators $\mathcal{O}_i(x)$ and $\gamma_{i \leftarrow j}$ is the anomalous dimension matrix regulating the energy evolution 
of $c_i$ at the desired perturbative order. 

Since the CP properties of the operators of Eq.~\eqref{eq:Lag1} are left unchanged along the renormalization group flow,
$\gamma_{i \leftarrow j}$ takes a block-diagonal form in the two distinct CP sectors:
\begin{align}
\mu\dv[]{}{\mu}
    \begin{pmatrix}
        \mathcal Y_S^{ij} \\ \mathcal C_g \\  \mathcal C_\gamma 
    \end{pmatrix}
    &=
    \begin{pmatrix}
        \gamma_{S_{ij} \leftarrow g} & \gamma_{S_{ij} \leftarrow \gamma} & \gamma_{S_{ij} \leftarrow S_{kl}}\\
        \gamma_{g \leftarrow g} & \gamma_{g \leftarrow \gamma} & \gamma_{g \leftarrow S_{kl}} \\
        \gamma_{\gamma \leftarrow g} & \gamma_{\gamma \leftarrow \gamma} & \gamma_{\gamma \leftarrow S_{kl}} 
    \end{pmatrix}
    \begin{pmatrix}
        \mathcal Y_S^{kl}\\ \mathcal C_g \\ \mathcal C_\gamma 
    \end{pmatrix} \,,
    \\
    \mu\dv[]{}{\mu}
    \begin{pmatrix}
         \mathcal Y_P^{ij}\\\tilde{\mathcal C}_g \\ \tilde{\mathcal C}_\gamma 
    \end{pmatrix}
    &=
    \begin{pmatrix}
        \gamma_{P_{ij} \leftarrow \tilde g} & \gamma_{P_{ij} \leftarrow \tilde \gamma} & \gamma_{P_{ij} \leftarrow P_{kl}}\\
        \gamma_{\tilde g \leftarrow \tilde g} & \gamma_{\tilde g \leftarrow \tilde \gamma} & \gamma_{\tilde g \leftarrow P_{kl}} \\
        \gamma_{\tilde \gamma \leftarrow \tilde g} & \gamma_{\tilde\gamma \leftarrow \tilde \gamma} & \gamma_{\tilde \gamma \leftarrow P_{kl}} 
    \end{pmatrix}
    \begin{pmatrix}\mathcal Y_P^{kl}\\
\tilde{\mathcal C}_g \\ \tilde{\mathcal C}_\gamma 
\end{pmatrix}\,.
\end{align}
\section{\label{sec:uvanomalousdim}
Ultraviolet anomalous dimensions}

Hereafter, we detail the computation of the ultraviolet anomalous dimensions relevant to ALP effective field theories through the on-shell method.
Since the master equation \eqref{eq:masterformula}
is only sensitive to the difference between the ultraviolet and infrared anomalous dimensions, the knowledge of the latter is required to obtain the UV anomalous dimension.
We report the computation of the IR anomalous dimensions in Appendix~\ref{app:IRanomalousdimensions}.

\subsection{Renormalization of ALP couplings}

We first analyze the renormalization of the ALP couplings of Eq.~\eqref{eq:Lag1}. 

\subsubsection{$\phi \bar f f$ and $\phi \bar f i \gamma_5 f$}
\label{section_ff}

\begin{figure}[htbp]
\centering
\includegraphics[width=\textwidth,page=14]{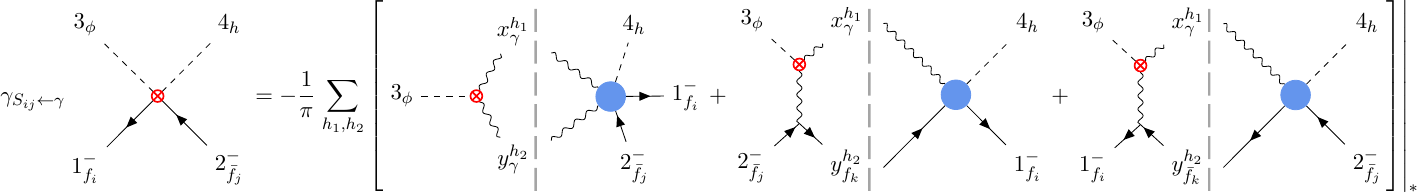}
\caption{Diagrammatic formula for computing $\gamma_{S_{ij}\leftarrow\gamma}$.}
\label{fig:mass_insertion}
\end{figure}

\paragraph{$\gamma_{S\leftarrow\gamma}$.}
The calculation of this anomalous dimension requires a fermion mass insertion, as can be inferred by dimensional analysis.
This can be achieved by renormalizing the operator
\begin{equation}
    \mathcal O_{hS_{ij}} = \frac{h}{v}\phi \bar f_i f_j
\end{equation}
instead of $\mathcal O_{S_{ij}}$ and by adding the Yukawa interaction $-y_ih\bar f_i f_i$ at the level of the lowest order Lagrangian.
Then, the master formula reads
\begin{equation}\label{eq:masterformulamassinsertion}
    \gamma_{S_{ij}\leftarrow\gamma} F_{hS_{ij}}|_*(1_{f_i}^-,2_{\bar f_j}^-,3_\phi,4_h) = -\frac{1}{\pi}(\mathcal M F_\gamma)|_*(1_{f_i}^-,2_{\bar f_j}^-,3_\phi,4_h)\,,
\end{equation}
whose diagram is shown in Fig.~\ref{fig:mass_insertion}.
%
%
On the left-hand side, the minimal form factor corresponding to $\mathcal O_{hS_{ij}}$ simply reads
\begin{equation}
    F_{hS_{ij}}|_*(1_{f_i}^-,2_{\bar f_j}^-,3_\phi,4_h) = \frac{1}{v}\agl{1}{2}\,,
\end{equation}
while, on the right-hand side, the convolution $(\mathcal M F_\gamma)|_*$ can be expanded as follows, taking into account all possible propagating states
\begin{align}\label{eq:M*F_gamma}
    (\mathcal M F_\gamma)|_*(1_{f_i}^-,2_{\bar f_j}^-,3_\phi,4_h) &= \sum_{h_1,h_2}\int d\text{LIPS}_2\,\bigg[
    \mathcal M|_*(1_{f_i}^-,2_{\bar f_j}^-,4_h;x_\gamma^{h_1},y_\gamma^{h_2}) F_\gamma|_*(x_\gamma^{h_1},y_\gamma^{h_2},3_\phi)\nonumber \\
    &\quad+\mathcal M|_*(1_{ f_i}^-,4_h;x_\gamma^{h_1},y_{ f_k}^{h_2}) F_\gamma|_*(x_\gamma^{h_1},y_{ f_k}^{h_2},2_{\bar f_j}^-,3_\phi)\nonumber\\
    &\quad+\mathcal M|_*(2_{\bar f_j}^-,4_h;x_\gamma^{h_1},y_{\bar f_k}^{h_2}) F_\gamma|_*(x_\gamma^{h_1},y_{\bar f_k}^{h_2},1_{f_i}^-,3_\phi)
    \bigg]\nonumber \\
    &= \int d\text{LIPS}_2\,(g_1+g_2+g_3) \,.
\end{align}

We can begin by noticing that we can neglect the first contribution to Eq.~\eqref{eq:M*F_gamma}.
In fact, since the form factor on the left-hand side of Eq.~\eqref{eq:M*F_gamma} involves more than three particles, it survives in the limit where we send to zero the off-shell momentum $q$ injected by the operator.
Therefore, we are allowed to set $q=0$ on both sides of the equation and to work fully on-shell.
This in turn implies that any form factor on the right-hand side involving less than four particles cannot contribute, since it vanishes if all the particles are massless and on-shell.
Therefore,
\begin{equation}
    g_1=\sum_{h_1,h_2}\mathcal M|_*(1_{f_i}^-,2_{\bar f_j}^-,4_h;x_\gamma^{h_1},y_\gamma^{h_2}) F_\gamma|_*(x_\gamma^{h_1},y_\gamma^{h_2},3_\phi)=0\,.
\end{equation}

Regarding the second contribution to Eq.~\eqref{eq:M*F_gamma}, it is given by the convolution between a non-minimal form factor and a four-point amplitude.
We can define their product as
\begin{equation}
    g_2=\sum_{h_1,h_2} \mathcal M|_*(1_{ f_i}^-,4_h;x_\gamma^{h_1},y_{ f_k}^{h_2}) F_\gamma|_*(x_\gamma^{h_1},y_{ f_k}^{h_2},2_{\bar f_j}^-,3_\phi)\,.
\end{equation}
The only helicity configuration that gives a non-zero result is $(h_1,h_2)=(-,+)$.
Indeed, $h_2$ must be the opposite of the helicity of the particle $2^-_{\bar f_j}$ as a consequence of the gauge interaction,
while, if $h_1=+$, the amplitude vanishes as can be inferred from the helicity selection rules~\cite{Grisaru:1977px,Grisaru:1976vm,Mangano:1990by,Cheung:2015aba,Parke:1985pn,Azatov:2016sqh}.

For the computation of $F_\gamma|_*(x_\gamma^{-},y_{ f_k}^{+},2_{\bar f_j}^-,3_\phi)$, 
one can exploit unitarity and locality in the form of the factorization of tree-level amplitudes~\cite{Benincasa:2007xk,Arkani-Hamed:2017jhn,Mangano:1990by,Dixon:1996wi},\footnote{Alternatively, one can use BCFW recurrence relations~\cite{Britto:2004ap,Britto:2005fq}.} which, in general, reads
\begin{equation}\label{eq:factorization}
    \mathcal M(1,\dotsc,n) \sim -\frac{1}{s_{1\dots m}}\sum_h\mathcal M(1,\dotsc,m; \ell^h) \mathcal M( \ell^h,m+1,\dotsc,n) 
\end{equation}
as $s_{1\dots m}=(p_1+\dotsb+ p_m)^2\to 0$, and relates the residues of higher-point amplitudes to products of lower-point ones.
Since $F_\gamma|_*(x_\gamma^{-},y_{ f_k}^{+},2_{\bar f_j}^-,3_\phi)$ has a single simple pole at $s_{2y}=0$ corresponding to the propagation of a virtual photon, we can exploit Eq.~\eqref{eq:factorization} to write it as
\begin{align}
    F_\gamma|_*(x_\gamma^{-},y_{ f_k}^{+},2_{\bar f_j}^-,3_\phi) &= -\frac{1}{s_{2y}}\sum_h F_\gamma|_*(x_\gamma^{-},3_\phi;\ell^h_\gamma)\mathcal M|_*(\ell^h_\gamma,y_{ f_k}^{+},2_{\bar f_j}^-)\nonumber \\
    &= -\frac{1}{\agl{2}{y}\sqr{y}{2}} F_\gamma|_*(x_\gamma^{-},3_\phi;\ell^+_\gamma)\mathcal M|_*(\ell^+_\gamma,y_{ f_k}^{+},2_{\bar f_j}^-)
\end{align}
and by using
\begin{align}
    F_\gamma|_*(x_\gamma^{-},3_\phi;\ell^+_\gamma) = \frac{2}{\Lambda}\agl{x}{\ell}^2\,, && 
    \mathcal M|_*(\ell^+_\gamma,y_{ f_k}^{+},2_{\bar f_j}^-) =- \sqrt 2 e Q_f \delta^{kj}\frac{\sqr{\ell}{y}^2}{\sqr{y}{2}}\,,
\end{align}
as well as $\agl{x}{\ell}\sqr{\ell}{y}=-\agl{x}{2}\sqr{2}{y}$, we can conclude that
\begin{equation}
    F_\gamma|_*(x_\gamma^{-},y_{ f_k}^{+},2_{\bar f_j}^-,3_\phi) = \frac{2\sqrt 2}{\Lambda}eQ_f \delta^{kj} \frac{\agl{2}{x}^2}{\agl{2}{y}}\,.
\end{equation}
Similar arguments can be applied to $\mathcal M|_*(1_{ f_i}^-,4_h;x_\gamma^{-},y_{ f_k}^{+})$ to find
\begin{equation}
    \mathcal M|_*(1_{ f_i}^-,4_h;x_\gamma^{-},y_{ f_k}^{+}) = \sqrt 2 e Q_f y_i \delta^{ik} \frac{\agl{1}{y}^2}{\agl{1}{x}\agl{x}{y}}\,,
\end{equation}
which leads to
\begin{equation}\label{eq:exprB}
    g_2= \frac{4}{\Lambda}e^2 Q_f^2 y_i\delta^{ij}\frac{\agl{1}{y}^2\agl{2}{x}^2}{\agl{1}{x}\agl{2}{y}\agl{x}{y}}\,.
\end{equation}

\subparagraph{Angular integration.}
In this case, it is convenient to exploit the hybrid parameterization of
Eq.~(\ref{eq:hybbrid}). Thus, we obtain
\begin{equation}
    g_2(z,t) = \frac{4}{\Lambda}e^2 Q_f^2 y_i\delta^{ij}\agl{1}{2}\frac{(r+tz)^2}{rt(1+t^2)(rt-z)}\,,
\end{equation}
where
\begin{equation}
    r = \frac{\agl{1}{2}}{\agl{2}{4}}\,.
\end{equation}
The contour integral over the unit circle in the complex plane can be computed by means of Cauchy's residue theorem
\begin{align}
    I_{g_2}(t) &= \oint_{|z|=1}\frac{dz}{2\pi i z}g_2(z,t)\nonumber \\
    &= \text{Res}_{z=0}\frac{g_2(z,t)}{z} + \Theta(1-|r|t)\text{Res}_{z=rt}\frac{g_2(z,t)}{z}\nonumber \\
    &= -\frac{4}{\Lambda}e^2 Q_f^2 y_i\delta^{ij}\agl{1}{2} \frac{-1+(1+t^2)^2 \Theta(1-|r|t)}{t^2(1+t^2)}\,,
\end{align}
where $\Theta$ denotes the Heaviside step function.
The remaining integral then leads to
\begin{align}
    \int d\text{LIPS}_2\, g_2 &= \frac{1}{8\pi}\int_0^\infty \frac{2t\,dt}{(1+t^2)^2}I_{g_2}(t)\nonumber \\
    &= \frac{e^2 Q_f^2}{4\pi\Lambda}y_i\delta^{ij}\agl{1}{2}\big[-3+2\log(1+|r|^2)\big]\nonumber  \\
    &= \frac{e^2 Q_f^2}{4\pi\Lambda}y_i\delta^{ij}\agl{1}{2}\bigg[-3+2\log\frac{s_{12}+s_{24}}{s_{24}}\bigg]\,,
\end{align}
since $|r|^2=s_{12}/s_{24}$.
Eventually, the third contribution to Eq.~\eqref{eq:M*F_gamma}
\begin{equation}
    g_3 = \sum_{h_1,h_2} \mathcal M|_*(2_{\bar f_j}^-,4_h;x_\gamma^{h_1},y_{\bar f_k}^{h_2}) F_\gamma|_*(x_\gamma^{h_1},y_{\bar f_k}^{h_2},1_{f_i}^-,3_\phi)
\end{equation}
can be simply related to $g_2$ by exchanging the external fermions labeled by $1$ and $2$ and adding a minus sign due to fermion reordering
\begin{align}
    \int d\text{LIPS}_2\,g_3 &= -\int d\text{LIPS}_2\,g_2\big|_{1\leftrightarrow 2}\nonumber\\
    & = \frac{e^2 Q_f^2}{4\pi\Lambda}y_i\delta^{ij}\agl{1}{2}\bigg[-3+2\log\frac{s_{14}+s_{12}}{s_{14}}\bigg]\,,
    \label{eq:g3_reorder}
\end{align}
yielding
\begin{align}
    (\mathcal M F_\gamma)|_*(1_{f_i}^-,2_{\bar f_j}^-,3_\phi,4_h) &= \int d\text{LIPS}_2\,(g_1+g_2+g_3) 
    \nonumber \\ &= -\frac{e^2Q_f^2}{2\pi\Lambda}y_i \delta^{ij} \agl{1}{2}\bigg[3-\log\frac{(s_{14}+s_{12})(s_{24}+s_{12})}{s_{14} s_{24}}\bigg]\nonumber \\
    &= -\frac{3e^2Q_f^2}{2\pi\Lambda}y_i \delta^{ij} \agl{1}{2}\,,
\end{align}
where $(s_{14}+s_{12})(s_{24}+s_{12})=s_{14}s_{24}$ follows from the on-shell condition $s_{14}+s_{24}+s_{12}=0$.
Thus, we have explicitly checked that only rational terms in the kinematic variables survive when we add all the contributions, as should be.

\subparagraph{Stokes integration.}
The calculation of this anomalous dimension is greatly simplified by the application of the Stokes theorem, which is exploited as follows.
Starting from the expression for $g_2$ in Eq.~\eqref{eq:exprB}, we can parameterize the internal helicity spinors $\lambda_x$ and $\lambda_y$ in terms of $\lambda_1$ and $\lambda_4$ as in Eq.~\eqref{eq:stokes_par},
leading to 
\begin{equation}
    g_2(z,\zb) = \frac{4}{\Lambda}e^2 Q_f^2 y_i\delta^{ij}\frac{(\agl{1}{2}-\zb \agl{2}{4})^2}{\zb (1+z\zb )(z\agl{1}{2}+\agl{2}{4})}\,.
\end{equation}
The rational part of its indefinite integral in the variable $\zb$, with the appropriate integration measure, is given by
\begin{align}
  G_{2\,\text{rat}}(z,\zb) 
  &= \text{Rat}  \int d \zb\, \frac{g_2(z,\zb)}{(1+z\zb)^2}  \nonumber \\
  &= \frac{2}{\Lambda}e^2 Q_f^2 y_i\delta^{ij} \frac{z(3+2z\zb)\agl{1}{2}-(1+2z\zb)\agl{2}{4}}{z^2(1+z\zb)^2}\, . 
\end{align}
Following the prescription outlined in Section~\ref{sec:phasespace_Stokes}, all logarithmic terms have been neglected as they are not related to the coefficient of the 2-point function.
Exploiting Cauchy's residue theorem, the integration in the $z$ variable localizes around the pole $z_0=0$ as
\begin{equation}
    \Res_{(z,\zb) = (0,0)}G_{2\,\text{rat}}(z,\zb) =  \frac{6}{\Lambda}e^2 Q_f^2 y_i\delta^{ij}\agl{1}{2} \, , 
\end{equation}
giving
\begin{equation}
   {\rm Rat}\int d\mathrm{LIPS}_2\, g_2 
   =  -\frac{3 e^2 Q_f^2}{4\pi\Lambda}y_i\delta^{ij}\agl{1}{2} \,.
   \label{eq:g2_contr}
\end{equation}
As explained in Eq.~\eqref{eq:g3_reorder}, the third contribution $g_3$ can be obtained from $g_2$, giving:
\begin{equation}
{\rm Rat}\int d\text{LIPS}_2\,(g_1+g_2+g_3) =  -\frac{3 e^2 Q_f^2}{2\pi\Lambda}y_i\delta^{ij}\agl{1}{2} \,.
   \label{eq:g3_contr}
\end{equation}

Finally, from Eq.~\eqref{eq:masterformulamassinsertion}, the final result reads
\begin{equation}
\gamma_{S_{ij}\leftarrow\gamma} = \frac{3e^2Q_f^2}{2\pi^2}\frac{m_i}{\Lambda}\delta^{ij}\,,
\label{eq:result_phiffb}
\end{equation}
since $m_i=vy_i$.

\paragraph{$\gamma_{S\leftarrow g}$.}
The calculation is completely analogous to the one just performed 
for $\gamma_{S\leftarrow \gamma}$.
The result is the same, provided that we substitute $e^2Q_f^2$ with $C_Fg_s^2c_f^2$.
\begin{equation}
    \gamma_{S_{ij}\leftarrow g} = C_F\frac{3g_s^2c_f^2}{2\pi^2}\frac{m_i}{\Lambda}\delta^{ij}\,.
\end{equation}

\begin{figure}[htbp]
\centering
\includegraphics[page=8]{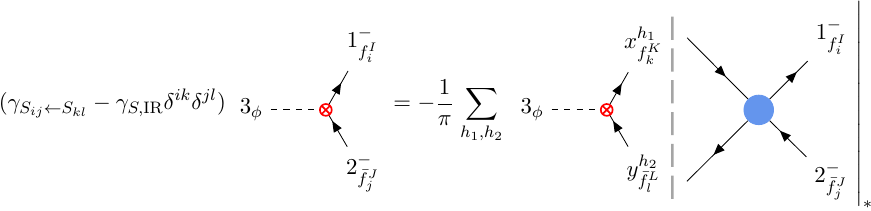}
\caption{Diagrammatic formula for computing $\gamma_{S_{ij}\leftarrow S_{kl}}$.}
\label{fig:gammaSS}
\end{figure}

\paragraph{$\gamma_{S\leftarrow S}$.}
The derivation of the diagonal element $\gamma_{S_{ij}\leftarrow S_{kl}}$ is more subtle since it requires the knowledge of the infrared anomalous dimension $\gamma_{S,\text{IR}}$, calculated in Appendix~\ref{sec:IRphiffb}.\footnote{
We observe that $F_{S}$ is nonvanishing only if the fermions have the same helicity, while $\gamma_{\text{S,IR}}$ does not depend on the helicities, in general. However, in Appendix~\ref{sec:IRphiffb}, $\gamma_{S,\text{IR}}$ is computed with the energy-momentum tensor, and, in this case, choosing opposite-helicity fermions is the only option, because otherwise $F_T=0$.} 
The master formula reads
\begin{equation}
    (\gamma_{S_{ij}\leftarrow S_{kl}}-\gamma_{S,\text{IR}}\delta^{ik}\delta^{jl})F_{S_{ij}}|_*(1_{f_i^I}^-,2_{\bar f_j^J}^-,3_\phi) = -\frac{1}{\pi}(\mathcal M F_{S_{kl}})|_*(1_{f_i^I}^-,2_{\bar f_j^J}^-,3_\phi)
\end{equation}
as represented in Fig.~\ref{fig:gammaSS}.

The form factor associated with the Yukawa operator reads
\begin{equation}
    F_{S_{ij}}|_*(1_{f_i^I}^-,2_{\bar f_j^J}^-,3_\phi) = \delta_{IJ}\agl{1}{2}\,,
\end{equation}
while the convolution takes the form
\begin{equation}
    (\mathcal M F_{S_{kl}})|_*(1_{f_i^I}^-,2_{\bar f_j^J}^-,3_\phi) = \sum_{h_1,h_2}\int d\text{LIPS}_2\, 
    \mathcal M|_*(1_{f_i^I}^-,2_{\bar f_j^J}^-;x^{h_1}_{f_k^K},y^{h_2}_{\bar f_l^L})
    F_{S_{kl}}|_*(x^{h_1}_{f_k^K},y^{h_2}_{\bar f_l^L},3_\phi)\,.
\end{equation}
Out of these four contributions, the only one that does not vanish is given by the configuration $(h_1,h_2)=(-,-)$, where the amplitude
\begin{equation}
    \mathcal M|_*(1_{f_i^I}^-,2_{\bar f_j^J}^-;x^{-}_{f_k^K},y^{-}_{\bar f_l^L}) \delta_{KL}= -2(e^2Q_f^2+C_Fg_s^2c_f^2)\delta_{IJ}\delta^{ik}\delta^{jl}\frac{\agl{1}{2}\sqr{x}{y}}{\agl{1}{x}\sqr{x}{1}}
\end{equation}
is multiplied by
\begin{equation}
    F_{S_{kl}}|_*(x^{-}_{f_k^K},y^{-}_{\bar f_l^L},3_\phi) = \delta_{KL}\agl{x}{y}\,.
\end{equation}

\subparagraph{Angular integration.}
By making use of the angular parameterization of the phase space, the anomalous dimension is then given by
\begin{align}
    \gamma_{S_{ij}\leftarrow S_{kl}}&=\left[\gamma_{S,\text{IR}}-\frac{1}{4\pi^2}(e^2Q_f^2+C_Fg_s^2c_f^2)\int_0^{\pi/2}2\sin\theta\cos\theta \,d\theta\,\frac{1}{\sin^2\theta}\right]\delta^{ik}\delta^{jl}\nonumber \\
    &= \frac{1}{4\pi^2}(e^2Q_f^2+C_Fg_s^2c_f^2)\delta^{ik}\delta^{jl} 
    \int_0^{\pi/2}2\sin\theta\cos\theta \,d\theta\,\bigg[\frac{\cos^4\theta}{\sin^2\theta}[-1+2\cos(2\theta)]\nonumber\\&\quad +2(\cos^4\theta + \sin^4\theta)-\frac{1}{\sin^2\theta}\bigg]\nonumber \\
    &= -\frac{3}{8\pi^2}(e^2Q_f^2+C_Fg_s^2c_f^2)\delta^{ik}\delta^{jl} \,,
\end{align}
where we exploited the expression for $\gamma_{S,\text{IR}}$ provided in Eq.~\eqref{eq:gammaS_IR_angular}.

\subparagraph{Stokes integration.}
By making use of the Stokes parameterization of the phase space instead,
the integrand reads as
\begin{align}
    g(z,\zb) &=\mathcal M|_*(1_{f_i^I}^-,2_{\bar f_j^J}^-;x^{-}_{f_k^K},y^{-}_{\bar f_l^L})F_{S_{kl}}|_*(x^{-}_{f_k^K},y^{-}_{\bar f_l^L},3_\phi)
    \nonumber \\
    &=2 (e^2Q_f^2+C_Fg_s^2c_f^2)\delta_{IJ}\delta^{ik}\delta^{jl}\agl{1}{2} \frac{1+z \zb}{z \zb} 
\end{align}
and leads to an integral whose rational component is vanishing:
\begin{equation}
    \int d\zb \,\frac{g(z,\zb)}{(1+z\zb)^2} = 2 (e^2Q_f^2+C_Fg_s^2c_f^2)\delta_{IJ}\delta^{ik}\delta^{jl}\agl{1}{2} \frac{\log(\zb)-\log(1+z\zb)}{z}\,.
\end{equation}
This implies that
\begin{align}
    \gamma_{S_{ij}\leftarrow S_{kl}}&= \gamma_{S,\text{IR}} = -\frac{3}{8\pi^2}(e^2Q_f^2+C_Fg_s^2c_f^2)\delta^{ik}\delta^{jl} \,,
\end{align}
where we used the expression for $\gamma_{S,\text{IR}}$ reported in Eq.~\eqref{eq:gammaS_IR_stokes}.

\paragraph{$\gamma_{P\leftarrow \tilde \gamma}$, $\gamma_{P \leftarrow \tilde g}$, and $\gamma_{P\leftarrow P}$.}
\label{par:Phiff}
Regarding the operator $\phi \bar f_i i \gamma_5 f_j$, its anomalous dimensions can be directly obtained from those we have just calculated for $\phi \bar f_i  f_j$.
In fact, the amplitudes involved are the same and the only quantities that change are the form factors, which satisfy the identities
\begin{align}
    F_{P_{ij}}|_*(1_{f_i}^-,2_{\bar f_j}^-,3_\phi) &= -i F_{S_{ij}}|_*(1_{f_i}^-,2_{\bar f_j}^-,3_\phi)\,,\\
    F_{\tilde\gamma}|_* (1^-_\gamma,2^-_\gamma,3_\phi)&=iF_\gamma|_* (1^-_\gamma,2^-_\gamma,3_\phi)\,,\\
    F_{\tilde g}|_* (1^-_{g^a},2^-_{g^b},3_\phi)&=iF_g|_* (1^-_{g^a},2^-_{g^b},3_\phi)\,.
\end{align}
The first one can be understood from the fact that a single particle fermion state with helicity $\pm 1/2$ is an eigenvector of $\gamma_5$ with eigenvalue $\pm 1$.
Instead, the latter ones arise from the field-strength tensor and its dual which can be expressed as
\begin{align}
\label{eq:Obs1}
    F_{\mu\nu}=F^-_{\mu\nu}+F^+_{\mu\nu} \,, && 
    \tilde F_{\mu\nu}=i\left(F^-_{\mu\nu}-F^+_{\mu\nu}\right)\,,
\end{align}
where 
$F^{+}_{\mu\nu}$ and $F^{-}_{\mu\nu}$ are self-dual 
and anti-self-dual tensors, respectively, which read
\begin{equation}
    F^{\pm}_{\mu\nu} = \pm \frac{i}{2}\varepsilon_{\mu\nu\rho\sigma}F^{\pm \, \rho\sigma}\,,
\end{equation}
and create single particle photon states with helicity $\pm 1$.
Based on these observations, we can therefore infer that
\begin{align}
    \gamma_{P_{ij}\leftarrow \tilde\gamma} &= - \gamma_{S_{ij}\leftarrow \gamma} = - \frac{3e^2Q_f^2}{2\pi^2}\frac{m_i}{\Lambda}\delta^{ij} \,,\\
    \gamma_{P_{ij}\leftarrow \tilde g} &= - \gamma_{S_{ij}\leftarrow g} = - C_F\frac{3g_s^2c_f^2}{2\pi^2}\frac{m_i}{\Lambda}\delta^{ij} \,,\\
    \gamma_{P_{ij}\leftarrow P_{kl}} &= \gamma_{S_{ij}\leftarrow S_{kl}} = -\frac{3}{8\pi^2}(e^2Q_f^2+C_Fg_s^2c_f^2)\delta^{ik}\delta^{jl} \,.
\end{align}

\subsubsection{$\phi FF$ and $\phi F\tilde F$}
\label{section_FFtilde}
\begin{figure}[htbp]
\centering
\includegraphics[page=1]{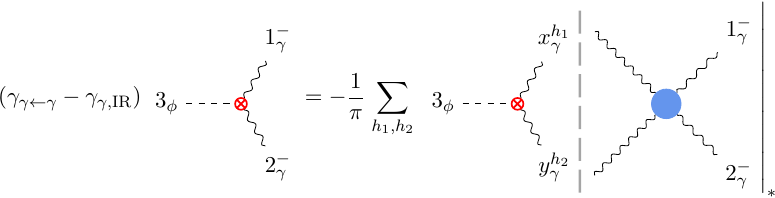}
\caption{Diagrammatic formula for computing $\gamma_{\gamma \leftarrow \gamma}$.}
\label{fig:gamma_gammagamma}
\end{figure}

\paragraph{$\gamma_{\gamma \leftarrow \gamma}$.}
The diagonal matrix element $\gamma_{\gamma \leftarrow \gamma}$ accounting for the multiplicative renormalization of the ALP effective operator $\phi FF$ is calculated with the master formula
\begin{equation}
    (\gamma_{\gamma \leftarrow \gamma} - \gamma_{\gamma,\text{IR}})F_\gamma|_* (1^-_\gamma,2^-_\gamma,3_\phi)= -\frac{1}{\pi}(\mathcal M F_\gamma)|_*(1^-_\gamma,2^-_\gamma,3_\phi)
\end{equation}
represented in Fig.~\ref{fig:gamma_gammagamma}.

The form factor on the left reads
\begin{equation}
    F_\gamma|_*(1^-_\gamma,2^-_\gamma,3_\phi) = -2\frac{\agl{1}{2}^2}{\Lambda}\,,
\end{equation}
while the convolution on the right is expanded as
\begin{equation}
    (\mathcal M F_\gamma)|_*(1^-_\gamma,2^-_\gamma,3_\phi) = \sum_{h_1,h_2}\int d\text{LIPS}_2\, \mathcal M|_*(1^-_\gamma,2^-_\gamma;x_\gamma^{h_1},y^{h_2}_\gamma)F_\gamma|_*(x_\gamma^{h_1},y^{h_2}_\gamma,3_\phi)\,.
\end{equation}
Since the 4-photon tree amplitude trivially vanishes for any choice of the helicities
\begin{equation}
    \mathcal M|_*(1^-_\gamma,2^-_\gamma;x_\gamma^{h_1},y^{h_2}_\gamma) = 0\,,
\end{equation}
we obtain
\begin{equation}
    \gamma_{\gamma \leftarrow \gamma} = \gamma_{\gamma,\text{IR}} = \frac{e^2}{6\pi^2} \sum_f Q_f^2\,,
\end{equation}
where we exploited the expression for $\gamma_{\gamma,\text{IR}}$ derived in Appendix~\ref{sec:IRphigammagamma}.
Here $f$ runs over all the fermions of the theory.
We can notice that we have successfully derived an expression that is equal to the anomalous dimension of $e^2$, namely $(\mu/e^2) \text{d}e^2/\text{d}\mu$.

\begin{figure}[htbp]
\centering
\includegraphics[page=2]{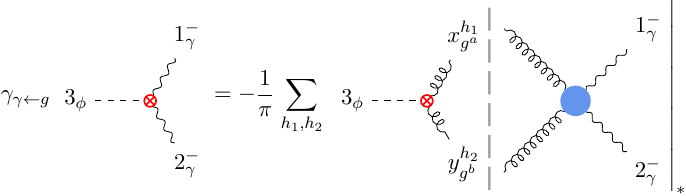}
\caption{Diagrammatic formula for computing $\gamma_{\gamma\leftarrow g}$.}
\label{fig:gamma_gammag}
\end{figure}

\paragraph{$\gamma_{\gamma\leftarrow g}$.}
The master formula associated with this matrix element reads
\begin{equation}
    \gamma_{\gamma\leftarrow g} F_\gamma|_*(1^-_\gamma,2^-_\gamma,3_\phi) = -\frac{1}{\pi}(\mathcal M F_g)|_*(1^-_\gamma,2^-_\gamma,3_\phi)
\end{equation}
and is represented in Fig.~\ref{fig:gamma_gammag}.

Also in this case the convolution on the right,
\begin{equation}
    (\mathcal M F_g)|_*(1^-_\gamma,2^-_\gamma,3_\phi) = \sum_{h_1,h_2}\int d\text{LIPS}_2\, \mathcal M|_*(1^-_\gamma,2^-_\gamma;x_{g^a}^{h_1},y^{h_2}_{g^b})F_g|_*(x_{g^a}^{h_1},y^{h_2}_{g^b},3_\phi)\,,
\end{equation}
vanishes since
\begin{equation}
    \mathcal M|_*(1^-_\gamma,2^-_\gamma;x_{g^a}^{h_1},y^{h_2}_{g^b}) = 0
\end{equation}
and leads to
\begin{equation}
    \gamma_{\gamma\leftarrow g} = 0\,.
\end{equation}

\begin{figure}[h!tbp]
\centering
\includegraphics[page=3]{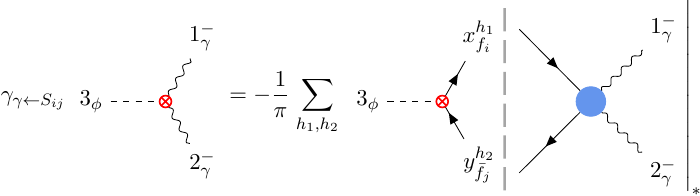}
\caption{Diagrammatic formula for computing $\gamma_{\gamma\leftarrow S_{ij}}$.}
\label{fig:gamma_gammaS}
\end{figure}

\paragraph{$\gamma_{\gamma\leftarrow S}$.}
The equation linked to this matrix element is given by
\begin{equation}
    \gamma_{\gamma\leftarrow S_{ij}} F_\gamma|_*(1^-_\gamma,2^-_\gamma,3_\phi) = -\frac{1}{\pi}(\mathcal M F_{S_{ij}})|_*(1^-_\gamma,2^-_\gamma,3_\phi)
\end{equation}
and is represented as in Fig.~\ref{fig:gamma_gammaS}.

From dimensional analysis, we expect
\begin{equation}
    \gamma_{\gamma\leftarrow S_{ij}}=0
\end{equation}
since $[\mathcal O_\gamma]-[\mathcal O_{S_{ij}}]=1$, which is in particular greater than $0$.
This is indeed consistent with the fact that the convolution
\begin{equation}
    (\mathcal M F_{S_{ij}})|_*(1^-_\gamma,2^-_\gamma,3_\phi) = \sum_{h_1,h_2}\int d\text{LIPS}_2\, \mathcal M|_*(1^-_\gamma,2^-_\gamma;x_{f_i}^{h_1},y^{h_2}_{\bar f_j})F_{S_{ij}}|_*(x_{f_i}^{h_1},y^{h_2}_{\bar f_j},3_\phi)
\end{equation}
is vanishing due to
\begin{equation}
    \mathcal M|_*(1^-_\gamma,2^-_\gamma;x_{f_i}^{h_1},y^{h_2}_{\bar f_j})=0
\end{equation}
for any choice of $h_1$ and $h_2$.

\paragraph{$\gamma_{\tilde\gamma \leftarrow \tilde\gamma}$, $\gamma_{\tilde\gamma \leftarrow \tilde g}$, and $\gamma_{\tilde\gamma \leftarrow P}$.}
\label{par:PhiFF}
The anomalous dimensions that contribute to the renormalization of $\phi F\tilde F$ are equal to those of $\phi FF$ up to signs that can be determined through the comments leading to Eq.~\eqref{eq:Obs1}.
\begin{align}
    \gamma_{\tilde\gamma \leftarrow \tilde\gamma} &= \gamma_{\gamma \leftarrow \gamma} = \frac{e^2}{6\pi^2}\sum_f Q_f^2 \,,\\
    \gamma_{\tilde\gamma \leftarrow \tilde g} &= \gamma_{\gamma \leftarrow  g} = 0\,,\\
    \gamma_{\tilde\gamma \leftarrow P_{ij}} &= -\gamma_{\gamma \leftarrow S_{ij}} = 0\,.
\end{align}

An interesting feature of the on-shell, unitarity-based method we are employing is that it makes some properties of the anomalous dimension matrix manifest. This is precisely the case for the operator $\phi FF$. Based on pure symmetry arguments, indeed, 
one would expect these operators to renormalize like the QED fine structure constant at one-loop level. The reason for this resides in the fact that the ALP is a pure SM gauge singlet, and hence $\phi FF$ is expected to renormalize as $FF$. In turn, as a consequence of Ward's identities, this equals the renormalization of $\alpha_{\text{em}}$. Such a property is however not manifestly apparent at the level of Feynman diagrams.
On the other hand, this property is immediately retrieved within the scope of the on-shell method.

\subsubsection{$\phi GG$ and $\phi G\tilde G$}
\label{sec:phiGG}

\begin{figure}[htbp]
\centering
\includegraphics[page=5]{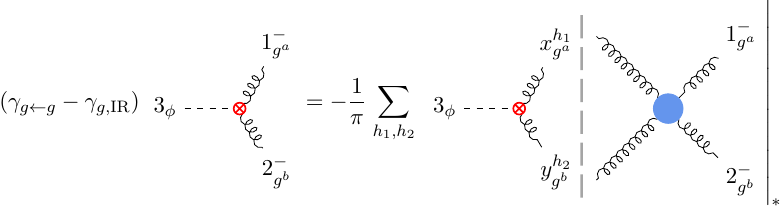}
\caption{Diagrammatic formula for computing $\gamma_{g\leftarrow g}$.}
\label{fig:gamma_gg}
\end{figure}

\paragraph{$\gamma_{g \leftarrow g}$.}
The multiplicative renormalization effect of the ALP effective operator $\phi GG$ is encoded in $\gamma_{g \leftarrow g}$, which can be derived from
\begin{equation}
     (\gamma_{g\leftarrow g}-\gamma_{g,\text{IR}}) F_g|_*(1^-_{g^a},2^-_{g^b},3_\phi) = -\frac{1}{\pi}(\mathcal M F_g)|_*(1^-_{g^a},2^-_{g^b},3_\phi)\,,
\end{equation}
schematized as in Fig.~\ref{fig:gamma_gg}.

The convolution is now given by
\begin{equation}
    (\mathcal M F_g)|_*(1^-_{g^a},2^-_{g^b},3_\phi) = \sum_{h_1,h_2}\int d\text{LIPS}_2 \, \mathcal M|_*(1^-_{g^a},2^-_{g^b};x^{h_1}_{g^c},y^{h_2}_{g^d})F_g|_*(x^{h_1}_{g^c},y^{h_2}_{g^d},3_\phi)\,,
\end{equation}
where the only contributing amplitude
\begin{equation}
    \mathcal M|_*(1^-_{g^a},2^-_{g^b};x^{-}_{g^c},y^{-}_{g^d})\delta^{cd} = -2C_Ag_s^2 \delta^{ab}\frac{\agl{1}{2}^4}{\agl{1}{x}\agl{x}{2}\agl{2}{y}\agl{y}{1}}
\end{equation}
is multiplied by
\begin{equation}
    F_g|_*(x^{-}_{g^c},y^{-}_{g^d},3_\phi) = -2\delta^{cd}\frac{\agl{x}{y}^2}{\Lambda}\,.
\end{equation}

\subparagraph{Angular integration.}
The calculation of the phase-space integral with the angular parameterization is as follows.
The product reads
\begin{equation}
    \mathcal M|_*(1^-_{g^a},2^-_{g^b};x^{-}_{g^c},y^{-}_{g^d})F_g|_*(x^{-}_{g^c},y^{-}_{g^d},3_\phi) = -4C_Ag_s^2 \delta^{ab} \frac{\agl{1}{2}^2}{\Lambda} \frac{1}{\cos^2\theta \sin^2\theta}
\end{equation}
yielding
\begin{equation}
    (\mathcal M F_g)|_*(1^-_{g^a},2^-_{g^b},3_\phi) = -4C_Ag_s^2 \delta^{ab} \frac{\agl{1}{2}^2}{\Lambda} \frac{1}{16\pi}\int_0^{\pi/2}2 \sin\theta\cos\theta\,d\theta\, \frac{1}{\cos^2\theta \sin^2\theta}\,.
\end{equation}
Therefore, by making use of the expression for $\gamma_{g,\text{IR}}$ derived in Appendix~\ref{sec:IRphigg} and reported in Eq.~\eqref{eq:gammaIRgluon_angular}, we obtain
\begin{align}
    \gamma_{g\leftarrow g} &= \gamma_{g,\text{IR}} - C_A\frac{g_s^2}{8\pi^2}  \int_0^{\pi/2}2\sin\theta\cos\theta \, d\theta \, \frac{1}{\cos^2\theta\sin^2\theta}\nonumber \\
    &= T_F\frac{g_s^2}{6\pi^2}\sum_f c_f^2 - C_A\frac{g_s^2}{8\pi^2}  \int_0^{\pi/2}2\sin\theta\cos\theta \, d\theta \, \frac{1 -\cos^8\theta- \sin^8\theta}{\cos^2\theta\sin^2\theta}\nonumber \\
    &= -\frac{g_s^2}{8\pi^2}\bigg(\frac{11}{3}C_A -\frac{4}{3}T_F\sum_fc_f^2\bigg)\,,
\end{align}
which is equal to the anomalous dimension of $g_s^2$, namely $(\mu/g_s^2)\text{d}g_s^2/\text{d}\mu$, since $\sum_f c_f^2$ denotes the number of quarks. 

\subparagraph{Stokes integration.}
The calculation of the phase-space integral with the Stokes parameterization is as follows.
The product reads
\begin{equation}
    \mathcal M|_*(1^-_{g^a},2^-_{g^b};x^{-}_{g^c},y^{-}_{g^d})F_g|_*(x^{-}_{g^c},y^{-}_{g^d},3_\phi) = -4C_Ag_s^2 \delta^{ab} \frac{\agl{1}{2}^2}{\Lambda} \frac{(1+z\zb)^2}{z\zb }
\end{equation}
and it is zero after performing the Stokes integration.
Therefore, also in this case, the anomalous dimension is given by $\gamma_{g,\text{IR}}$ derived in Appendix~\ref{sec:IRphigg} and reported in Eq.~\eqref{eq:gammaIRgluon_stokes}:
\begin{equation}
    \gamma_{g\leftarrow g} = \gamma_{g,\text{IR}}
    = -\frac{g_s^2}{8\pi^2}\bigg(\frac{11}{3}C_A -\frac{4}{3}T_F\sum_fc_f^2\bigg)\,.
\end{equation}

\begin{figure}[htbp]
\centering
\includegraphics[page=4]{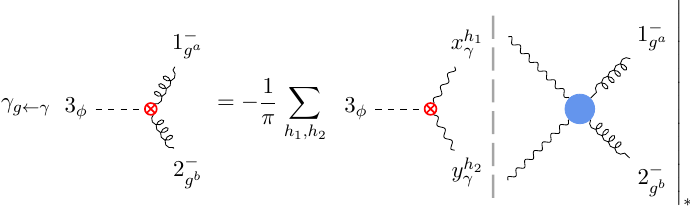}
\caption{Diagrammatic formula for computing $\gamma_{g\leftarrow \gamma}$.}
\label{fig:gamma_ggamma}
\end{figure}

\paragraph{$\gamma_{g\leftarrow \gamma}$.}
The master formula for computing $\gamma_{g\leftarrow \gamma}$ is
\begin{equation}
    \gamma_{g\leftarrow \gamma} F_g|_*(1^-_{g^a},2^-_{g^b},3_\phi) = -\frac{1}{\pi}(\mathcal M F_\gamma)|_*(1^-_{g^a},2^-_{g^b},3_\phi)
\end{equation}
and its diagrammatic expression is reported in Fig.~\ref{fig:gamma_ggamma}.

On the left, the form factor associated with $\phi GG$ reads
\begin{equation}
    F_g|_*(1^-_{g^a},2^-_{g^b},3_\phi) = -2\delta^{ab}\frac{\agl{1}{2}^2}{\Lambda}
\end{equation}
and the convolution on the right is expanded as
\begin{equation}
    (\mathcal M F_\gamma)|_*(1^-_{g^a},2^-_{g^b},3_\phi) = \sum_{h_1,h_2}\int d\text{LIPS}_2 \, \mathcal M|_*(1^-_{g^a},2^-_{g^b};x^{h_1}_\gamma,y^{h_2}_\gamma)F_\gamma|_*(x^{h_1}_\gamma,y^{h_2}_\gamma,3_\phi)\,.
\end{equation}
Since the amplitudes trivially vanish
\begin{equation}
    \mathcal M|_*(1^-_{g^a},2^-_{g^b};x^{h_1}_\gamma,y^{h_2}_\gamma) = 0\,,
\end{equation}
we obtain
\begin{equation}
    \gamma_{g\leftarrow \gamma} = 0\,.
\end{equation}

\begin{figure}[htbp]
\centering
\includegraphics[page=6]{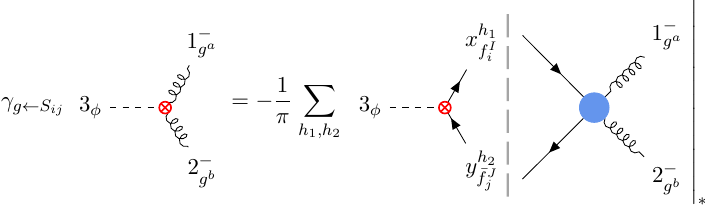}
\caption{Diagrammatic formula for computing $\gamma_{g\leftarrow S_{ij}}$.}
\label{fig:gamma_gS}
\end{figure}

\paragraph{$\gamma_{g\leftarrow S}$.}
The formula corresponding to $\gamma_{g\leftarrow S_{ij}}$ is
\begin{equation}
    \gamma_{g\leftarrow S_{ij}} F_g|_*(1^-_{g^a},2^-_{g^b},3_\phi) = -\frac{1}{\pi}(\mathcal M F_{S_{ij}})|_*(1^-_{g^a},2^-_{g^b},3_\phi)
\end{equation}
and is reported diagrammatically in Fig.~\ref{fig:gamma_gS}. 

Also in this case, analogously to $\gamma_{\gamma\leftarrow S_{ij}}$, we expect
\begin{equation}
    \gamma_{g\leftarrow S_{ij}}=0
\end{equation}
because $[\mathcal O_g]-[\mathcal O_{S_{ij}}]=1>0$.
This is indeed consistent with the fact that the convolution
\begin{equation}
    (\mathcal M F_{S_{ij}})|_*(1^-_{g^a},2^-_{g^b},3_\phi) = \sum_{h_1,h_2}\int d\text{LIPS}_2\, \mathcal M|_*(1^-_{g^a},2^-_{g^b};x_{f_i^I}^{h_1},y^{h_2}_{\bar f_j^J})F_{S_{ij}}|_*(x_{f_i^I}^{h_1},y^{h_2}_{\bar f_j^J},3_\phi)
\end{equation}
vanishes due to
\begin{equation}
    \mathcal M|_*(1^-_{g^a},2^-_{g^b};x_{f_i^I}^{h_1},y^{h_2}_{\bar f_j^J})=0
\end{equation}
for any choice of $h_1$ and $h_2$.

\paragraph{$\gamma_{\tilde g \leftarrow \tilde\gamma}$, $\gamma_{\tilde g \leftarrow \tilde g}$, and $\gamma_{\tilde g \leftarrow P}$.}
\label{par:PhiGG}
Once again, the anomalous dimensions that contribute to the renormalization of $\phi G\tilde G$ are equal to those of $\phi GG$ up to signs that can be determined through the comments leading to Eq. \eqref{eq:Obs1}:
\begin{align}
    \gamma_{\tilde g \leftarrow \tilde \gamma } &= \gamma_{ g \leftarrow  \gamma } = 0\,,\\
    \gamma_{\tilde g \leftarrow \tilde g } &= \gamma_{ g \leftarrow  g } = -\frac{g_s^2}{8\pi^2}\bigg(\frac{11}{3}C_A -\frac{4}{3}T_F\sum_fc_f^2\bigg) \,,\\
    \gamma_{\tilde g \leftarrow P_{ij} } &= -\gamma_{ g \leftarrow  S_{ij} } = 0\,.
\end{align}

\subsubsection{Renormalization group equations}

In this section, we have successfully reproduced some known results in the literature regarding the renormalization of the CP-violating ALP Lagrangian \cite{Bauer:2021mvw, Chala:2020wvs, DasBakshi:2023lca}. Here we report a summary of our results:
\begin{gather}
    \mu\dv[]{\mathcal C_\gamma}{\mu} = \frac{e^2}{6\pi^2}\mathcal C_\gamma \sum_f Q_f^2\,,
    \qquad
    \mu\dv[]{\tilde{\mathcal C}_\gamma}{\mu} = \frac{e^2}{6\pi^2}\tilde{\mathcal C}_\gamma \sum_f Q_f^2\,,
    \\
    \mu\dv[]{\mathcal C_g}{\mu} = -\frac{g_s^2}{8\pi^2} \bigg(\frac{11}{3} C_A - \frac{4}{3}T_F \sum_f c_f^2\bigg) \mathcal C_g\,,
    \qquad
    \mu\dv[]{\tilde{\mathcal C}_g}{\mu} = -\frac{g_s^2}{8\pi^2} \bigg(\frac{11}{3} C_A - \frac{4}{3}T_F \sum_f c_f^2\bigg) \tilde{\mathcal C}_g \,,
    \\
    \mu\dv[]{\mathcal Y_S^{ij}}{\mu} = -\frac{3}{8\pi^2}\left(e^2Q_f^2+C_Fg_s^2c_f^2 \right)\mathcal Y_S^{ij} + \frac{3}{2\pi^2}\frac{m_i}{\Lambda}\left(e^2Q_f^2 \mathcal C_\gamma + C_F g_s^2 c_f^2 \mathcal C_g\right)\delta^{ij}\,,\\
    \mu\dv[]{\mathcal Y_P^{ij}}{\mu} = -\frac{3}{8\pi^2}\left(e^2Q_f^2 +C_Fg_s^2c_f^2 \right)\mathcal Y_P^{ij} - \frac{3}{2\pi^2}\frac{m_i}{\Lambda}\left(e^2Q_f^2 \tilde{\mathcal C}_\gamma + C_F g_s^2 c_f^2 \tilde{\mathcal C}_g\right)\delta^{ij}\,.
\end{gather}

\subsection{Renormalization of SM effective operators below the EW scale}
\label{subsec:LEFT}

The phenomenological consequences of the ALP-SM interactions encoded in Eq.~\eqref{eq:Lag1} are rich and diverse. Of particular interest among these are the indirect effects on precision observables that are induced by the virtual exchange of an ALP. Such precision observables entail not only CP-violating probes, such as the electric dipole moment of particles, nucleons, nuclei and molecules, but also CP-conserving ones, as for instance the anomalous magnetic moment of leptons~\cite{Marciano:2016yhf,DiLuzio:2020oah}. 
Being the impact on these physical observables generated at the quantum level, a natural expectation is that their size can be determined by the leading logarithms that emerge from 
the solution of the RGEs. This expectation is rooted in the large separation of scales between the energies at which experiments are performed and those at which the effective Lagrangian is defined. 

The resulting CP-even $SU(3)_c \times U(1)_{\text{em}}$ invariant Lagrangian, $\mathcal{L}^{\text{even}}_{\text{CP}}$, that is generated by integrating out the ALP at one-loop level reads
\begin{align}
\label{eq:CPELag}
 \mathcal{L}^{\text{even}}_{\text{CP}} &= 
\frac{c_{\rm M}^{ij}}{\Lambda} \, \bar{f}_i \sigma^{\mu\nu} f_j \, F_{\mu\nu} +  \frac{c_{\rm CM}^{ij}}{\Lambda} \, \bar{f}_i \sigma^{\mu\nu} T^a f_j \, G^a_{\mu\nu} + \frac{D_G}{3\Lambda^2} f^{abc} G_\mu^{a, \nu} G_{\nu}^{b, \rho} G_{\rho}^{c, \mu} \nonumber \\
& \quad + 
\frac{c_{LL1}^{ijkl}}{\Lambda^2}(\bar f_{Li} \gamma^\mu f_{Lj})(\bar f'_{Lk}\gamma_\mu f'_{Ll}) + \frac{c_{LL8}^{ijkl}}{\Lambda^2}(\bar f_{Li} \gamma^\mu T^a f_{Lj})(\bar f'_{Lk}\gamma_\mu T^a f'_{Ll})\nonumber \\
& \quad +
\frac{c_{RR1}^{ijkl}}{\Lambda^2}(\bar f_{Ri} \gamma^\mu f_{Rj})(\bar f'_{Rk}\gamma_\mu f'_{Rl}) +
\frac{c_{RR8}^{ijkl}}{\Lambda^2}(\bar f_{Ri} \gamma^\mu T^a f_{Rj})(\bar f'_{Rk}\gamma_\mu T^a f'_{Rl})\nonumber \\
& \quad +
\frac{c_{LR1}^{ijkl}}{\Lambda^2}(\bar f_{Li} \gamma^\mu f_{Lj})(\bar f'_{Rk}\gamma_\mu f'_{Rl}) +
\frac{c_{LR8}^{ijkl}}{\Lambda^2}(\bar f_{Li} \gamma^\mu T^a f_{Lj})(\bar f'_{Rk}\gamma_\mu T^a f'_{Rl})
\,. 
\end{align}
The corresponding CP-odd Lagrangian, $\mathcal{L}^{\text{odd}}_{\text{CP}}$, is given by
\begin{align}
\label{eq:CPOLag}
 \mathcal{L}^{\text{odd}}_{\text{CP}} = 
 \frac{c_{\rm E}^{ij}}{\Lambda} \, \bar{f}_i \sigma^{\mu\nu} i\gamma_5 f_j \, F_{\mu\nu} +  \frac{c_{\rm CE}^{ij}}{\Lambda} \, \bar{f}_i \sigma^{\mu\nu} i\gamma_5 T^a f_j \, G^a_{\mu\nu} + \frac{d_G}{3\Lambda^2} f^{abc} G_\mu^{a, \nu} G_{\nu}^{b, \rho} \tilde G_{\rho}^{c, \mu}\,.
\end{align}
Notice that in the above Lagrangians we have neglected operators which emerge by integrating out the ALP at tree level, such as $(\bar{f} f)( \bar{f}' f')$, $GGGG$, etc.
In fact, in this case, RGE effects are fully accounted for by evaluating the effective ALP couplings of Eqs.~\eqref{eq:LagA} and \eqref{eq:LagB} at the ALP mass scale.

The objective of this section is to evaluate the Wilson coefficients of the above Lagrangians that are generated
by running effects from $\Lambda$ down to the ALP mass scale.

\subsubsection{$GG\tilde G$ and $GGG$}
\label{sec:GGG}

\begin{figure}[htbp]
\centering
\includegraphics[page=9]{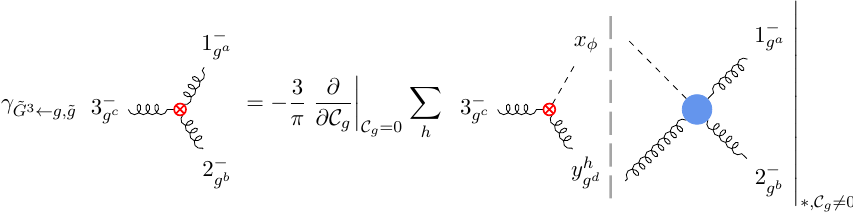}
\caption{Diagrammatic formula for computing $\gamma_{\tilde G^3\leftarrow g,\tilde g}$.}
\label{fig:gammatildeG3}
\end{figure}

\paragraph{$\gamma_{\tilde G^3\leftarrow g,\tilde g}$.}
We define the Weinberg dimension-six operator as
\begin{equation}
    \mathcal O_{\tilde G^3} = \frac{1}{3}f^{abc}G_\mu^{a, \nu} G_{\nu}^{b, \rho} \tilde G_{\rho}^{c, \mu}\,.
\end{equation}
The renormalization of the corresponding Wilson coefficient induced by the operators $\phi GG$ and $\phi G\tilde G$ at one-loop order can be evaluated from
\begin{equation}\label{eq:mastertildeG^3}
    \gamma_{\tilde G^3\leftarrow g,\tilde g}F_{\tilde G^3}|_*(1^-_{g^a},2^-_{g^b},3^-_{g^c})=-\frac{1}{\pi}\left. \frac{\partial}{\partial \mathcal C_g} \right|_{\mathcal C_g = 0} (\mathcal M F_{\tilde g})|_{*,\mathcal C_g \neq  0}(1^-_{g^a},2^-_{g^b},3^-_{g^c})\,,
\end{equation}
which diagrammatically reads as in Fig.~\ref{fig:gammatildeG3}.

The minimal form factor of $\mathcal O_{\tilde G^3}$ reads
\begin{equation}
    F_{\tilde G^3}|_*(1^-_{g^a},2^-_{g^b},3^-_{g^c})=\frac{\sqrt 2}{\Lambda^2}f^{abc}\agl{1}{2}\agl{2}{3}\agl{3}{1}\,,
\end{equation}
while the convolution can be written as
\begin{equation}
    (\mathcal M F_{\tilde g})|_{*,\mathcal C_g \neq  0}(1^-_{g^a},2^-_{g^b},3^-_{g^c}) = 3 \sum_{h}\int d\text{LIPS}_2\, \mathcal{M}|_{*,\mathcal C_g \neq  0}(1^-_{g^a},2^-_{g^b};x_\phi,y^h_{g^d})
    F_{\tilde g}|_*(x_\phi,y^h_{g^d},3^-_{g^c})\,,
\end{equation}
where the factor $3$ accounts for all the permutations of the external gluons.
The only helicity $h$ that gives a nonzero contribution is the negative one, since $F_{\tilde g}|_*(x_\phi,y^+_{g^d},3^-_{g^c}) = 0$.
The amplitude with $\mathcal C_g\neq 0$ and all the other Wilson coefficients turned off and the minimal form factor of $\mathcal O_{\tilde g}$ are, respectively, given by
\begin{align}
    \mathcal{M}|_{*,\mathcal C_g \neq  0}(1^-_{g^a},2^-_{g^b};x_\phi,y^-_{g^d}) &= 2i\sqrt{2}g_s \frac{\mathcal C_g}{\Lambda}f^{abd}\frac{ \agl{1}{2}^3 }{ \agl{1}{y} \agl{2}{y}}\,,\\
    F_{\tilde g}|_*(x_\phi,y^-_{g^d},3^-_{g^c}) &= -\frac{2i}{\Lambda}\delta^{cd}\agl{3}{y}^2\,.
\end{align}

\subparagraph{Angular integration.}
Using the angular parameterization for the phase-space integral the amplitude reads:
\begin{equation}\label{eq:amplitudeCg}
    \mathcal{M}|_{*,\mathcal C_g \neq  0}(1^-_{g^a},2^-_{g^b};x_\phi,y^-_{g^d}) = -2i\sqrt{2}g_s \frac{\mathcal C_g}{\Lambda}f^{abd}\agl{1}{2}\frac{1}{ \cos\theta\sin\theta}e^{i\phi}\,.
\end{equation}
The integration in the azimuthal angle $\phi$ only involves
\begin{align}
    \int_0^{2\pi}\frac{d\phi}{2\pi}\,F_{\tilde g}|_*(x_\phi,y^-_{g^d},3^-_{g^c}) e^{i\phi} &= 
    -\frac{2i}{\Lambda}\delta^{cd}\int_0^{2\pi}\frac{d\phi}{2\pi}\, (\agl{3}{1}e^{-i\phi}\sin\theta + \agl{3}{2}\cos\theta)^2e^{i\phi}
    \nonumber \\&=\frac{4i}{\Lambda}\delta^{cd}\agl{2}{3}\agl{3}{1}\cos\theta\sin\theta\,,\label{eq:phiintegral}
\end{align}
where $\int_0^{2\pi} d\phi\,e^{in\phi}=2\pi\delta_{0n}$ has been used.
Therefore, the $\theta$ dependences of Eqs.~\eqref{eq:amplitudeCg} and \eqref{eq:phiintegral} cancel each other, and we are left with a trivial integral in $\theta$, which leads to
\begin{align}
    (\mathcal M F_{\tilde g})|_{*,\mathcal C_g \neq  0}(1^-_{g^a},2^-_{g^b},3^-_{g^c}) &= 3\times 8\sqrt 2 g_s \frac{\mathcal C_g}{\Lambda^2}f^{abc}\agl{1}{2}\agl{2}{3} \agl{3}{1} \frac{1}{8\pi}\int_0^{\pi/2}2\sin\theta\cos\theta\,d\theta \nonumber \\
    &= \frac{3\sqrt{2}}{\pi\Lambda^2} g_s\mathcal  C_gf^{abc} \agl{1}{2}\agl{2}{3} \agl{3}{1}\,.
\end{align}

\subparagraph{Stokes integration.}
Using the Stokes parameterization for the phase-space integral the amplitude reads as
\begin{equation}
    \mathcal{M}|_{*,\mathcal C_g \neq  0}(1^-_{g^a},2^-_{g^b};x_\phi,y^-_{g^d}) = 2i\sqrt{2}g_s \frac{\mathcal C_g}{\Lambda}f^{abd}\agl{1}{2}\frac{1+z\zb}{ z}\,,
\end{equation}
which implies
\begin{equation}
    (\mathcal M F_{\tilde g})|_{*,\mathcal C_g \neq  0}(1^-_{g^a},2^-_{g^b},3^-_{g^c}) =  \frac{3\sqrt{2}}{\pi\Lambda^2} g_s\mathcal  C_gf^{abc} \agl{1}{2}\agl{2}{3} \agl{3}{1}\,.
\end{equation}
Finally, from the master formula of Eq.~\eqref{eq:mastertildeG^3}, we obtain
\begin{equation}
\label{eq:gammatildeG3}
    \gamma_{\tilde G^3\leftarrow g,\tilde g} = -\frac{3g_s}{\pi^2}\,.
\end{equation}

\paragraph{$\gamma_{G^3\leftarrow g, g}$ and $\gamma_{G^3\leftarrow \tilde g, \tilde g}$.}
\label{par:GGG}
The beta function associated with the Wilson coefficient of the CP-even operator
\begin{equation}
    \mathcal O_{G^3} = \frac{1}{3}f^{abc}G_\mu^{a, \nu} G_{\nu}^{b, \rho} G_{\rho}^{c, \mu}
\end{equation}
receives contributions from double insertions of the operators $\phi GG$ and $\phi G\tilde G$.
The corresponding anomalous dimensions $\gamma_{ G^3\leftarrow g, g}$ and $\gamma_{ G^3\leftarrow \tilde g,\tilde g}$ can be computed via
\begin{align}
    \gamma_{ G^3\leftarrow g, g}F_{ G^3}|_*(1^-_{g^a},2^-_{g^b},3^-_{g^c})&=-\frac{1}{\pi}\left. \frac{\partial}{\partial \mathcal C_g} \right|_{\mathcal C_g = 0} (\mathcal M F_{g})|_{*,\mathcal C_g \neq  0}(1^-_{g^a},2^-_{g^b},3^-_{g^c})\,, \\
    \gamma_{ G^3\leftarrow \tilde g,\tilde g}F_{ G^3}|_*(1^-_{g^a},2^-_{g^b},3^-_{g^c})&=-\frac{1}{\pi}\left. \frac{\partial}{\partial \mathcal{\tilde C}_g} \right|_{\mathcal{\tilde C}_g = 0} (\mathcal M F_{\tilde g})|_{*,\mathcal{\tilde C}_g \neq  0}(1^-_{g^a},2^-_{g^b},3^-_{g^c})\,,
\end{align}
respectively, and both can be straightforwardly related to the anomalous dimension $\gamma_{\tilde G^3\leftarrow g,\tilde g}$ in Eq.~\eqref{eq:gammatildeG3}.
Indeed, by taking into account that
\begin{align}
    F_{ G^3}|_*(1^-_{g^a},2^-_{g^b},3^-_{g^c}) &= -i F_{\tilde G^3}|_*(1^-_{g^a},2^-_{g^b},3^-_{g^c})\,, \\
    F_{g}|_*(x_\phi,y^-_{g^d},3^-_{g^c}) &=-i F_{\tilde g}|_*(x_\phi,y^-_{g^d},3^-_{g^c})\,,\\
    \frac{\partial}{\partial \mathcal{\tilde C}_g}\mathcal{M}|_{*,\mathcal{\tilde C}_g \neq  0}(1^-_{g^a},2^-_{g^b};x_\phi,y^-_{g^d})&= i \frac{\partial}{\partial \mathcal{C}_g} \mathcal{M}|_{*,\mathcal C_g \neq  0}(1^-_{g^a},2^-_{g^b};x_\phi,y^-_{g^d})\,,
\end{align}
we obtain
\begin{equation}
\label{eq:GGG_AD}
    \gamma_{ G^3\leftarrow g, g} =- \gamma_{ G^3\leftarrow \tilde g, \tilde g} = \gamma_{ \tilde G^3\leftarrow g, \tilde g} = -\frac{3g_s}{\pi^2}\,.
\end{equation}

\subsubsection{
$\bar f \sigma\!\cdot\! F i\gamma_5 f$
and $\bar f \sigma\!\cdot\! F f$}

\begin{figure}[htbp]
\centering
\includegraphics[width=\linewidth,page=10]{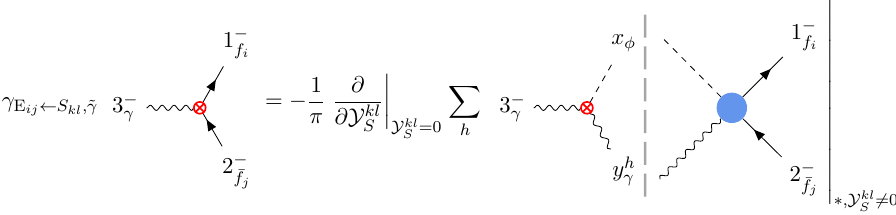}
\caption{Diagrammatic formula for computing $\gamma_{\txt{E}_{ij}\leftarrow S_{kl},\tilde\gamma}$.}
\label{fig:gammaEDM}
\end{figure}

\paragraph{$\gamma_{\txt{E}\leftarrow S,\tilde\gamma}$.}
The first anomalous dimension $\gamma_{\txt{E}_{ij}\leftarrow S_{kl},\tilde\gamma}$ of the electric dipole operator
\begin{equation}
    \mathcal O_{\text{E}_{ij}} = 
    \bar f_i \sigma^{\mu\nu}i\gamma_5 f_j F_{\mu\nu}\,,
\end{equation}
is induced by the ALP operators $\phi \bar f_k f_l$ and $\phi F\tilde F$.
The corresponding master formula reads
\begin{equation}
    \gamma_{\txt{E}_{ij}\leftarrow S_{kl},\tilde\gamma} F_{\text{E}_{ij}}|_*(1^-_{f_i},2^-_{\bar f_j},3^-_\gamma) = -\frac{1}{\pi}\left.\frac{\partial}{\partial \mathcal Y_S^{kl}}\right|_{\mathcal Y_S^{kl}=0}(\mathcal M F_{\tilde\gamma})|_{*,\mathcal Y_S^{kl}\neq 0}(1^-_{f_i},2^-_{\bar f_j},3^-_\gamma)\,,
\end{equation}
whose diagrammatic expression is provided in Fig.~\ref{fig:gammaEDM}.

On the left-hand side we have the form factor of the electric dipole operator
\begin{equation}
    F_{\text{E}_{ij}}|_*(1^-_{f_i},2^-_{\bar f_j},3^-_\gamma) = -\frac{2i \sqrt{2}}{\Lambda} \agl{1}{3}\agl{2}{3}\,,
\end{equation}
while, on the right-hand side, the convolution reads
\begin{equation}
    (\mathcal M F_{\tilde\gamma})|_{*,\mathcal Y_S^{kl}\neq 0}(1^-_{f_i},2^-_{\bar f_j},3^-_\gamma) = \sum_{h} \int d\text{LIPS}_2 \, \mathcal M|_{*,\mathcal Y_S^{kl}\neq 0}(1^-_{f_i},2^-_{\bar f_j};x_\phi,y^h_\gamma) F_{\tilde\gamma}|_*(x_\phi,y^h_\gamma,3_\gamma^-)\,,
\end{equation}
where the only non-vanishing amplitude is
\begin{equation}
    \mathcal M|_{*,\mathcal Y_S^{kl}\neq 0}(1^-_{f_i},2^-_{\bar f_j};x_\phi,y^-_\gamma) =  -\sqrt 2 eQ_f \mathcal Y_S^{kl}\delta^{ik}\delta^{jl} \frac{\agl{1}{2}^2}{\agl{1}{y}\agl{y}{2}}
\end{equation}
and is multiplied by
\begin{equation}
    F_{\tilde\gamma}|_*(x_\phi,y^-_\gamma,3_\gamma^-) = -\frac{2i}{\Lambda}\agl{3}{y}^2\,.
\end{equation}

\subparagraph{Angular integration.}
The calculation of the phase-space integral with the angular parameterization is as follows.
The amplitude reads
\begin{equation}
    \mathcal M|_{*,\mathcal Y_S^{kl}\neq 0}(1^-_{f_i},2^-_{\bar f_j};x_\phi,y^-_\gamma) =  -\sqrt 2 eQ_f \mathcal Y_S^{kl}\delta^{ik}\delta^{jl} \frac{1}{\cos\theta \sin\theta}e^{i\phi}
\end{equation}
and thus the integration in the azimuthal angle $\phi$ only involves
\begin{equation}
    \int_0^{2\pi}\frac{d\phi}{2\pi}\,F_{\tilde\gamma}|_*(x_\phi,y^-_\gamma,3_\gamma^-)e^{i\phi} = - \frac{4i}{\Lambda}\agl{2}{3}\agl{1}{3}\cos\theta\sin\theta\,.
\end{equation}
Then, the remaining integral is simply
\begin{align}
    (\mathcal M F_{\tilde\gamma})|_{*,\mathcal Y_S^{kl}\neq 0}(1^-_{f_i},2^-_{\bar f_j},3^-_\gamma) &=  \frac{4 i\sqrt 2}{\Lambda}  eQ_f \mathcal Y_S^{kl}\delta^{ik}\delta^{jl}  \agl{2}{3}\agl{1}{3} \frac{1}{8\pi}\int_0^{\pi/2}2\sin\theta\cos\theta\,d\theta \nonumber \\
    &=  \frac{ i\sqrt 2}{2\pi\Lambda}  eQ_f \mathcal Y_S^{kl}\delta^{ik}\delta^{jl}  \agl{2}{3}\agl{1}{3}\,,
\end{align}
which leads to
\begin{equation}\label{eq:gammaE<-Sgammatilde}
    \gamma_{\txt{E}_{ij}\leftarrow S_{kl},\tilde\gamma} =  \frac{eQ_f }{4\pi^2}\delta^{ik}\delta^{jl}\,.
\end{equation}

\subparagraph{Stokes integration.}
Using the Stokes parameterization the amplitude reads
\begin{equation}
    \mathcal M|_{*,\mathcal Y_S^{kl}\neq 0}(1^-_{f_i},2^-_{\bar f_j};x_\phi,y^-_\gamma) =  \sqrt 2 eQ_f \mathcal Y_S^{kl}\delta^{ik}\delta^{jl} \frac{1+z \zb}{z} \ , 
\end{equation}
and combining it with the form factor we get
\begin{equation}
    (\mathcal M F_{\tilde\gamma})|_{*,\mathcal Y_S^{kl}\neq 0}(1^-_{f_i},2^-_{\bar f_j},3^-_\gamma)
    =  \frac{ i\sqrt 2}{2\pi\Lambda}  eQ_f \mathcal Y_S^{kl}\delta^{ik}\delta^{jl}  \agl{2}{3}\agl{1}{3}\,,
\end{equation}
which leads to Eq.~\eqref{eq:gammaE<-Sgammatilde}.

\paragraph{$\gamma_{\txt{E}\leftarrow P,\gamma}$.}
The second anomalous dimension $\gamma_{\txt{E}_{ij}\leftarrow P_{kl},\gamma}$, corresponding to the insertion of the ALP operators $\phi \bar f_k i\gamma_5 f_l$ and $\phi FF$, can be obtained from the master formula
\begin{equation}
    \gamma_{\txt{E}_{ij}\leftarrow P_{kl},\gamma} F_{\text{E}_{ij}}|_*(1^-_{f_i},2^-_{\bar f_j},3^-_\gamma) = -\frac{1}{\pi}\left.\frac{\partial}{\partial \mathcal Y_P^{kl}}\right|_{\mathcal Y_P^{kl}=0}(\mathcal M F_{\gamma})|_{*,\mathcal Y_P^{kl}\neq 0}(1^-_{f_i},2^-_{\bar f_j},3^-_\gamma)
\end{equation}
and since these identities hold
\begin{align}
    F_{\gamma}|_*(x_\phi,y^-_\gamma,3_\gamma^-) &= -i F_{\tilde\gamma}|_*(x_\phi,y^-_\gamma,3_\gamma^-) \,,
    \label{eq:F_gamma_identity}\\
    \frac{\partial}{\partial \mathcal Y_P^{kl}}\mathcal M|_{*,\mathcal Y_P^{kl}\neq 0}(1^-_{f_i},2^-_{\bar f_j};x_\phi,y^-_\gamma)&=-i \frac{\partial}{\partial \mathcal Y_S^{kl}} \mathcal M|_{*,\mathcal Y_S^{kl}\neq 0}(1^-_{f_i},2^-_{\bar f_j};x_\phi,y^-_\gamma)\,,\label{eq:amplitudeyukawaidentity}
\end{align}
we can relate it to $\gamma_{\txt{E}_{ij}\leftarrow S_{kl},\tilde\gamma}$, concluding that
\begin{equation}
    \gamma_{\txt{E}_{ij}\leftarrow P_{kl},\gamma} = - \gamma_{\txt{E}_{ij}\leftarrow S_{kl},\tilde\gamma}= -\frac{eQ_f }{4\pi^2}\delta^{ik}\delta^{jl}\,.
\end{equation}

\paragraph{$\gamma_{\text{M}\leftarrow P,\tilde \gamma}$ and $\gamma_{\text{M}\leftarrow S,\gamma}$.}
The magnetic dipole operator is defined as
\begin{equation}
    \mathcal O_{\text{M}_{ij}} = \bar f_i \sigma^{\mu\nu}f_j F_{\mu\nu}
\end{equation}
and the corresponding anomalous dimensions $\gamma_{\txt{M}_{ij}\leftarrow S_{kl},\gamma}$ and $\gamma_{\txt{M}_{ij}\leftarrow P_{kl},\tilde\gamma}$ can be obtained from the master formulae
\begin{align}
    \gamma_{\txt{M}_{ij}\leftarrow S_{kl},\gamma} F_{\text{M}_{ij}}|_*(1^-_{f_i},2^-_{\bar f_j},3^-_\gamma) &= -\frac{1}{\pi}\left.\frac{\partial}{\partial \mathcal Y_S^{kl}}\right|_{\mathcal Y_S^{kl}=0}(\mathcal M F_{\gamma})|_{*,\mathcal Y_S^{kl}\neq 0}(1^-_{f_i},2^-_{\bar f_j},3^-_\gamma)\,, \\
    \gamma_{\txt{M}_{ij}\leftarrow P_{kl},\tilde\gamma} F_{\text{M}_{ij}}|_*(1^-_{f_i},2^-_{\bar f_j},3^-_\gamma) &= -\frac{1}{\pi}\left.\frac{\partial}{\partial \mathcal Y_P^{kl}}\right|_{\mathcal Y_P^{kl}=0}(\mathcal M F_{\tilde\gamma})|_{*,\mathcal Y_P^{kl}\neq 0}(1^-_{f_i},2^-_{\bar f_j},3^-_\gamma)\,.
\end{align}
If we exploit the identity
\begin{equation}
    F_{\text{M}_{ij}}|_*(1^-_{f_i},2^-_{\bar f_j},3^-_\gamma) =i F_{\text{E}_{ij}}|_*(1^-_{f_i},2^-_{\bar f_j},3^-_\gamma)\,,
\end{equation}
as well as those in Eqs.~\eqref{eq:F_gamma_identity} and \eqref{eq:amplitudeyukawaidentity}, we can relate both of them to $\gamma_{\txt{E}_{ij}\leftarrow S_{kl},\tilde\gamma}$, concluding that
\begin{equation}
    \gamma_{\txt{M}_{ij}\leftarrow P_{kl},\tilde\gamma}=
    \gamma_{\txt{M}_{ij}\leftarrow S_{kl},\gamma} = -
    \gamma_{\txt{E}_{ij}\leftarrow S_{kl},\tilde\gamma} = -\frac{eQ_f }{4\pi^2}\delta^{ik}\delta^{jl}\,.
\end{equation}

\subsubsection{
$\bar f \sigma\!\cdot\! G i\gamma_5 f$
and $\bar f \sigma\!\cdot\! G f$
}
\label{sec:chromoelectric_dipole_moment}

\paragraph{$\gamma_{\text{CE}\leftarrow S,\tilde g}$ and $\gamma_{\text{CE}\leftarrow P, g}$.}
The renormalization group equations for the chromoelectric dipole operator
\begin{equation}
    \mathcal O_{\text{CE}_{ij}} = \bar f_i \sigma^{\mu\nu}i\gamma_5 T^a f_j G_{\mu\nu}^a
\end{equation}
are captured by the anomalous dimensions $\gamma_{\txt{CE}_{ij}\leftarrow S_{kl},\tilde g}$ and $\gamma_{\txt{CE}_{ij}\leftarrow P_{kl}, g}$.
The former corresponds to the insertion of the ALP operators $\phi \bar f_k f_l$ and $\phi G\tilde G$, while the latter is induced by $\phi \bar f_k i \gamma_5 f_l$ and $\phi G G$.
They can be easily derived from $\gamma_{\txt{E}_{ij}\leftarrow S_{kl},\tilde \gamma}$ and $\gamma_{\txt{E}_{ij}\leftarrow P_{kl}, \gamma}$, respectively, by replacing $eQ_f$ with $g_s c_f$:
\begin{equation}
    \gamma_{\txt{CE}_{ij}\leftarrow S_{kl},\tilde g} =- \gamma_{\txt{CE}_{ij}\leftarrow P_{kl}, g}=  \frac{g_sc_f }{4\pi^2}\delta^{ik}\delta^{jl}\,.
\end{equation}

\paragraph{$\gamma_{\text{CM}\leftarrow S, g}$ and $\gamma_{\text{CM}\leftarrow P,\tilde g}$.}
Finally, concerning the chromomagnetic dipole operator
\begin{equation}
    \mathcal O_{\text{CM}_{ij}} = \bar f_i \sigma^{\mu\nu} T^a f_j G_{\mu\nu}^a\,,
\end{equation}
we can straightforwardly compute its anomalous dimensions from $\gamma_{\txt{M}_{ij}\leftarrow P_{kl},\tilde\gamma}$ and $\gamma_{\txt{M}_{ij}\leftarrow S_{kl},\gamma}$, by following the same prescription used for the case of the chromoelectric dipole operator:
\begin{equation}
    \gamma_{\txt{CM}_{ij}\leftarrow S_{kl}, g}=\gamma_{\txt{CM}_{ij}\leftarrow P_{kl},\tilde g}= -\frac{g_sc_f }{4\pi^2}\delta^{ik}\delta^{jl}\,.
\end{equation}

\subsubsection{Four-fermion operators}

\paragraph{$\gamma_{LL1 \leftarrow \gamma, \gamma}$, $\gamma_{LL1 \leftarrow \tilde \gamma, \tilde \gamma}$, $\gamma_{RR1 \leftarrow \gamma, \gamma}$, and $\gamma_{RR1 \leftarrow \tilde \gamma, \tilde \gamma}$.} 

\begin{figure}[htbp]
\centering
\includegraphics[width=\linewidth,page=15]{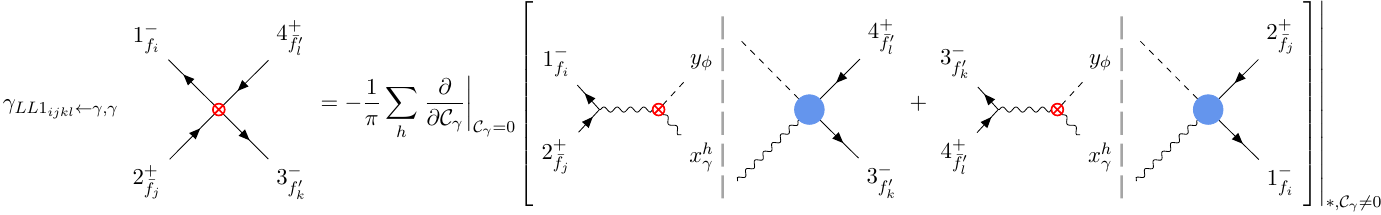}
\caption{Diagrammatic formula for computing $\gamma_{LL1_{ijkl} \leftarrow \gamma, \gamma}$.}
\label{fig:4fermion}
\end{figure}

Having defined the 4-fermion operator
\begin{equation}
    \mathcal O_{LL1_{ijkl}} = (\bar f_{Li} \gamma^\mu f_{Lj})(\bar f'_{Lk}\gamma_\mu f'_{Ll})\,,
\end{equation}
whose minimal form factor is
\begin{equation}
    F_{LL1_{ijkl}}|_*(1^-_{f_i},2^+_{\bar f_j}, 3^-_{f'_k},4^+_{\bar f'_l}) = \frac{2}{\Lambda^2}\agl{1}{3}\sqr{4}{2}
\end{equation}
for $f\neq f'$,
we first analyze its anomalous dimension as induced by two ALP-photon couplings, i.e.,~$\gamma_{LL1_{ijkl} \leftarrow \gamma, \gamma}$ and $\gamma_{LL1_{ijkl} \leftarrow \tilde \gamma, \tilde \gamma}$. The master formula for the calculation of the former is
\begin{equation}
    \gamma_{LL1_{ijkl} \leftarrow \gamma, \gamma}  F_{LL1_{ijkl}}|_*(1^-_{f_i},2^+_{\bar f_j}, 3^-_{f'_k},4^+_{\bar f'_l}) = -\frac{1}{\pi}
    \left.\frac{\partial}{\partial \mathcal C_\gamma}\right|_{\mathcal C_\gamma = 0}
    (\mathcal M F_\gamma)|_{*,\mathcal C_\gamma \neq 0} (1^-_{f_i},2^+_{\bar f_j}, 3^-_{f'_k},4^+_{\bar f'_l})
\end{equation}
whose diagram is shown in Fig.~\ref{fig:4fermion}.

As for $\gamma_{LL1_{ijkl} \leftarrow \tilde\gamma, \tilde\gamma}$, $\gamma_{RR1_{ijkl} \leftarrow \gamma, \gamma}$, and $\gamma_{RR1_{ijkl} \leftarrow \tilde\gamma, \tilde\gamma}$, they are equal to $\gamma_{LL1_{ijkl} \leftarrow \gamma, \gamma}$ --- including the sign --- given the aforementioned considerations.

The convolution on the right can be expanded as
\begin{align}
    (\mathcal M F_\gamma)|_{*,\mathcal C_\gamma \neq 0} (1^-_{f_i},2^+_{\bar f_j}, 3^-_{f'_k},4^+_{\bar f'_l}) &= \sum_h \int d \text{LIPS}_2 \,\bigg[ \mathcal M|_{*,\mathcal C_\gamma \neq 0}(3^-_{f'_k},4^+_{\bar f'_l}; x_{\gamma}^h , y_\phi) F_{\gamma}|_*(1^-_{f_i},2^+_{\bar f_j},x_{\gamma}^h , y_\phi) \nonumber \\
    &\quad +
    \mathcal M|_{*,\mathcal C_\gamma \neq 0}(1^-_{f_i},2^+_{\bar f_j}; x_{\gamma}^h , y_\phi) F_{\gamma}|_*(3^-_{f'_k},4^+_{\bar f'_l},x_{\gamma}^h , y_\phi) 
    \bigg]
\end{align}
where, for the $h=-$ configuration, we have
\begin{align}
    \mathcal M|_{*,\mathcal C_\gamma \neq 0}(3^-_{f'_k},4^+_{\bar f'_l}; x_{\gamma}^- , y_\phi) &= \frac{2\sqrt{2}}{\Lambda}e Q_{f'}\mathcal C_\gamma\delta^{kl} \frac{\sqr{4}{x}^2}{\sqr{4}{3}}\,,\\
    F_{\gamma}|_*(1^-_{f_i},2^+_{\bar f_j},x_{\gamma}^- , y_\phi) &= \frac{2\sqrt{2}}{\Lambda}e Q_{f}\delta^{ij} \frac{\agl{1}{x}^2}{\agl{1}{2}}  \,.
\end{align}
Thus, the only integral to perform is 
\begin{equation}
    \int d \text{LIPS}_2 \, \agl{1}{x}^2 \sqr{4}{x}^2 = \frac{1}{24\pi}\agl{1}{3}^2 \agl{4}{3}^2
\end{equation}
that, after accounting for the $h=+$ configuration and permuting the fermions, leads to the following expression for the convolution, valid for $f \neq f'$:
\begin{align}
    (\mathcal M F_\gamma)|_{*,\mathcal C_\gamma \neq 0} (1^-_{f_i},2^+_{\bar f_j}, 3^-_{f'_k},4^+_{\bar f'_l}) = - \frac{4e^2}{3\pi \Lambda^2}Q_f Q_{f'} \mathcal C_\gamma \delta^{ij} \delta^{kl} \agl{1}{3}\sqr{4}{2}\,.
\end{align}

Instead, for $f = f'$, the above expression holds if we substitute $\delta^{ij} \delta^{kl} \to \delta^{ij} \delta^{kl} + \delta^{il} \delta^{kj}$, and since the form factor on the left of the master formula acquires an additional factor of $2$, the anomalous dimension elements can be compactly written as
\begin{align}
    \gamma_{LL1_{ijkl} \leftarrow \gamma,\gamma } &= 
    \gamma_{LL1_{ijkl} \leftarrow \tilde\gamma, \tilde\gamma} = 
    \gamma_{RR1_{ijkl} \leftarrow \gamma,\gamma } =
    \gamma_{RR1_{ijkl} \leftarrow \tilde\gamma, \tilde\gamma}\nonumber \\
    &= (1-\delta_{ff'})\frac{2e^2}{3\pi^2 } Q_f Q_{f'}  \delta^{ij} \delta^{kl} +   \delta_{ff'}\frac{e^2}{3\pi^2 } Q_f^2  \delta^{ij} \delta^{kl}\,.\label{eq:LL1<-gammagamma}
\end{align}

\paragraph{$\gamma_{LL1\leftarrow g,g}$, $\gamma_{LL1\leftarrow \tilde g, \tilde g}$, $\gamma_{RR1\leftarrow g,g}$, and $\gamma_{RR1\leftarrow \tilde g, \tilde g}$.}
These anomalous dimensions are nonvanishing only if $f = f'$ given that in this case the operator $(\bar f_{Li} \gamma^\mu T^a f_{Lj})(\bar f_{Lk} \gamma_\mu T^a f_{Ll})$ can be written as a linear combination of $\mathcal O_{LL1}$ operators, i.e.,
\begin{equation}
    (\bar f_{Li} \gamma^\mu T^a f_{Lj})(\bar f_{Lk} \gamma_\mu T^a f_{Ll}) = \frac{1}{2}\!\left[(\bar f_{Lk} \gamma^\mu  f_{Lj})(\bar f_{Li} \gamma_\mu f_{Ll})-\frac{1}{N_c}
    (\bar f_{Li} \gamma^\mu  f_{Lj})(\bar f_{Lk} \gamma_\mu f_{Ll})
    \right]\!\,,
\end{equation}
as a consequence of the identity $T^a_{IJ}T^a_{KL} = \frac{1}{2}(\delta_{KJ}\delta_{IL}-\frac{1}{N_c}\delta_{IJ}\delta_{KL})$.
This implies that
\begin{align}
    \gamma_{LL1_{ijkl} \leftarrow g,g } &= 
    \gamma_{LL1_{ijkl} \leftarrow \tilde g, \tilde g} = 
    \gamma_{RR1_{ijkl} \leftarrow g,g } =
    \gamma_{RR1_{ijkl} \leftarrow \tilde g, \tilde g}\nonumber \\
    &= \delta_{ff'}\frac{g_s^2}{6\pi^2}c_f^2 \bigg(
    \delta^{kj}\delta^{il}-\frac{1}{N_c}\delta^{ij}\delta^{kl}
    \bigg)\,.
\end{align}

\paragraph{$\gamma_{LL8\leftarrow g,g}$, $\gamma_{LL8\leftarrow \tilde g, \tilde g}$, $\gamma_{RR8\leftarrow g,g}$, and $\gamma_{RR8\leftarrow \tilde g, \tilde g}$.}
For these anomalous dimensions we can recycle the results already obtained. Indeed, since we work with a basis where the operators $(\bar f_{Li} \gamma^\mu T^a f_{Lj})(\bar f'_{Lk} \gamma_\mu T^a f'_{Ll})$ and $(\bar f_{Ri} \gamma^\mu T^a f_{Rj})(\bar f'_{Rk} \gamma_\mu T^a f'_{Rl})$ are considered only for $f\neq f'$, we can extract the term of Eq.~\eqref{eq:LL1<-gammagamma} proportional to $(1-\delta_{ff'})$ and apply the substitution $e^2 Q_f Q_{f'} \to g_s^2 c_f c_{f'}$:
\begin{align}
    \gamma_{LL8_{ijkl} \leftarrow g,g } &= 
    \gamma_{LL8_{ijkl} \leftarrow \tilde g, \tilde g} = 
    \gamma_{RR8_{ijkl} \leftarrow g,g } =
    \gamma_{RR8_{ijkl} \leftarrow \tilde g, \tilde g}= \frac{2g_s^2}{3\pi^2 } c_f c_{f'}  \delta^{ij} \delta^{kl}\,.
\end{align}

\paragraph{$\gamma_{LR1 \leftarrow \gamma,\gamma}$, $\gamma_{LR1 \leftarrow \tilde \gamma,\tilde \gamma}$, $\gamma_{LR8\leftarrow g,g}$, and $\gamma_{LR8\leftarrow \tilde g, \tilde g}$.}
As for the mixed-chirality 4-fermion operators, the calculation of their anomalous dimensions is completely analogous to that corresponding to the same-chirality 4-fermion operators.
Regarding $\gamma_{LR1_{ijkl} \leftarrow \gamma,\gamma}$ and $\gamma_{LR1_{ijkl} \leftarrow \tilde \gamma,\tilde \gamma}$ we have
\begin{equation}
    \gamma_{LR1_{ijkl} \leftarrow \gamma,\gamma} = \gamma_{LR1_{ijkl} \leftarrow \tilde \gamma,\tilde \gamma} = \frac{2e^2}{3\pi^2}Q_f Q_{f'} \delta^{ij} \delta^{kl}\,,
\end{equation}
while for $\gamma_{LR8_{ijkl}\leftarrow g,g}$ and $\gamma_{LR8_{ijkl} \leftarrow \tilde g, \tilde g}$
\begin{equation}
    \gamma_{LR8_{ijkl}\leftarrow g,g} = \gamma_{LR8_{ijkl} \leftarrow \tilde g, \tilde g} = \frac{2g_s^2}{3\pi^2}c_f c_{f'} \delta^{ij} \delta^{kl}\,.
\end{equation}

\subsubsection{Renormalization group equations}

In the following, we summarize our results for
the renormalization group equations of the complete set of SM effective operators below the EW scale as induced by the presence of a CP-violating ALP:
\begin{gather}
    \mu \dv[]{d_G}{\mu} = - \frac{3g_s}{\pi^2}\mathcal{C}_g \tilde{\mathcal{C}}_g\,,
    \qquad
    \mu \dv[]{D_G}{\mu} = -\frac{3g_s}{2\pi^2}\left(\mathcal{C}_g^2 - \tilde{\mathcal{C}}_g^2\right)\,,
    \\
    \mu\dv[]{c_{\rm E}^{ij}}{\mu} =  \frac{e Q_f}{4 \pi^2} \left( \mathcal{Y}_S^{ij} \tilde{\mathcal{C}}_\gamma-\mathcal{Y}_P^{ij} \mathcal{C}_\gamma  \right) \,,
    \qquad
    \mu\dv[]{c_{\rm M}^{ij}}{\mu} = - \frac{e Q_f}{4 \pi^2} \left(\mathcal{Y}_P^{ij} \tilde{\mathcal{C}}_\gamma  + \mathcal{Y}_S^{ij} \mathcal{C}_\gamma\right) \,,
    \\
    \mu\dv[]{c_{\rm CE}^{ij}}{\mu} = \frac{g_s c_f}{4 \pi^2} \left( \mathcal{Y}_S^{ij} \tilde{\mathcal{C}}_g-\mathcal{Y}_P^{ij} \mathcal{C}_g  \right) \,,
    \qquad
    \mu\dv[]{c_{\rm CM}^{ij}}{\mu} = -\frac{g_s c_f}{4 \pi^2} \left(\mathcal{Y}_P^{ij} \tilde{\mathcal{C}}_g  + \mathcal{Y}_S^{ij} \mathcal{C}_g\right) \,,\\
    \begin{aligned}
    \mu\dv[]{c_{LL1}^{ijkl}}{\mu} &=  \delta_{ff'}\bigg[\frac{e^2Q_f^2}{6\pi^2}\left(\mathcal C_\gamma^2 + \tilde{\mathcal C}_\gamma^2\right)\delta^{ij}\delta^{kl}+\frac{g_s^2c_f^2}{12\pi^2}\left(\mathcal C_g^2 + \tilde{\mathcal C}_g^2\right)\bigg(\delta^{il}\delta^{kj}-\frac{1}{N_c}\delta^{ij}\delta^{kl}\bigg) \bigg] \\&\quad + (1-\delta_{ff'})\frac{e^2}{3\pi^2}Q_f Q_{f'}\left(\mathcal C_\gamma^2 + \tilde{\mathcal C}_\gamma^2\right)\delta^{ij}\delta^{kl}\,, 
    \end{aligned}\\
    \begin{aligned}
    \mu\dv[]{c_{RR1}^{ijkl}}{\mu} &= \delta_{ff'}\bigg[\frac{e^2Q_f^2}{6\pi^2}\left(\mathcal C_\gamma^2 + \tilde{\mathcal C}_\gamma^2\right)\delta^{ij}\delta^{kl}+\frac{g_s^2c_f^2}{12\pi^2}\left(\mathcal C_g^2 + \tilde{\mathcal C}_g^2\right)\bigg(\delta^{il}\delta^{kj}-\frac{1}{N_c}\delta^{ij}\delta^{kl}\bigg) \bigg] \\&\quad+ (1-\delta_{ff'})\frac{e^2}{3\pi^2}Q_f Q_{f'}\left(\mathcal C_\gamma^2 + \tilde{\mathcal C}_\gamma^2\right)\delta^{ij}\delta^{kl}\,, 
    \end{aligned}\\
    \mu\dv[]{c_{LL8}^{ijkl}}{\mu} = \frac{g_s^2}{3\pi^2}c_f c_{f'}\left(\mathcal C_g^2 + \tilde{\mathcal C}_g^2\right)\delta^{ij}\delta^{kl}\,,
    \qquad
    \mu\dv[]{c_{RR8}^{ijkl}}{\mu} = \frac{g_s^2}{3\pi^2}c_f c_{f'}\left(\mathcal C_g^2 + \tilde{\mathcal C}_g^2\right)\delta^{ij}\delta^{kl}\,,\\
    \mu\dv[]{c_{LR1}^{ijkl}}{\mu} = \frac{e^2}{3\pi^2}Q_f Q_{f'}\left(\mathcal C_\gamma^2 + \tilde{\mathcal C}_\gamma^2\right)\delta^{ij}\delta^{kl}\,,
    \qquad
    \mu\dv[]{c_{LR8}^{ijkl}}{\mu} = \frac{g_s^2}{3\pi^2}c_f c_{f'}\left(\mathcal C_g^2 + \tilde{\mathcal C}_g^2\right)\delta^{ij}\delta^{kl}\,.
\end{gather}

Our general findings improve and extend previous results in the literature \cite{Marciano:2016yhf,Galda:2023qjx,DiLuzio:2020oah}.
In particular, we reproduce the results of \cite{Galda:2023qjx} in the limit of a CP-odd ALP, where $\mathcal C_\gamma = \mathcal C_g = \mathcal Y_S^{ij} = 0$. Moreover, we also agree with the results of \cite{Marciano:2016yhf,DiLuzio:2020oah} where only the running of $d_G$, $c_{\text{E}}^{ij}$, and $c_{\text{CE}}^{ij}$ was computed.

\subsection{Renormalization of SM effective operators above the EW scale}

In the previous sections, we focused on RGE effects below the EW scale as they are expected to be dominant in the low-energy phenomenology of CP-violating ALPs \cite{Marciano:2016yhf,DiLuzio:2020oah}. Indeed, as long as these particles are considerably lighter than the EW scale, their phenomenology is mainly driven by operators in Eq.~\eqref{eq:Lag1} that are invariant under the unbroken SM gauge group $SU(3)_c\times U(1)_\text{em}$. The interactions of such particles with heavier states play a subdominant role on the low-energy observables that constitute the main phenomenological probes of such a NP candidate \cite{DiLuzio:2020oah,DiLuzio:2023cuk}. However, in the sight of embedding a CP-violating ALP in a UV-complete model, it is important to keep track of its interactions with the SM fields in the unbroken phase.
These are encoded in the following Lagrangian: 
\begin{align}
\mathcal{L}_\text{EFT} &= \tilde{\mathcal{C}}_{g}\frac{\phi}{\Lambda} G\tilde{G} + \tilde{\mathcal{C}}_{W}\frac{\phi}{\Lambda} W\tilde{W}+ \tilde{\mathcal{C}}_{B}\frac{\phi}{\Lambda} B\tilde{B} +\, \mathcal{C}_{g}\frac{\phi}{\Lambda} GG + \mathcal{C}_{W}\frac{\phi}{\Lambda} WW+ \mathcal{C}_{B}\frac{\phi}{\Lambda} BB \nonumber \\
&\quad + \frac{\phi}{\Lambda} \left(\bar{q}_p\tilde{\mathcal{Y}}^{pr}_u\tilde{H}u_r+ \bar{q}_p\tilde{\mathcal{Y}}^{pr}_d H d_r+ \bar{\ell}_p\tilde{\mathcal{Y}}^{pr}_e He_r + \text{h.c.}\right) + \mathcal{L}_{\text{SM}}\,,
\end{align}
where $\tilde{\mathcal{Y}}_{f}$ with $f = u,d,e$ are generic $3 \times 3$ matrices and where 
\begin{align}
\mathcal{L}_\text{SM} &= -\frac{1}{4}G_{\mu\nu}^aG^{a\mu\nu}-\frac{1}{4}W_{\mu\nu}^IW^{I\mu\nu}-\frac{1}{4}B_{\mu\nu}B^{\mu\nu}+(D_\mu H)^\dagger (D^\mu H)-\lambda \left(H^\dagger H-\frac{v^2}{2}\right)^2 \nonumber \\
&\quad + i (\bar \ell \gamma^\mu D_\mu \ell + \bar q \gamma^\mu D_\mu q + \bar e \gamma^\mu D_\mu e + \bar u \gamma^\mu D_\mu u + \bar d \gamma^\mu D_\mu d) \nonumber \\
&\quad - (Y_e^{pr} \bar \ell_p e_r H + Y_u^{pr} \bar q_p u_r \tilde H + Y_d^{pr} \bar q_p d_r H + \text{h.c.})\,,
\end{align}
with $D_\mu = \partial_\mu - i g_1 y B_\mu - i g_2 \frac{\sigma^I}{2}W_\mu^I-i g_s T^a G^a_\mu$.
The hypercharge assignments are $y_\ell = -1/2$, $y_e = -1$, $y_q=1/6$, $y_u=2/3$, $y_d = -1/3$, and $y_H = 1/2$.
The impact of the virtual exchange of a CP-violating ALP on the SMEFT operators as induced by the Lagrangian above can be readily computed by making use of the on-shell method. In this section we report our results, under the notation 
\begin{equation}
\mathcal L_{\text{SMEFT}}^{(6)} = \frac{1}{\Lambda^2} \sum_i C_i \,Q_i \,,
\end{equation}
where the $C_i$ denote the Wilson coefficients of the SMEFT in the Warsaw basis \cite{Grzadkowski:2010es}.

Our results reproduce those of \cite{Galda:2021hbr}, relative to a CP-odd ALP, and generalize them to include interactions of a CP-even ALP.
In particular, as we will discuss systematically in the following, in our framework of a CP-violating ALP we are forced to introduce and then renormalize dimension-six CP-odd operators, which were not accounted for before.

\subsubsection{Class $X^3$}

Operators in this class include 
\begin{align}
Q_G &= f^{abc} G_\mu^{a\nu}G_\nu^{b \rho}G_\rho^{c\mu}\,,  & Q_{\tilde{G}} &= f^{abc} \tilde G_\mu^{a\nu}G_\nu^{b \rho}G_\rho^{c\mu}\,,  \\
Q_W &= \epsilon^{IJK} W_\mu^{I\nu}W_\nu^{J \rho}W_\rho^{K\mu}\,,  & Q_{\tilde{W}} &= \epsilon^{IJK} \tilde{W}_\mu^{I\nu}W_\nu^{J \rho}W_\rho^{K\mu}\,. 
\end{align}

The computation of the RGEs for this class of operators proceeds in complete analogy to the renormalization of the same operators we have considered in Sec.~\ref{subsec:LEFT}. We report here the results:
\begin{align}
\mu\dv[]{}{\mu}{C}_G &= \frac{g_s}{2\pi^2}   \left(\tilde{\mathcal{C}}_{G}^2 - \mathcal{C}^2_{G} \right)\,, & \mu\dv[]{}{\mu}{C}_{\tilde{G}} &= -\frac{g_s}{\pi^2}  \,\tilde{\mathcal{C}}_{G}\mathcal{C}_{G}\,,  \\
\mu\dv[]{}{\mu}{C}_W &=\frac{g_2}{2\pi^2}   \left(\tilde{\mathcal{C}}_{W}^2 - \mathcal{C}^2_{W} \right)\,, & \mu\dv[]{}{\mu}{C}_{\tilde{W}} &= -\frac{g_2}{\pi^2} \, \tilde{\mathcal{C}}_{W}\mathcal{C}_{W}  \,.
\end{align}

\subsubsection{Class $X^2 H^2$}
Operators in this class include
\begin{align}
Q_{HG} &= H^\dagger H \, G_{\mu\nu}^a G^{a\mu\nu}\,, & Q_{H\tilde{G}}&= H^\dagger H\,\tilde{G}_{\mu\nu}^aG^{a\mu\nu}\,, \\
Q_{HW} &= H^\dagger H \, W_{\mu\nu}^I W^{I\mu\nu}\,, &Q_{H\tilde{W}}&= H^\dagger H\,\tilde{W}_{\mu\nu}^I W^{I\mu\nu}\,,  \\
Q_{HB} &= H^\dagger H \, B_{\mu\nu}B^{\mu\nu}\,, &Q_{H\tilde{B}}&= H^\dagger H\,\tilde{B}_{\mu\nu}B^{\mu\nu} \,,\\
Q_{HWB} &= H^\dagger \sigma^I H \, W_{\mu\nu}^I B^{\mu\nu}\,, & Q_{H\tilde W B}&= H^\dagger\sigma^I  H\,\tilde{W}_{\mu\nu}^I B^{\mu\nu} \,.
\end{align}

The RGEs for the operators in this class can in principle receive contributions from all the unitarity cuts in Fig.~\ref{fig:classX2H2}.
\begin{figure}[h]
\centering
	\begin{tikzpicture}
	\begin{feynman}[small]
		\node[red, dot](a);
		\vertex[right =  of a](b);
		\vertex[left =  of a](d);
		\vertex[below = 1.3 of a](f);
		\vertex[left = of f](g);
		\vertex[right = of f](h);
		\diagram*{
			(b) -- [scalar](a),
			(a) -- [photon](d),
			(f) -- [photon](a),
			(g) -- [charged scalar](f),
			(f) -- [charged scalar](h),
		};
	\end{feynman}
	\end{tikzpicture}
	\,
	\begin{tikzpicture}
	\draw[dashed,gray,thick] (0.0, -0.7) -- (0.0, 0.7);
	\end{tikzpicture}
	\,
	\begin{tikzpicture}
	\begin{feynman}[small]
		\node[red, dot](a);
		\vertex[left = of a](b);
		\vertex[right = of a](d);
		\vertex[below = 1.3 of a](f);
		\vertex[left = of f](g);
		\vertex[right = of f](h);
		\diagram*{
			(b) -- [scalar](a),
			(a) -- [photon](d),
			(f) -- [photon](a),
			(g) -- [charged scalar](f),
			(f) -- [charged scalar](h),
		};
	\end{feynman} 
	\end{tikzpicture}
	\qquad 
	\begin{tikzpicture}
	\begin{feynman}[small]
		\vertex(a);
		\vertex[right =  of a](b);
		\vertex[left =  of a](d);
		\vertex[below = 1.3 of a](f);
		\vertex[left = of f](g);
		\vertex[right = of f](h);
		\diagram*{
			(b) -- [photon](a),
			(a) -- [charged scalar](d),
			(f) -- [charged scalar](a),
			(g) -- [charged scalar](f),
			(f) -- [photon](h),
		};
	\end{feynman}
	\end{tikzpicture}
	\,
	\begin{tikzpicture}
	\draw[dashed,gray,thick] (0.0, -0.7) -- (0.0, 0.7);
	\end{tikzpicture}
	\,
	\begin{tikzpicture}
	\begin{feynman}[small]
		\node[red, dot](a);
		\vertex[left =  of a](b);
		\vertex[right = of a](d);
		\node[red, dot, below = 1.3 of a](f);
		\vertex[left = of f](g);
		\vertex[right = of f](h);
		\diagram*{
			(b) -- [photon](a),
			(a) -- [photon](d),
			(f) -- [scalar](a),
			(g) -- [photon](f),
			(f) -- [photon](h),
		};
	\end{feynman} 
	\end{tikzpicture}
    \qquad 
	\begin{tikzpicture}
	\begin{feynman}[small]
		\vertex(a);
		\vertex[above right =  of a](b);
		\vertex[ above left =  of a](d);
		\vertex[ below left = of a](g);
		\vertex[ below right = of a](h);
		\diagram*{
			(b) -- [photon](a),
			(a) -- [charged scalar](d),
			(g) -- [charged scalar](a),
			(a) -- [photon](h),
		};
	\end{feynman}
	\end{tikzpicture}
	\,
	\begin{tikzpicture}
	\draw[dashed,gray,thick] (0.0, -0.7) -- (0.0, 0.7);
	\end{tikzpicture}
	\,
	\begin{tikzpicture}
	\begin{feynman}[small]
		\node[red, dot](a);
		\vertex[left =  of a](b);
		\vertex[right =  of a](d);
		\node[red, dot, below = 1.3  of a](f);
		\vertex[left = of f](g);
		\vertex[right = of f](h);
		\diagram*{
			(b) -- [photon](a),
			(a) -- [photon](d),
			(f) -- [scalar](a),
			(g) -- [photon](f),
			(f) -- [photon](h),
		};
	\end{feynman} 
	\end{tikzpicture}
    \caption{Unitarity cuts that contribute to the RGEs of the class $X^2H^2$. Dashed lines with an arrow denote the hypercharge flow along a Higgs line, while dashed lines without arrows denote an ALP. Only the first cut contributes to the RGE due to helicity arguments (see main text).}
    \label{fig:classX2H2}
\end{figure}

However, it is easy to get convinced that only the leftmost unitarity cut can actually contribute.
Indeed, in order to have a non-zero contribution to the operator $H^2 X^2$, the two external gauge bosons must have the same helicity.
In addition to this, all the vector bosons involved in a dimension-5 ALP-SM vertex must have as well the same helicity. As a consequence, the two virtual gauge bosons in the second and third diagrams must have the same helicity.
While this is not a problem as far as the BSM amplitude is concerned, it is impossible to build a non-zero SM amplitude with two external Higgs fields and two gauge bosons with the same helicities.
Hence, only the first diagram can possibly provide a non-null contribution to the renormalization of the SMEFT operators in the class $X^2 H^2$. The results we find read:
\begin{align}
\mu \dv[]{}{\mu} {C}_{HG} &= 0\,, & \mu \dv[]{}{\mu} {C}_{H\tilde{G}} &= 0\,,  \\
\mu \dv[]{}{\mu} {C}_{HW} &= -\frac{g_2^2}{8\pi^2}  \left(\tilde{\mathcal{C}}_{W}^2 - \mathcal{C}^2_{W} \right)\,, & \mu \dv[]{}{\mu} {C}_{H\tilde{W}} &=  \frac{g_2^2}{4\pi^2}\,\mathcal{C}_{W} \tilde{\mathcal{C}}_{W}\,,  \\
\mu \dv[]{}{\mu} {C}_{HB} &= -\frac{g_1^2}{8\pi^2}  \left(\tilde{\mathcal{C}}_{B}^2 - \mathcal{C}^2_{B} \right)\,, & \mu \dv[]{}{\mu} {C}_{H\tilde{B}} &= \frac{g_1^2}{4\pi^2} \, \mathcal{C}_{B} \tilde{\mathcal{C}}_{B}\,,    \\
\mu \dv[]{}{\mu} {C}_{HWB} &= -\frac{g_1 g_2}{4\pi^2}  \left(\tilde{\mathcal{C}}_{W}\tilde{\mathcal{C}}_{B} - \mathcal{C}_{W}\mathcal{C}_{B} \right)\,,
 &
 \mu \dv[]{}{\mu} {C}_{H\tilde{W}B} &= \frac{g_1 g_2}{4\pi^2}\left(\mathcal{C}_{W} \tilde{\mathcal{C}}_{B}+ \mathcal{C}_{B} \tilde{\mathcal{C}}_{W}\right)\,.
\end{align}

\subsubsection{Classes $H^6$ and $H^4D^2$}

The operators belonging to these classes are given by: 
\begin{align}
Q_H &= (H^\dagger H)^3\,, \\
Q_{H\Box} &= (H^\dagger H)\Box(H^\dagger H)\,, \\
Q_{HD} &= (H^\dagger D^\mu H)^*(H^\dagger D_\mu H)\,.
\end{align}

The relevant unitarity cuts for this class of operators are reported in Fig.~\ref{fig:classH6}. 

\begin{figure}[h]  
\centering
	\begin{tikzpicture}
	\begin{feynman}[small]
		\vertex(a);
		\vertex[above left = of a](b);
		\vertex[below left = of a](c);
		\node[red, dot, right = of a ](d);
		\vertex[above right = of d](e);
		\vertex[below right = of d](f);
		\diagram*{
			(a) -- [charged scalar](b),
			(c) -- [charged scalar](a),
			(a) -- [photon](d),
			(d) -- [scalar](e),
			(d) -- [photon](f),
		};
	\end{feynman}
	\end{tikzpicture}
	\,
	\begin{tikzpicture}
	\draw[dashed,gray,thick] (0.0, -0.7) -- (0.0, 0.7);
	\end{tikzpicture}
	\,
	\begin{tikzpicture}
	\begin{feynman}[small]
		\node[red, dot](a);
		\vertex[above left = of a](b);
		\vertex[below left = of a](c);
		\vertex[right = of a ](d);
		\vertex[above right = of d](e);
		\vertex[below right = of d](f);
		\diagram*{
			(a) -- [scalar](b),
			(c) -- [photon](a),
			(a) -- [photon](d),
			(d) -- [charged scalar](e),
			(f) -- [charged scalar](d),
		};
	\end{feynman} 
	\end{tikzpicture}
    \qquad \qquad \qquad 
    \begin{tikzpicture}
	\begin{feynman}[small]
		\vertex(a);
		\vertex[above left = of a](b);
		\vertex[below left = of a](c);
        \vertex[above left = 0.5 of a](a1);
        \vertex[below left = 0.5 of a1](a2);
        \vertex[ above right = 0.5 of a1](a3);
		\node[red, dot, right = of a ](d);
		\vertex[above right = of d](e);
		\vertex[below right = of d](f);
		\diagram*{
			(a) -- [charged scalar](a1),
            (a1) -- [charged scalar](b),
			(c) -- [charged scalar](a),
			(a) -- [photon](d),
			(d) -- [scalar](e),
			(d) -- [photon](f),
            (a1) -- [charged scalar](a2),
            (a3) -- [charged scalar](a1),
		};
	\end{feynman}
	\end{tikzpicture}
	\,
	\begin{tikzpicture}
	\draw[dashed,gray,thick] (0.0, -0.7) -- (0.0, 0.7);
	\end{tikzpicture}
	\,
	\begin{tikzpicture}
	\begin{feynman}[small]
		\node[red, dot](a);
		\vertex[above left= of a](b);
		\vertex[below left= of a](c);
		\vertex[right = of a ](d);
		\vertex[above right = of d](e);
		\vertex[below right = of d](f);
		\diagram*{
			(a) -- [scalar](b),
			(c) -- [photon](a),
			(a) -- [photon](d),
			(d) -- [charged scalar](e),
			(f) -- [charged scalar](d),
		};
	\end{feynman} 
	\end{tikzpicture}
\caption{Unitarity cuts that contribute to the RGEs of the classes $H^6$ and $H^4 D^2$. Dashed lines with an arrow denote the hypercharge flow along a Higgs line, while dashed lines without arrows denote an ALP.}
\label{fig:classH6}
\end{figure}

The corresponding RGEs read:
\begin{align}
\mu \dv[]{}{\mu} {C}_{H} &= \frac{\lambda  g_2^2}{3\pi^2}  \left(\mathcal{C}_{W}^2+ \tilde{\mathcal{C}}_{W}^2 \right)\,,  \\
\mu \dv[]{}{\mu} {C}_{H\Box} &= \frac{g_1^2}{24\pi^2}   \left(\mathcal{C}_{B}^2+ \tilde{\mathcal{C}}_{B}^2\right)+ \frac{g_2^2}{8\pi^2}  \left(\mathcal{C}_{W}^2+ \tilde{\mathcal{C}}_{W}^2\right)\,, \\
\mu \dv[]{}{\mu} {C}_{HD} &= \frac{g_1^2}{6\pi^2}   \left (\mathcal{C}_{B}^2+ \tilde{\mathcal{C}}_{B}^2 \right)\,.
\end{align}

\subsubsection{Class $\psi^2 H X$}

In this class one finds the following dipole operators, accompanied with their Hermitian conjugate: 
\begin{align}
Q_{uB}^{pr} &= (\bar{q}_p\sigma^{\mu\nu} u_r)\,\tilde H\,B_{\mu\nu}\,,  &Q_{uW}^{pr} &= (\bar{q}_p\sigma^{\mu\nu} \sigma^I u_r)\,\tilde H\,W^I_{\mu\nu}\,, \\
 Q_{dB}^{pr} &= (\bar{q}_p\sigma^{\mu\nu} d_r)\,H\,B_{\mu\nu}\,,  & Q_{dW}^{pr} &= (\bar{q}_p\sigma^{\mu\nu} \sigma^I d_r)\,H\,W^I_{\mu\nu} \,,  \\
Q_{eB}^{pr} &= (\bar{\ell}_p\sigma^{\mu\nu} e_r)\,H\,B_{\mu\nu}\,, &Q_{eW}^{pr} &= (\bar{\ell}_p\sigma^{\mu\nu} \sigma^I e_r)\,H\,W^I_{\mu\nu}\,, \\
Q_{uG}^{pr} &= (\bar{q}_p\sigma^{\mu\nu} T^a u_r)\,\tilde H\,G^a_{\mu\nu}\,, & Q_{dG}^{pr} &= (\bar{q}_p\sigma^{\mu\nu} T^a d_r)\,H\,G^a_{\mu\nu}\,.
\end{align}

The computations for this class of RGEs proceed in complete analogy to the case of dipole operators in Sec.~\ref{subsec:LEFT} and we report here the results:
\begin{align}
    \mu \dv[]{}{\mu}  C_{uB}^{pr} &=\frac{5 g_1}{48\pi^2}  \tilde{\mathcal{Y}}_u^{pr} ({\mathcal{C}}_{B} + i  \tilde{\mathcal{C}}_{B})\,,
    &
    \mu \dv[]{}{\mu}  C_{uW}^{pr} &=\frac{g_2}{16\pi^2}  
 \tilde{\mathcal{Y}}_u^{pr} ({\mathcal{C}}_{W} + i  \tilde{\mathcal{C}}_{W})\,,  \\
 \mu \dv[]{}{\mu}  C_{dB}^{pr} &= -\frac{g_1}{48\pi^2}  \tilde{\mathcal{Y}}_d^{pr} ({\mathcal{C}}_{B} + i  \tilde{\mathcal{C}}_{B})\,,
    &
    \mu \dv[]{}{\mu}  C_{dW}^{pr} &= \frac{g_2}{16\pi^2} 
 \tilde{\mathcal{Y}}_d^{pr} ({\mathcal{C}}_{W} + i  \tilde{\mathcal{C}}_{W})\,,  \\
 \mu \dv[]{}{\mu}  C_{eB}^{pr} &= -\frac{3g_1}{16\pi^2}  \tilde{\mathcal{Y}}_e^{pr} ({\mathcal{C}}_{B} + i  \tilde{\mathcal{C}}_{B})\,,
    &
    \mu \dv[]{}{\mu}  C_{eW}^{pr} &= \frac{g_2}{16\pi^2} 
 \tilde{\mathcal{Y}}_e^{pr} ({\mathcal{C}}_{W} + i  \tilde{\mathcal{C}}_{W})\,,  \\
 \mu \dv[]{}{\mu}  C_{uG}^{pr} &= \frac{g_s }{4\pi^2} 
 \tilde{\mathcal{Y}}_u^{pr} ({\mathcal{C}}_{G} +i  \tilde{\mathcal{C}}_{G})\,, &
 \mu \dv[]{}{\mu}  C_{dG}^{pr} &= \frac{g_s}{4\pi^2}  
 \tilde{\mathcal{Y}}_d^{pr} ({\mathcal{C}}_{G} +i  \tilde{\mathcal{C}}_{G})\,.
\end{align}

\subsubsection{Class $\psi^2 H^2 D$}

In this class we find the following set of operators:
\begin{align}
Q_{H\ell}^{(1)pr} &= (H^\dagger i \overset{\leftrightarrow}{D}_\mu H)(\bar{\ell}_p \gamma^\mu \ell_r)\,,  &Q_{H \ell}^{(3)pr} &= (H^\dagger i \overset{\leftrightarrow}{D}{}^I_\mu H)(\bar{\ell}_p \sigma^I \gamma^\mu \ell_r) \,, \\
Q_{Hq}^{(1)pr} &= (H^\dagger i \overset{\leftrightarrow}{D}_\mu H)(\bar{q}_p \gamma^\mu q_r)\,,  &Q_{Hq}^{(3)pr} &= (H^\dagger i \overset{\leftrightarrow}{D}{}^I_\mu H)(\bar{q}_p \sigma^I \gamma^\mu q_r)\,, \\
Q_{He}^{pr} &= (H^\dagger i \overset{\leftrightarrow}{D}_\mu H)(\bar{e}_p \gamma^\mu e_r)\,,  &Q_{Hu}^{pr} &= (H^\dagger i \overset{\leftrightarrow}{D}_\mu H)(\bar{u}_p  \gamma^\mu u_r)\,, \\
Q_{Hd}^{pr} &= (H^\dagger i \overset{\leftrightarrow}{D}_\mu H)(\bar{d}_p \gamma^\mu d_r)\,,  &Q_{Hud}^{pr} &= i(\tilde{H}^\dagger D_\mu H)(\bar{u}_p  \gamma^\mu d_r) \,.
\end{align}

The unitarity cuts relevant to the computations of the RGEs for this class of operators are reported in Fig.~\ref{fig:classespsi2}. 

\begin{figure}[h]
\centering
	\begin{tikzpicture}
	\begin{feynman}[small]
		\node[red, dot](a);
		\vertex[above left = of a](b);
		\vertex[below left = of a](c);
		\vertex[below right = of a](e);
		\vertex[above right = of a](f);
		\diagram*{
			(b) -- [charged scalar](a),
			(c) -- [fermion](a),
			(a) -- [fermion](e),
			(a) -- [scalar](f),
		};
	\end{feynman}
	\end{tikzpicture}
	\,
	\begin{tikzpicture}
	\draw[dashed,gray,thick] (0.0, -0.7) -- (0.0, 0.7);
	\end{tikzpicture}
	\,
	\begin{tikzpicture}
	\begin{feynman}[small]
		\node[red, dot](a);
		\vertex[left =  of a](b);
		\vertex[right =  of a](d);
		\vertex[below = 1.3 of a](f);
		\vertex[left = of f](g);
		\vertex[right = of f](h);
		\diagram*{
			(b) -- [scalar](a),
			(a) -- [photon](d),
			(f) -- [photon](a),
			(g) -- [fermion](f),
			(f) -- [fermion](h),
		};
	\end{feynman} 
	\end{tikzpicture}
        \quad \quad
    \begin{tikzpicture}
    	\begin{feynman}[small]
		\node[red, dot](a);
		\vertex[above left = of a](b);
		\vertex[below left = of a](c);
		\vertex[above right = of a](e);
		\vertex[below right = of a](f);
		\diagram*{
			(b) -- [charged scalar](a),
			(c) -- [fermion](a),
			(a) -- [fermion](e),
			(a) -- [scalar](f),
		};
	\end{feynman}
	\end{tikzpicture}
	\,
	\begin{tikzpicture}
	\draw[dashed,gray,thick] (0.0, -0.7) -- (0.0, 0.7);
	\end{tikzpicture}
	\,
	\begin{tikzpicture}
	\begin{feynman}[small]
		\node[red, dot](a);
		\vertex[above left = of a](b);
		\vertex[below left = of a](c);
		\vertex[above right = of a](e);
		\vertex[below right = of a](f);
		\diagram*{
			(b) -- [fermion](a),
			(a) -- [scalar](c),
			(a) -- [charged scalar](e),
			(a) -- [fermion](f),
		};
	\end{feynman} 
	\end{tikzpicture}
    \quad \quad 
    \begin{tikzpicture}
	\begin{feynman}[small]
		\vertex(a);
		\vertex[above left = of a](b);
		\vertex[below left = of a](c);
		\node[red, dot, right = of a ](d);
		\vertex[above right = of d](e);
		\vertex[below right = of d](f);
		\diagram*{
			(a) -- [fermion](b),
			(c) -- [fermion](a),
			(a) -- [photon](d),
			(d) -- [scalar](e),
			(d) -- [photon](f),
		};
	\end{feynman}
	\end{tikzpicture}
	\,
	\begin{tikzpicture}
	\draw[dashed,gray,thick] (0.0, -0.7) -- (0.0, 0.7);
	\end{tikzpicture}
	\,
	\begin{tikzpicture}
	\begin{feynman}[small]
		\node[red, dot](a);
		\vertex[above left = of a](b);
		\vertex[below left = of a](c);
		\vertex[right = of a ](d);
		\vertex[above right = of d](e);
		\vertex[below right = of d](f);
		\diagram*{
			(a) -- [scalar](b),
			(c) -- [photon](a),
			(a) -- [photon](d),
			(d) -- [charged scalar](e),
			(f) -- [charged scalar](d),
		};
	\end{feynman} 
	\end{tikzpicture}
    \caption{Unitarity cuts that contribute to the RGEs of the classes $\psi^2 XH$ and $\psi^2 H^2 D$ (right). Dashed lines with an arrow denote the hypercharge flow along a Higgs line, while dashed lines without arrows denote an ALP.}
    \label{fig:classespsi2}
\end{figure}

The corresponding RGEs read:
\begin{align}
\mu \dv[]{}{\mu} {C}_{H\ell}^{(1)pr} &=\frac{1}{16\pi^2}\left[ \frac{1}{4}(\tilde{\mathcal{Y}}_e\tilde{\mathcal{Y}}_e^\dagger)^{pr} - \frac{4}{3} \,g_1^2 \,(\tilde{\mathcal{C}}_{B}^2+ \mathcal{C}_{B}^2)\delta^{pr}\right] \,,  \\
\mu \dv[]{}{\mu} {C}_{H\ell}^{(3)pr} &=\frac{1}{16\pi^2}\left[ \frac{1}{4}(\tilde{\mathcal{Y}}_e\tilde{\mathcal{Y}}_e^\dagger)^{pr} + \frac{4}{3}\, g_2^2\,(\tilde{\mathcal{C}}_{W}^2+ \mathcal{C}_{W}^2)\delta^{pr}\right] \,,  \\
\mu \dv[]{}{\mu} {C}_{He}^{pr} &=\frac{1}{16\pi^2}\left[ -\frac{1}{2}(\tilde{\mathcal{Y}}_e^\dagger \tilde{\mathcal{Y}}_e)^{pr} - \frac{8}{3} \,g_1^2 \,(\tilde{\mathcal{C}}_{B}^2+ \mathcal{C}_{B}^2)\delta^{pr}\right] \,,  \\
\mu \dv[]{}{\mu} {C}_{Hq}^{(1)pr} &=\frac{1}{16\pi^2}\left[ \frac{1}{4}(\tilde{\mathcal{Y}}_d\tilde{\mathcal{Y}}_d^\dagger-\tilde{\mathcal{Y}}_u\tilde{\mathcal{Y}}_u^\dagger)^{pr} + \frac{4}{9} \,g_1^2 \,(\tilde{\mathcal{C}}_{B}^2+ \mathcal{C}_{B}^2) \delta^{pr}\right] \,, \\
\mu \dv[]{}{\mu} {C}_{Hq}^{(3)pr} &=\frac{1}{16\pi^2}\left[ \frac{1}{4}(\tilde{\mathcal{Y}}_d\tilde{\mathcal{Y}}_d^\dagger+\tilde{\mathcal{Y}}_u\tilde{\mathcal{Y}}_u^\dagger)^{pr} + \frac{4}{3}\, g_2^2\,(\tilde{\mathcal{C}}_{W}^2+ \mathcal{C}_{W}^2) \delta^{pr}\right] \,,  \\
\mu \dv[]{}{\mu} {C}_{Hd}^{pr} &=\frac{1}{16\pi^2}\left[ -\frac{1}{2}(\tilde{\mathcal{Y}}_d^\dagger \tilde{\mathcal{Y}}_d)^{pr} - \frac{8}{9} \,g_1^2 \,(\tilde{\mathcal{C}}_{B}^2+ \mathcal{C}_{B}^2)\delta^{pr}\right] \,,  \\
\mu \dv[]{}{\mu} {C}_{Hu}^{pr} &=\frac{1}{16\pi^2}\left[ \frac{1}{2}(\tilde{\mathcal{Y}}_u^\dagger \tilde{\mathcal{Y}}_u)^{pr} + \frac{16}{9} \,g_1^2 \,(\tilde{\mathcal{C}}_{B}^2+ \mathcal{C}_{B}^2)\delta^{pr}\right] \,,  \\
\mu \dv[]{}{\mu} {C}_{Hud}^{pr} &= -\frac{1}{16\pi^2}(\tilde{\mathcal{Y}}_u^\dagger \tilde{\mathcal{Y}}_d)^{pr}\,.
\end{align}

\subsubsection{Class $\psi^2 H^3$}

The three operators in this class read:
\begin{align}
Q_{eH}^{pr} = (H^\dagger H)(\bar{\ell}_p e_r H)\,, \qquad Q_{uH}^{pr} = (H^\dagger H)(\bar{q}_p u_r \tilde{H})\,, \qquad
Q_{dH}^{pr} = (H^\dagger H)(\bar{q}_p d_r H)\,.
\end{align}

The unitarity cuts relevant to the computation of the RGEs for such a class of operators are depicted in Fig. \ref{fig:classpsi2H3}. 

\begin{figure}[h]
\centering
    \begin{tikzpicture}
	\begin{feynman}[small]
		\node[red, dot](a);
		\vertex[above left = of a](b);
		\vertex[below left = of a](c);
		\vertex[below right = of a](e);
		\vertex[above right = of a](f);
		\diagram*{
			(b) -- [charged scalar](a),
			(c) -- [fermion](a),
			(a) -- [fermion](e),
			(a) -- [scalar](f),
		};
	\end{feynman}
	\end{tikzpicture}
	\,
	\begin{tikzpicture}
	\draw[dashed,gray,thick] (0.0, -0.7) -- (0.0, 0.7);
	\end{tikzpicture}
	\,
	\begin{tikzpicture}
	\begin{feynman}[small]
		\node[red, dot](a);
		\vertex[left = of a](b);
		\vertex[above = of a](c);
		\vertex[right = of a](d);
		\vertex[below = 1.3 of a](f);
		\vertex[left = of f](g);
		\vertex[right = of f](h);
		\diagram*{
			(b) -- [scalar](a),
			(a) -- [fermion](c),
			(a) -- [charged scalar](d),
			(f) -- [fermion](a),
			(g) -- [fermion](f),
			(h) -- [charged scalar](f),
		};
	\end{feynman} 
	\end{tikzpicture}
	\qquad\qquad\qquad\qquad
    \begin{tikzpicture}
	   \begin{feynman}[small]
		    \vertex(a);
		      \vertex[above left = of a](b);
		      \vertex[below left = of a](c);
            \vertex[above left = 0.5 of a](a1);
            \vertex[above right = 0.5 of a1](a2);
		      \node[red, dot, right = of a ](d);
		      \vertex[above right = of d](e);
		      \vertex[below right = of d](f);
		      \diagram*{
			     (a) -- [fermion](a1),
              (a1) -- [fermion](b),
			     (c) -- [fermion](a),
			     (a) -- [photon](d),
			     (d) -- [scalar](e),
			     (d) -- [photon](f),
              (a1) -- [charged scalar](a2),
		};
	\end{feynman}
	\end{tikzpicture}
	\,
	\begin{tikzpicture}
	\draw[dashed,gray,thick] (0.0, -0.7) -- (0.0, 0.7);
	\end{tikzpicture}
	\,
	\begin{tikzpicture}
	\begin{feynman}[small]
		\node[red, dot](a);
		\vertex[above left = of a](b);
		\vertex[below left = of a](c);
		\vertex[right = of a ](d);
		\vertex[above right = of d](e);
		\vertex[below right = of d](f);
		\diagram*{
			(a) -- [scalar](b),
			(c) -- [photon](a),
			(a) -- [photon](d),
			(d) -- [charged scalar](e),
			(f) -- [charged scalar](d),
		};
	\end{feynman} 
	\end{tikzpicture}
    \caption{Unitarity cuts that contribute to the RGEs of the class $\psi^2 H^3$. Dashed lines with an arrow denote the hypercharge flow along a Higgs line, while dashed lines without arrows denote an ALP.}
    \label{fig:classpsi2H3}
\end{figure}

The associated RGEs read:
\begin{equation}
 \mu \dv[]{}{\mu} {{C}}_{fH}^{pr} =-\frac{1}{16\pi^2}\left[ 2  ( \tilde{\mathcal{Y}}_f {Y}_f^\dagger \tilde{\mathcal{Y}}_f)^{pr} + \frac{1}{2} (\tilde{\mathcal{Y}}_f \tilde{\mathcal{Y}}_f^\dagger {Y}_f)^{pr} + \frac{1}{2} ({Y}_f \tilde{\mathcal{Y}}_f^\dagger \tilde{\mathcal{Y}}_f)^{pr} - \frac{4}{3}g_2^2 Y_f^{pr}
 (\tilde{\mathcal{C}}_{W}^2 +  \mathcal{C}_{W}^2)\right]
\end{equation}
with $f = e, u, d$.

\subsubsection{Four-fermion operators}

In this section we report the RGEs relative to four-fermion SMEFT operators. The computations mimic those already performed for the case of the LEFT. Representative unitarity cuts are reported in Fig.~\ref{fig:class4fermions}.

\begin{figure}[h]
\centering
	\begin{tikzpicture}
	\begin{feynman}[small]
		\vertex(a);
		\vertex[above left = of a](b);
		\vertex[below left = of a](c);
		\node[red, dot, right = of a ](d);
		\vertex[above right = of d](e);
		\vertex[below right = of d](f);
		\diagram*{
			(a) -- [fermion](b),
			(c) -- [fermion](a),
			(a) -- [photon](d),
			(d) -- [scalar](e),
			(d) -- [photon](f),
		};
	\end{feynman}
	\end{tikzpicture}
	\,
	\begin{tikzpicture}
	\draw[dashed,gray,thick] (0.0, -0.7) -- (0.0, 0.7);
	\end{tikzpicture}
	\,
	\begin{tikzpicture}
	\begin{feynman}[small]
		\node[red, dot](a);
		\vertex[above left = of a](b);
		\vertex[below left = of a](c);
		\vertex[right = of a ](d);
		\vertex[above right = of d](e);
		\vertex[below right = of d](f);
		\diagram*{
			(a) -- [scalar](b),
			(c) -- [photon](a),
			(a) -- [photon](d),
			(d) -- [fermion](e),
			(f) -- [fermion](d),
		};
	\end{feynman} 
	\end{tikzpicture}
\qquad \qquad \qquad \qquad
    	\begin{tikzpicture}
	\begin{feynman}[small]
		\node[red, dot](a);
		\vertex[above right = of a](b);
		\vertex[below left = of a](c);
		\vertex[above left = of a](e);
		\vertex[below right = of a](f);
		\diagram*{
			(a) -- [scalar](b),
			(c) -- [fermion](a),
			(a) -- [fermion](e),
			(a) -- [charged scalar](f),
		};
	\end{feynman}
	\end{tikzpicture}
	\,
	\begin{tikzpicture}
	\draw[dashed,gray,thick] (0.0, -0.7) -- (0.0, 0.7);
	\end{tikzpicture}
	\,
	\begin{tikzpicture}
	\begin{feynman}[small]
		\node[red, dot](a);
		\vertex[above right = of a](b);
		\vertex[below left = of a](c);
		\vertex[above left = of a](e);
		\vertex[below right = of a](f);
		\diagram*{
			(a) -- [fermion](b),
			(c) -- [charged scalar](a),
			(a) -- [scalar](e),
			(f) -- [fermion](a),
		};
	\end{feynman} 
	\end{tikzpicture}
\caption{Unitarity cuts that contribute to the RGEs of four-fermion operators. The left diagram contributes to the classes of operators $(\bar{L}L)(\bar{L}L)$, $(\bar{R}R)(\bar{R}R)$, and $(\bar{R}R)(\bar{L}L)$, while the one on the right contributes to $(\bar{R}R)(\bar{L}L)$, $(\bar{R}L)(\bar{L}R)$, and $(\bar{L}R)(\bar{L}R)$. Dashed lines with an arrow denote the hypercharge flow along a Higgs line, while dashed lines without arrows denote an ALP.}
\label{fig:class4fermions}
\end{figure}

\paragraph{Class $(\bar{L}L)(\bar{L}L)$.}

The operators in this class read:
\begin{align}
Q_{\ell \ell}^{prst} &= (\bar{\ell}_p \gamma^\mu \ell_r)(\bar{\ell}_s \gamma_\mu \ell_t)\,,&&\\ Q_{q q}^{(1)prst} &= (\bar{q}_p \gamma^\mu q_r)(\bar{q}_s \gamma_\mu q_t)\,,  &Q_{q q}^{(3)prst} &= (\bar{q}_p \gamma^\mu \sigma^I q_r)(\bar{q}_s \gamma_\mu \sigma^I q_t)\,,   \\
Q_{q \ell}^{(1)prst} &= (\bar{\ell}_p \gamma^\mu \ell_r)(\bar{q}_s \gamma_\mu q_t)\,,  &Q_{q \ell}^{(3)prst} &= (\bar{\ell}_p \gamma^\mu \sigma^I \ell_r)(\bar{q}_s \gamma_\mu \sigma^I q_t)\,,
\end{align}
while the corresponding RGEs are given by
\begin{align}
    \mu \dv[]{}{\mu}{C}_{\ell\ell}^{prst} &= \frac{1}{24\pi^2}\left[ g_2^2(2 \delta^{pt}\delta^{sr}- \delta^{pr}\delta^{st})(\tilde{\mathcal{C}}_{W}^2 + \mathcal{C}_{W}^2)+ g_1^2\, \delta^{pr}\delta^{st}\,(\tilde{\mathcal{C}}_{B}^2 + \mathcal{C}_{B}^2)\right]\, ,  \\
    \mu \dv[]{}{\mu}{C}_{qq}^{(1)prst} &=\frac{1}{24\pi^2}\left[ g_s^2\bigg(\delta^{pt}\delta^{sr}-\frac{2}{N_c} \delta^{pr}\delta^{st}\bigg)(\tilde{\mathcal{C}}_{G}^2 + \mathcal{C}_{G}^2)+ \frac{1}{9}g_1^2 \,\delta^{pr}\delta^{st}\,(\tilde{\mathcal{C}}_{B}^2 + \mathcal{C}_{B}^2)\right]\, , \\
    \mu \dv[]{}{\mu}{C}_{qq}^{(3)prst} &=\frac{1}{24\pi^2}\left[ g_s^2\,\delta^{pt}\delta^{sr}\,(\tilde{\mathcal{C}}_{G}^2 + \mathcal{C}_{G}^2)+ g_2^2\,\delta^{pr}\delta^{st}\,(\tilde{\mathcal{C}}_{W}^2 + \mathcal{C}_{W}^2)\right] \,, \\
    \mu \dv[]{}{\mu}{C}_{\ell q}^{(1)prst} &= -\frac{g_1^2}{36\pi^2}\,  \delta^{pr}\delta^{st}\,(\tilde{\mathcal{C}}_{B}^2 + \mathcal{C}_{B}^2)\, , \\
    \mu \dv[]{}{\mu}{C}_{\ell q}^{(3)prst} &=\frac{g_2^2}{12\pi^2}  \,\delta^{pr}\delta^{st}\,(\tilde{\mathcal{C}}_{W}^2 + \mathcal{C}_{W}^2)\,.
\end{align}

\paragraph{Class $(\bar{R}R)(\bar{R}R)$.}

The operators in this class read:
\begin{align}
Q_{ee}^{prst} &= (\bar{e}_p \gamma^\mu e_r)(\bar{e}_s \gamma_\mu e_t)\,, && \\ Q_{uu}^{prst} &= (\bar{u}_p \gamma^\mu u_r)(\bar{u}_s \gamma_\mu u_t)\,, & Q_{dd}^{prst} &= (\bar{d}_p \gamma^\mu d_r)(\bar{d}_s \gamma_\mu d_t)\,, \\
Q_{eu}^{prst} &= (\bar{e}_p \gamma^\mu e_r)(\bar{u}_s \gamma_\mu u_t)\,, & Q_{ed}^{prst} &= (\bar{e}_p \gamma^\mu e_r)(\bar{d}_s \gamma_\mu d_t)\,,\\
Q_{ud}^{(1)prst} &= (\bar{u}_p \gamma^\mu u_r)(\bar{d}_s \gamma_\mu d_t)\,, & Q_{ud}^{(8)prst} &= (\bar{u}_p \gamma^\mu T^a u_r)(\bar{d}_s \gamma_\mu T^a d_t)\,,
\end{align}
while the corresponding RGEs are given by
\begin{align}
    \mu \dv[]{}{\mu}{C}_{ee}^{prst} &=\frac{1}{6\pi^2}  g_1^2\, \delta^{pr}\delta^{st}\,(\tilde{\mathcal{C}}_{B}^2 + \mathcal{C}_{B}^2)\, , \\
    \mu \dv[]{}{\mu}{C}_{uu}^{prst} &=\frac{1}{108\pi^2}\left[  8g_1^2\, \delta^{pr}\delta^{st}\,(\tilde{\mathcal{C}}_{B}^2 + \mathcal{C}_{B}^2) +9g_s^2\bigg(\delta^{pt}\delta^{sr}-\frac{1}{N_c} \delta^{pr}\delta^{st}\bigg)(\tilde{\mathcal{C}}_{G}^2 + \mathcal{C}_{G}^2)\right]\, , \\
    \mu \dv[]{}{\mu}{C}_{dd}^{prst} &=\frac{1}{108\pi^2}\left[  2g_1^2\, \delta^{pr}\delta^{st}\,(\tilde{\mathcal{C}}_{B}^2 + \mathcal{C}_{B}^2) +9g_s^2\bigg(\delta^{pt}\delta^{sr}-\frac{1}{N_c} \delta^{pr}\delta^{st}\bigg)(\tilde{\mathcal{C}}_{G}^2 + \mathcal{C}_{G}^2)\right]\, , \\
    \mu \dv[]{}{\mu}{C}_{eu}^{prst} &=  -\frac{2}{9\pi^2}g_1^2\, \delta^{pr}\delta^{st}\,(\tilde{\mathcal{C}}_{B}^2 + \mathcal{C}_{B}^2)\, , \\
    \mu \dv[]{}{\mu}{C}_{ed}^{prst} &=  \frac{1}{9\pi^2}g_1^2\, \delta^{pr}\delta^{st}\,(\tilde{\mathcal{C}}_{B}^2 + \mathcal{C}_{B}^2)\, , \\
    \mu \dv[]{}{\mu}{C}_{ud}^{(1)prst} &=  -\frac{2}{27\pi^2}g_1^2\, \delta^{pr}\delta^{st}\,(\tilde{\mathcal{C}}_{B}^2 + \mathcal{C}_{B}^2)\, , \\ 
    \mu \dv[]{}{\mu}{C}_{ud}^{(8)prst} &=  \frac{1}{3\pi^2}g_s^2\,\delta^{pr}\delta^{st}\,(\tilde{\mathcal{C}}_{G}^2 + \mathcal{C}_{G}^2) \,.
\end{align}

\paragraph{Class $(\bar{L}L)(\bar{R}R)$.}

The operators in this class read:
\begin{align}
Q_{\ell e}^{prst} &= (\bar{\ell}_p \gamma^\mu \ell_r)(\bar{e}_s \gamma_\mu e_t)\,, & Q_{\ell u}^{prst} &= (\bar{\ell}_p \gamma^\mu \ell_r)(\bar{u}_s \gamma_\mu u_t)\,, \\
Q_{\ell d}^{prst}&= (\bar{\ell}_p \gamma^\mu \ell_r)(\bar{d}_s \gamma_\mu d_t)\,, & Q_{qe}^{prst} &= (\bar{q}_p \gamma^\mu q_r)(\bar{e}_s \gamma_\mu e_t)\,,\\
Q_{qu}^{(1)prst} &= (\bar{q}_p \gamma^\mu q_r)(\bar{u}_s \gamma_\mu u_t)\,, & Q_{qu}^{(8)prst} &= (\bar{q}_p \gamma^\mu T^a q_r)(\bar{u}_s \gamma_\mu T^au_t)\,,\\
Q_{qd}^{(1)prst} &= (\bar{q}_p \gamma^\mu q_r)(\bar{d}_s \gamma_\mu d_t)\,, & Q_{qd}^{(8)prst} &= (\bar{q}_p \gamma^\mu T^a q_r)(\bar{d}_s \gamma_\mu T^a d_t)\,,
\end{align}
while the corresponding RGEs are given by
\begin{align}
\mu \dv[]{}{\mu}{C}_{\ell e}^{prst} &= \frac{1}{16\pi^2}\left[\tilde{\mathcal{Y}}_e^{pt}\,\tilde{\mathcal{Y}}_e^{\dagger sr} + \frac{8}{3}\,g_1^2\,\delta^{pr}\delta^{st} \,(\tilde{\mathcal{C}}_{B}^2 + \mathcal{C}_{B}^2)\right] \, ,  
\\
\mu \dv[]{}{\mu}{C}_{\ell u}^{prst} &= -\frac{1}{9\pi^2}\,g_1^2 \,\delta^{pr}\delta^{st}\, (\tilde{\mathcal{C}}_{B}^2 + \mathcal{C}_{B}^2) \, ,  
\\
\mu \dv[]{}{\mu}{C}_{\ell d}^{prst} &= \frac{1}{18\pi^2}\,g_1^2 \,\delta^{pr}\delta^{st}\, (\tilde{\mathcal{C}}_{B}^2 + \mathcal{C}_{B}^2) \, ,  
\\
\mu \dv[]{}{\mu}{C}_{qe}^{prst} &= -\frac{1}{18\pi^2}\,g_1^2\,\delta^{pr}\delta^{st} \, (\tilde{\mathcal{C}}_{B}^2 + \mathcal{C}_{B}^2) \, ,  
\\
\mu \dv[]{}{\mu}{C}_{qu}^{(1)prst} &=\frac{1}{16\pi^2}\left[\frac{1}{N_c}\,\tilde{\mathcal{Y}}_u^{pt}\,\tilde{\mathcal{Y}}_u^{\dagger sr} + \frac{16}{27}\,g_1^2\,\delta^{pr}\delta^{st} \,(\tilde{\mathcal{C}}_{B}^2 + \mathcal{C}_{B}^2)\right] \, ,  
\\
\mu \dv[]{}{\mu}{C}_{qu}^{(8)prst} &=\frac{1}{8\pi^2}\left[\tilde{\mathcal{Y}}_u^{pt}\,\tilde{\mathcal{Y}}_u^{\dagger sr} + \frac{8}{3}\,g_s^2\,\delta^{pr}\delta^{st} \,(\tilde{\mathcal{C}}_{G}^2 + \mathcal{C}_{G}^2)\right] \, ,  
\\
\mu \dv[]{}{\mu}{C}_{qd}^{(1)prst} &=\frac{1}{16\pi^2}\left[\frac{1}{N_c}\,\tilde{\mathcal{Y}}_d^{pt}\,\tilde{\mathcal{Y}}_d^{\dagger sr} - \frac{8}{27}\,g_1^2\,\delta^{pr}\delta^{st} \,(\tilde{\mathcal{C}}_{B}^2 + \mathcal{C}_{B}^2)\right] \, ,  
\\
\mu \dv[]{}{\mu}{C}_{qd}^{(8)prst} &=\frac{1}{8\pi^2}\left[\tilde{\mathcal{Y}}_d^{pt}\,\tilde{\mathcal{Y}}_d^{\dagger sr} + \frac{8}{3}\,g_s^2\,\delta^{pr}\delta^{st} \,(\tilde{\mathcal{C}}_{G}^2 + \mathcal{C}_{G}^2)\right] \, .
\end{align}

\paragraph{Classes $(\bar{L}R)(\bar{R}L)$ and $(\bar{L}R)(\bar{L}R)$.}

In the class $(\bar{L}R)(\bar{R}L)$ there is only one operator: 
\begin{equation}
Q_{\ell e dq}^{prst} = (\bar{\ell}_p^j e_r)(\bar{d}_sq_{tj})\,,
\end{equation}
whereas in the class $(\bar{L}R)(\bar{L}R)$ one has the following operators:
\begin{align}
Q^{(1)prst}_{quqd}&=(\bar{q}_p^j u_r)\epsilon_{jk}(\bar{q}_s^k d_t)\,,  &Q^{(8)prst}_{quqd}&=(\bar{q}_p^j T^au_r)\epsilon_{jk}(\bar{q}_s^k T^a d_t)\,, \\
Q^{(1)prst}_{\ell equ}&=(\bar{\ell}_p^j e_r)\epsilon_{jk}(\bar{q}_s^k u_t)\,,  &Q^{(3)prst}_{\ell equ}&=(\bar{\ell}_p^j \sigma_{\mu\nu}e_r)\epsilon_{jk}(\bar{q}_s^k \sigma^{\mu\nu} u_t) \,.
\end{align}
The corresponding RGEs are found to be:
\begin{align}
\mu\dv[]{}{\mu}{C}_{\ell edq}^{prst} &= -\frac{1}{8\pi^2}\,\tilde{\mathcal{Y}}_e^{pr}\,\tilde{\mathcal{Y}}_d^{\dagger st} \, ,  && \\
\mu\dv[]{}{\mu}{C}_{quqd}^{(1)prst} &=-\frac{1}{8\pi^2}\,\tilde{\mathcal{Y}}_u^{pr}\,\tilde{\mathcal{Y}}_d^{\dagger st} \, ,  &
\mu\dv[]{}{\mu}{C}_{quqd}^{(8)prst} &=0\, ,  \\
\mu\dv[]{}{\mu}{C}_{\ell equ}^{(1)prst} &= -\frac{1}{8\pi^2}\,\tilde{\mathcal{Y}}_e^{pr}\,\tilde{\mathcal{Y}}_u^{\dagger st} \, ,  &
\mu\dv[]{}{\mu}{C}_{\ell equ}^{(3)prst} &=0\, . 
\end{align}

\section{Comparison between on-shell and standard methods}\label{sec:comparison}

In this work, we have shown how to compute anomalous dimensions via the on-shell method.
Its advantages over standard diagrammatic techniques are numerous and diverse, and it is our purpose to illustrate some of them in this section.
In order to do so, we will consider explicit examples from our previous computations. 

The first reason why we find the on-shell method to be particularly efficient in computing RGEs resides in the significant simplification of the calculations to be performed.
Indeed, working with on-shell quantities often leads to naturally simple expressions for the amplitudes to be considered, without any complication emerging from unphysical degrees of freedom.
Unitarity, on the other hand, allows one to extract information about loop quantities from lower-order ones.

These computational advantages of on-shell methods compared to standard ones become more and more relevant as the loop order is raised, when the inherently recursive structure of the method --- a direct consequence of unitarity --- drastically reduces the number of amplitudes to be computed.
Moreover, further simplifications occur when dealing with a large number of non-Abelian gauge bosons.
Their presence generally renders computations with standard techniques lengthy and computationally expensive: checks for gauge invariance have to be performed, and the eventual cancellation of different Lorentz and gauge structures is often non-trivial.

\begin{figure}[h!tb]
\centering
\includegraphics[width = \textwidth]{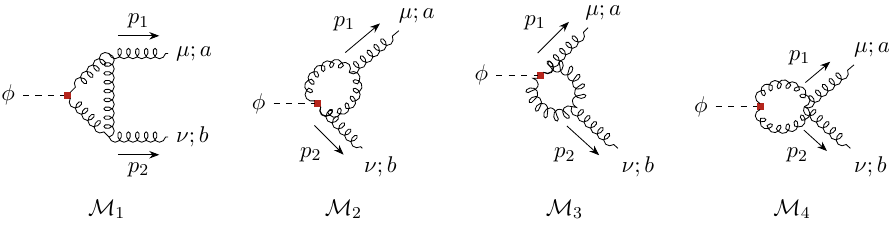}
\caption{Feynman diagrams contributing to the renormalization of the vertex $\phi GG$.}
\label{fig:PhiGGdiagB}
\end{figure}

This is clearly shown by the computation of the anomalous dimension for the operator $\phi GG$, which requires the evaluation of the Feynman diagrams of Fig.~\ref{fig:PhiGGdiagB}.
The remaining diagrams in Fig.~\ref{fig:PhiGGdiagBNotDiv} are either null or give rise to no divergences.
\begin{figure}[h!tb]
\centering
\begin{tikzpicture}
	\begin{feynman}[small]
	\node [BrickRed, square dot] (b);
    \vertex [above left=1.5cm of b] (a) {$\phi$};
    \vertex [below left=0.75cm of b] (c) ;
    \vertex [above right=1.5cm of b] (d) {\(\mu;a\)};
	\vertex [below right =1.5cm of b] (e) {\(\nu;b\)} ;
    \diagram*{
    (a) -- [scalar] (b),
    (b) -- [gluon, momentum=\(p_2\)] (e),
    (b) -- [gluon, momentum=\(p_1\)] (d),
    (b) -- [gluon, half left] (c),
    (c) -- [gluon, half left] (b),
    };
	\end{feynman}
	\end{tikzpicture}
\quad
\begin{tikzpicture}
	\begin{feynman}[small]
	\vertex (a) {\(\phi\)};
	\node [BrickRed, square dot, right=  of a] (b);
	\vertex [above right = of b] (c);
	\vertex [below right=of b] (d);
	\vertex [right =  of c] (e) {\(\mu;a\)} ;
	\vertex [right =  of d] (f) {\(\nu;b\)} ;
	\diagram*{
		(a) -- [scalar] (b),
		(c) -- [fermion] (b),
		(b) -- [fermion] (d),
		(d) -- [gluon, momentum'=\(p_2\)] (f),
		(d) -- [fermion] (c),
		(c) -- [gluon, momentum=\(p_1\)] (e),
	};
	\end{feynman}
	\end{tikzpicture}
    \quad
    \begin{tikzpicture}
	\begin{feynman}[small]
	\vertex (a) {\(\phi\)};
	\node [BrickRed, square dot, right=  of a] (b);
	\vertex [above right = of b] (c);
	\vertex [below right=of b] (d);
	\vertex [right =  of c] (e) {\(\mu;a\)} ;
	\vertex [right =  of d] (f) {\(\nu;b\)} ;
	\diagram*{
		(a) -- [scalar] (b),
		(c) -- [anti fermion] (b),
		(b) -- [anti fermion] (d),
		(d) -- [gluon, momentum'=\(p_2\)] (f),
		(d) -- [anti fermion] (c),
		(c) -- [gluon, momentum=\(p_1\)] (e),
	};
	\end{feynman}
	\end{tikzpicture}
\caption{Feynman diagrams that do not contribute to the renormalization of the vertex $\phi G G $ because they are either identically zero or because they give rise to no UV divergences.}
\label{fig:PhiGGdiagBNotDiv}
\end{figure}
We find the following divergent terms for the diagrams in Fig.~\ref{fig:PhiGGdiagB}:
\begin{align}
    i {\mathcal{M}_{1}}^{ab}_{\mu\nu} &= i \frac{\alpha_s}{3\pi \epsilon} \frac{\mathcal{C}_g}{\Lambda} C_A \delta^{ab}\left[ (23 + 6 \xi_G) p_{1\nu} p_{2 \mu} - (37 + 6 \xi_G) p_1\cdot p_2 g_{\mu\nu}\right]\,,\\
    i {\mathcal{M}_{2}}^{ab}_{\mu\nu} &= i \frac{\alpha_s}{2 \pi \epsilon} \frac{\mathcal{C}_g}{\Lambda} C_A \delta^{ab}(5 +  \xi_G) \left(-p_{1\nu} p_{2 \mu} + p_1\cdot p_2 g_{\mu\nu} \right)\,, \\
    i {\mathcal{M}_{3}}^{ab}_{\mu\nu} &= i \frac{\alpha_s}{2 \pi \epsilon} \frac{\mathcal{C}_g}{\Lambda} C_A \delta^{ab}(5 +  \xi_G) \left(-p_{1\nu} p_{2 \mu} + p_1\cdot p_2 g_{\mu\nu} \right)\,,\\
    i {\mathcal{M}_{4}}^{ab}_{\mu\nu} &=
    i \frac{\alpha_s}{3 \pi \epsilon} \frac{\mathcal{C}_g}{\Lambda} C_A \delta^{ab}\left(p_{1\nu} p_{2 \mu} + 13 p_1\cdot p_2 g_{\mu\nu} \right)
    \,,
\end{align}
which add up to
\begin{equation}
i \mathcal{M}^{ab}_{\mu\nu} = i \frac{\alpha_s}{\pi \epsilon}\frac{\mathcal{C}_g}{\Lambda}C_A \delta^{ab}  (3+ \xi_G) \left(p_{1\nu} p_{2\mu} - p_1 \cdot p_2 g_{\mu\nu}\right)\,.
\end{equation}
Therefore, 
from the Feynman rule for $\phi GG$
\begin{equation}
    \includegraphics[valign=c,page=1]{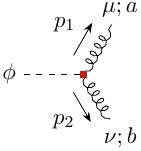}
    = 4i \frac{\mathcal C_g}{\Lambda}\delta^{ab}(p_{1\nu} p_{2\mu} - p_1 \cdot p_2 g_{\mu\nu})\,,
\end{equation}
we can identify the renormalization factor of $\mathcal C_g$ to be
\begin{equation}
     Z_{\mathcal C_g} = 1 - \frac{\alpha_s}{4\pi\epsilon} (3+\xi_G) C_A \,.
\end{equation}
Finally, by exploiting the expression for the renormalization factor of the gluon field
\begin{eqnarray}
    Z_g =1+ \frac{\alpha_s}{4\pi \epsilon}\bigg[\bigg(\frac{13}{3}-\xi_G\bigg)C_A-\frac{8}{3}T_F \sum_f c_f^2 \bigg]
\end{eqnarray}
and $\mu\dv[]{}{\mu}\alpha_s=-\epsilon \alpha_s + \dots$, we obtain the following RGE:
\begin{equation}
    \frac{1}{\mathcal{C}_g} \mu \dv[]{\mathcal C_g}{\mu} =  -  \frac{1}{Z_{\mathcal C_g}} \mu \dv[]{Z_{\mathcal C_g}}{\mu} +  \frac{1}{Z_g} \mu \dv[]{Z_g}{\mu} = - \frac{\alpha_s}{2\pi} \bigg[ \frac{11}{3} C_A - \frac{4}{3}T_F \sum_f c_f^2 \bigg]\,.
\end{equation}
These results reproduce the ones obtained with the on-shell method, but at the expense of computing a relatively large number of one-loop diagrams with non-trivial Lorentz and gauge-dependent structures.
In the on-shell method, no gauge dependence is present at any level of the computation, which requires only the convolution of one tree-level amplitude with a single form factor. 

Additionally, the on-shell method allowed us to manifest some hidden structures of the computation which are jeopardized in the standard approach.
Indeed, owing to general symmetry arguments, one would expect the operator $\phi GG$ to renormalize precisely as $GG$, and, hence, just like the IR anomalous dimension associated with a pair of photons (whose long-distance dynamics is clearly dictated by their kinetic term).
The on-shell method formalizes this property in a rather elegant way: a simple inspection of the form of few tree-level amplitudes directly allows us to solidly derive such property.

Another interesting example is given by the renormalization of the Weinberg $GG\tilde G$ operator.
Its Feynman rule in momentum space is given by
\begin{equation}
    \includegraphics[valign=c,page=2]{figures/FeynRules.pdf}
    \begin{aligned}
    &= -\frac{2}{3}\frac{d_G}{\Lambda^2}f^{abc}[\varepsilon_{\mu\nu\rho\alpha}(p_1^\alpha p_2\cdot p_3+p_2^\alpha p_3\cdot p_1+p_3^\alpha p_1\cdot p_2)
    \\
    &+\varepsilon_{\mu\nu\alpha\beta}(p_1-p_2)_\rho p_1^\alpha p_2^\beta  + \varepsilon_{\nu\rho\alpha\beta}(p_2-p_3)_\mu p_2^\alpha p_3^\beta + \varepsilon_{\rho\mu\alpha\beta}(p_3-p_1)_\nu p_3^\alpha p_1^\beta 
    ]\,.
    \end{aligned}
\end{equation}

Within the standard diagrammatic framework, extracting its anomalous dimension requires computing different diagrams, which we can conveniently classify as triangle and bubble diagrams; see Fig.~\ref{fig:Wienbergdiagrams}. 

\begin{figure}[h!tb]
\centering
\begin{tikzpicture}
	\begin{feynman}[small]
	\vertex (a) {\(\mu;a\)};
	\node [NavyBlue, square dot, right= 1.5cm of a] (b);
	\node [BrickRed, square dot, above right=of b] (c);
	\vertex [below right=of b] (d);
	\vertex [right=1.7cm of c] (e) {\(\nu;b\)} ;
	\vertex [right=1.3cm of d] (f) {\(\rho;c\)} ;
	\vertex (g) [below =of a] {\(\mathcal{M}_{1}^\triangle\)};
	\diagram*{
		(a) -- [gluon, rmomentum=\(p_1\)] (b),
		(c) -- [scalar] (b),
		(b) -- [gluon] (d),
		(d) -- [gluon, momentum'=\(p_3\)] (f),
		(d) -- [gluon] (c),
		(c) -- [gluon, momentum=\(p_2\)] (e),
	};
	\end{feynman}
	\end{tikzpicture}
\qquad
\begin{tikzpicture}
	\begin{feynman}[small]
	\vertex (a) {\(\mu;a\)};
	\node [BrickRed, square dot, right= 1.5cm of a] (b);
	\node [NavyBlue, square dot, above right=of b] (c);
	\vertex [below right=of b] (d);
	\vertex [right=1.7cm of c] (e) {\(\nu;b\)} ;
	\vertex [right=1.3cm of d] (f) {\(\rho;c\)} ;
	\vertex (g) [below =of a] {\(\mathcal{M}_{2}^\triangle\)};
	\diagram*{
		(a) -- [gluon, rmomentum=\(p_1\)] (b),
		(c) -- [scalar] (b),
		(b) -- [gluon] (d),
		(d) -- [gluon, momentum'=\(p_3\)] (f),
		(d) -- [gluon] (c),
		(c) -- [gluon, momentum=\(p_2\)] (e),
	};
	\end{feynman}
	\end{tikzpicture}
    \\
    \begin{tikzpicture}[baseline = (aux)]
	\begin{feynman}[small]
	\vertex (a) {\(\mu;a\)};
    \vertex [above =0.7cm of a] (aux)  ;
	\node [NavyBlue, square dot, right=1.5cm of a] (b);
	\node [BrickRed, square dot,  right=1.3 cm of b] (c);
	\vertex [above right=1.5cm of c] (e) {\(\nu;b\)} ;
	\vertex [below right=1.5cm of c] (f) {\(\rho;c\)} ;
	\vertex (g) [below = of a] {\(\mathcal{M}_{1}^\bigcirc\)};
	\diagram*{
		(a) -- [gluon, rmomentum=\(p_1\)] (b),
		(c) -- [scalar, half left] (b),
		(b) -- [gluon, half left] (c),
		(c) -- [gluon, momentum'=\(p_3\)] (f),
		(c) -- [gluon, momentum'=\(p_2\)] (e),
	};
	\end{feynman}
	\end{tikzpicture}
    \qquad
    \begin{tikzpicture}[baseline = (aux)]
	\begin{feynman}[small]
	\vertex (a) {\(\mu;a\)};
    \vertex [above =0.7cm of a] (aux)  ;
	\node [BrickRed, square dot, right=1.5cm of a] (b);
	\node [NavyBlue, square dot,  right=1.3 cm of b] (c);
	\vertex [above right=1.5cm of c] (e) {\(\nu;b\)} ;
	\vertex [below right=1.5cm of c] (f) {\(\rho;c\)} ;
	\vertex (g) [below = of a] {\(\mathcal{M}_{2}^\bigcirc\)};
	\diagram*{
		(a) -- [gluon, rmomentum=\(p_1\)] (b),
		(c) -- [scalar, half left] (b),
		(b) -- [gluon, half left] (c),
		(c) -- [gluon, momentum'=\(p_3\)] (f),
		(c) -- [gluon, momentum'=\(p_2\)] (e),
	};
	\end{feynman}
	\end{tikzpicture}
\caption{Triangle and bubble diagrams contributing to the renormalization of the three-gluon Weinberg operator. 
The eight additional amplitudes with the three external gluons permuted are omitted.}
\label{fig:Wienbergdiagrams}
\end{figure}

The divergences associated with the first class of 3-point diagrams 
are 
($d=4-\epsilon$)
\begin{align}
    i {\mathcal M_{1}^\triangle}_{\mu\nu\rho}^{abc} &= \frac{1}{\epsilon}\frac{\mathcal C_g \tilde{\mathcal C}_g}{3\pi^2 \Lambda^2}g_sf^{abc}\big[-\varepsilon_{\mu\nu\rho\alpha}(p_1\cdot p_2+5p_2\cdot p_3)p_1^\alpha 
    +\varepsilon_{\mu\nu\alpha\beta}(2p_1+p_3)_\rho p_1^\alpha p_2^\beta \nonumber \\
    &\quad
    -4 \varepsilon_{\mu\nu\alpha\beta}p_{2\rho} p_1^\alpha p_3^\beta 
    + \varepsilon_{\mu\rho\alpha\beta}(p_1+5p_3)_\nu p_1^\alpha p_2^\beta - 4 g_{\nu\rho}\varepsilon_{\mu\alpha\beta\gamma}p_1^\alpha p_2^\beta p_3^\gamma
    \big]\,,\\
    i {\mathcal M_{2}^\triangle}_{\mu\nu\rho}^{abc} &= \frac{1}{\epsilon}\frac{\mathcal C_g \tilde{\mathcal C}_g}{3\pi^2 \Lambda^2}g_sf^{abc}\big[
    \varepsilon_{\mu\nu\rho\alpha}p_1^2 p_2^\alpha
    -4 \varepsilon_{\mu\nu\rho\alpha}p_2^\alpha  p_1\cdot p_3 
    + \varepsilon_{\mu\nu\alpha\beta}(3p_1+p_3)_\rho p_1^\alpha p_2^\beta \nonumber \\
    &\quad
    - 5 \varepsilon_{\mu\nu\alpha\beta}p_{1\rho} p_2^\alpha p_3^\beta 
    +\varepsilon_{\nu\rho\alpha\beta}(4p_3-p_1)_\mu p_1^\alpha p_2^\beta
    -5 g_{\mu\rho} \varepsilon_{\nu\alpha\beta\gamma} p_1^\alpha p_2^\beta p_3^\gamma
    \big]\,.
\end{align}
By taking into account the permutations of the three external gluons, the sum of these diagrams amounts to 
\begin{equation}
i {\mathcal{M}^{\triangle}}_{\mu\nu\rho}^{abc} = \frac{1}{\epsilon}\frac{2\mathcal C_g \tilde{\mathcal C}_g}{\pi^2 \Lambda^2}g_s f^{abc} \big[
2\varepsilon_{\mu\nu\alpha\beta}p_{2\rho} 
+ 2\varepsilon_{\mu\rho\alpha\beta}p_{3\nu}
+\varepsilon_{\nu\rho\alpha\beta}(p_3-p_2)_\mu
\big] p_2^\alpha p_3^\beta\,,
\end{equation}
where we assumed the energy momentum conservation $p_1 = - (p_2+p_3)$, the transversality conditions for gluons, and
$p_2^2=p_3^2=p_2\cdot p_3 = 0$. 

The divergences associated with the 2-point diagrams 
read instead 
\begin{align}
    i {\mathcal M_{1}^\bigcirc}_{\mu\nu\rho}^{abc} &= -\frac{1}{\epsilon}\frac{\mathcal C_g \tilde{\mathcal C}_g}{3\pi^2\Lambda^2}g_sf^{abc}(3m_\phi^2 -p_1^2)\varepsilon_{\mu\nu\rho\alpha}p_1^\alpha \,,
    \\
    i {\mathcal M_{2}^\bigcirc}_{\mu\nu\rho}^{abc} &= -\frac{1}{\epsilon}\frac{\mathcal C_g \tilde{\mathcal C}_g}{3\pi^2\Lambda^2}g_sf^{abc}
    \big[
    \varepsilon_{\mu\nu\rho\alpha}[(3m_\phi^2+p_1^2)p_1^\alpha 
    +3p_1^2 (p_2+p_3)^\alpha ]\nonumber \\
    &\quad -3\varepsilon_{\nu\rho\alpha\beta}p_{1\mu}p_1^\alpha(p_2+p_3)^\beta
    \big]\,,
\end{align}
which are not of the desired form of the Feynman rule of $GG\tilde G$ and can be only interpreted as pertaining to the renormalization of the $G\tilde G$ operator.
Indeed, the Feynman rule of the $G\tilde G$ operator is proportional to $p_1+p_2+p_3$, which has to vanish for on-shell gluons, as it is indeed the case for these bubble contributions.
Moreover, we find that tadpole diagrams are identically vanishing. 

As a consequence, the RGE associated with the Wilson coefficient $d_G$ is
\begin{equation}
    \mu\dv[]{}{\mu}d_G = -\frac{3g_s}{\pi^2}\mathcal C_g \tilde{\mathcal C}_g\,.
\end{equation}
This reproduces the result previously reported in Eq.~\eqref{eq:gammatildeG3},
but at the price of computing more diagrams with different Lorentz structures.
On the other hand, the on-shell method only required the calculation of one form factor and one amplitude, yielding the same result in a more transparent and elegant way.

Yet another advantage of the on-shell method is that it directly allows us to relate the anomalous dimension for a given operator to the one of its CP-counterpart (such as $\phi FF$ to $\phi F \tilde F$, see Sec.~\ref{section_FFtilde}, or $\phi \bar f f$ to $ \phi \bar f i \gamma_5 f$, see Sec.~\ref{section_ff}).
In the previous example, for instance, the knowledge of the anomalous dimension for the operator $\phi GG$ allowed us to immediately infer the one for the operator $\phi G\tilde G$ (Sec.~\ref{sec:phiGG}), and similarly for $GGG$ and $GG \tilde{G}$ (Sec.~\ref{sec:GGG}).
Such a duality is not manifest by working in the standard approach, where CP-dual operators possess entirely different Lorentz structures at the level of Feynman rules and no similarity in the pattern of cancellations among gauge-dependent terms is present, despite the common diagrammatic structure.
Such a property is instead manifest within the framework of on-shell methods, where the presence of the same external degrees of freedom naturally suggests similarities between amplitudes related to CP-dual operators. 
\section{\label{sec:conclusions}Conclusions}

On-shell amplitude techniques have proven to be very effective for computing the renormalization group equations of quantum field theories~\cite{Caron-Huot:2016cwu,EliasMiro:2020tdv,Baratella:2020lzz,Jiang:2020mhe,Bern:2020ikv,Baratella:2020dvw,AccettulliHuber:2021uoa,EliasMiro:2021jgu,Baratella:2022nog,Machado:2022ozb}. 
In particular, the on-shell method~\cite{Caron-Huot:2016cwu} relates anomalous dimensions with unitarity cuts. 
As recently discussed in~\cite{Bresciani:2023jsu}, this method can be easily applied also to describe mixings among operators with different dimensions 
and to capture leading mass effects, which are of paramount importance in several phenomenological studies.

In this work, we have extensively applied the above techniques~\cite{Caron-Huot:2016cwu,Bresciani:2023jsu} to the one-loop renormalization of CP-violating interactions, both above and below the electroweak scale,  of an Axion-Like Particle (ALP) with SM fields, reproducing and extending previous results~\cite{Marciano:2016yhf,DiLuzio:2020oah,Bauer:2020jbp,Chala:2020wvs,DasBakshi:2023lca,Galda:2021hbr,Galda:2023qjx}.

In particular, we  first derived the anomalous dimensions for ALP couplings with fermions, $\phi \bar f f$ and $\phi \bar f i \gamma_5 f$, which require a fermion mass insertion. This allowed us to apply the on-shell method~\cite{Caron-Huot:2016cwu} supplemented by the Higgs low-energy theorem to keep track of leading mass effects while still working in a massless formalism~\cite{Bresciani:2023jsu}. 

Then, we considered the renormalization of ALP couplings to photons and gluons, $\phi FF$ and $\phi GG$, along with their CP counterparts, $\phi F \tilde F$ and $\phi G \tilde G$, recovering the well-known result that they renormalize precisely as $FF$ and $GG$ and, hence, just like their related gauge couplings squared. 
The on-shell method shows this property in a simple and elegant way by just inspecting few tree-level amplitudes.
Moreover, we evaluated the RGEs of operators up to dimension-6 emerging after integrating-out the ALP at one-loop level,  
such as the Weinberg operator $GG\tilde G$ and $GGG$, the (chromo-)magnetic and (chromo-)electric dipole moments, 
i.e.,~$\bar f \sigma\!\cdot\! F f$, $\bar f \sigma\!\cdot\! G f$,
$\bar f \sigma\!\cdot\! F i\gamma_5 f$, and $\bar f \sigma\!\cdot\! G i\gamma_5 f$.

A detailed derivation of the anomalous dimension matrix was carried out both with on-shell and standard techniques, 
aiming to closely compare their virtues and shortcomings.

We found that on-shell methods are computationally advantageous as compared to standard ones, owing to the
significantly lower, as well as less challenging, number of required contributions to be computed.
Moreover, the presence of a large number of non-Abelian gauge bosons generally renders calculations with standard techniques lengthy and computationally expensive: checks for gauge invariance were performed, often involving non-trivial cancellation of different Lorentz and gauge structures.
Remarkably, the on-shell method makes more manifest the connection between the anomalous dimension of operators related by symmetries. 
For instance, the knowledge of the anomalous dimension for the operator $\phi GG$ allowed us to immediately 
infer the one for the CP-dual operator $\phi G\tilde G$: this duality is hidden within the standard approach, where CP-dual operators have different Lorentz structures at the level of Feynman rules, and no similarity in the pattern of cancellations among gauge-dependent terms is present.

Our results were obtained by systematically evaluating all phase-space cut-integrals adopting two different integration techniques, namely by angular integration~\cite{Zwiebel:2011bx}, and via Stokes' theorem~\cite{Mastrolia:2009dr}. The latter, motivated by unitarity, offers the advantage of projecting directly the rational coefficient of the 2-point functions out of the double-cut of one-loop integrals, therefore simplifying the evaluation of anomalous dimensions.  
It would be interesting to extend Stokes' integration method for the evaluation of multi-particle phase-space integrals, and to apply it for the evaluation of anomalous dimensions at higher orders by unitarity cuts, which we plan for future developments.

\acknowledgments
We wish to thank Manoj Mandal for interesting discussions during the various stages of the project. 
This work received funding by the INFN Iniziative Specifiche 
AMPLITUDES and APINE
and from the European Union’s Horizon 2020 research and innovation programme under the Marie Sklodowska-Curie grant agreements n. 860881 – HIDDeN, n. 101086085 – ASYMMETRY. This work was also partially supported by the Italian MUR Departments of Excellence grant 2023-2027 “Quantum Frontiers”.
The work of P.P. is supported by the European Union – Next Generation EU
and by the Italian Ministry of University and Research (MUR) via the 
PRIN 2022 project n. 2022K4B58X – AxionOrigins. 
G.B. research is supported by the European Research Council, under grant
ERC–AdG–88541, and by Università Italo-Francese, under grant Vinci.
G.L. gratefully acknowledges financial support by the Swiss National Science Foundation (Project No.\
TMCG-2\_213690).

\appendix
\section{Notation and Conventions}
\label{app:notation}

\paragraph{Spinor helicity.}
In this article, amplitudes and form factors have been expressed in terms of contractions of the fundamental two-dimensional spinors $\lambda_{\alpha}$ and $\tilde\lambda^{\dot\alpha}$ that transform in the $(1/2,0)$ and $(0,1/2)$ representations of $SL(2,\mathbb{C})$, respectively.
The spinor decomposition of a light-like four-momentum $p_{\mu}$ of an outgoing particle is given by
\begin{equation}
    p_{\alpha \dot\alpha} = p_\mu \sigma^{\mu}_{\alpha \dot\alpha} = \lambda_\alpha \tilde \lambda_{\dot \alpha}\,,
\end{equation}
where $\sigma^\mu = (\mathbf{1},\vec \sigma)$ and $\sigma^i$ are the Pauli matrices.
The Lorentz-invariant antisymmetric contractions are
\begin{align}
    \agl{i}{j} = \lambda^\alpha_i \lambda_{j\,\alpha} = \epsilon_{\alpha\beta}\lambda_i^\alpha \lambda_j^\beta\,,
    &&
    \sqr{i}{j} = \tilde \lambda_{i\,\dot\alpha}\tilde\lambda_j^{\dot\alpha} = -\epsilon_{\dot\alpha \dot\beta}\tilde\lambda_{i}^{\dot\alpha} \tilde\lambda_j^{\dot\beta}\,,
\end{align}
where we used the following convention for the two-dimensional Levi-Civita tensor: $\epsilon^{12} = \epsilon^{\dot 1 \dot 2} = -\epsilon_{12} =-\epsilon_{\dot 1 \dot 2} = 1$.
The Mandelstam invariants are then $s_{ij} = (p_i+p_j)^2 = \agl{i}{j}\sqr{j}{i}$.
In this formalism, polarization vectors are written as
\begin{align}
    \varepsilon_\mu^-(p) = \frac{\langle p\,\sigma_\mu \, q]}{\sqrt{2}\sqr{p}{q}}\,, &&
    \varepsilon_\mu^+(p) = \frac{\langle q\,\sigma_\mu \, p]}{\sqrt{2}\agl{q}{p}}\,,
\end{align}
where $q$ is a reference momentum such that $\sqr{p}{q}, \agl{q}{p}\neq 0$, while Dirac fermion spinors are
\begin{align}
    u_+(p) &= v_-(p) = 
    \begin{pmatrix}
        \lambda_\alpha \\ 0
    \end{pmatrix}
    \,,
    &
    u_-(p) &= v_+(p) = 
    \begin{pmatrix}
        0 \\ \tilde\lambda^{\dot\alpha}
    \end{pmatrix}
    \,, 
    \\
    \bar u_+(p) &= \bar v_-(p) = 
    \begin{pmatrix}
        0 & \tilde\lambda_{\dot\alpha}
    \end{pmatrix}
    \,,
    &
    \bar u_-(p) &= \bar v_+(p) = 
    \begin{pmatrix}
        \lambda^\alpha & 0
    \end{pmatrix}
    \,.
\end{align}
In order to flip the momentum of a particle, we used $\lambda_{-p}=i\lambda_p$ and $\tilde \lambda_{-p}=i\tilde \lambda_p$.
Accordingly, when a fermion is exchanged from the outgoing to the incoming state, the amplitude is multiplied by $(-i)$, that is, $\mathcal M(X;\bar f) = -i \mathcal M(X+f)$.
Spinor manipulations have been handled in \texttt{Mathematica} through the package \texttt{S@M}~\cite{Maitre:2007jq}.

\paragraph{Gauge group conventions.}
The conventions used for the invariants of the adjoint and fundamental representations of the gauge group $SU(N_c)$ are summarized as
\begin{align}
    f^{acd}f^{bcd} &= C_A\delta^{ab}\,, &
    C_A &= N_c = 3\,, \\
    T^a_{IK}T^a_{KJ} &= C_F \delta_{IJ}\,, & C_F &=\frac{N_c^2-1}{2N_c}=\frac{4}{3}\,, \\
    \text{Tr}(T^a T^b) &= T_F \delta^{ab}\,, &
    T_F &= \frac{1}{2}\,.
\end{align}
The covariant derivative is taken to be $D_\mu f = (\partial_\mu - i e Q_f A_\mu - i g_s c_f G_\mu^a T^a)f$, and, accordingly, the $SU(3)_c$ field strength tensor is $G^a_{\mu\nu} = \partial_\mu G_\nu^a - \partial_\nu G_\mu^a + g_s f^{abc} G_\mu^b G_\nu^c$.
The coefficient $c_f$ takes the value $1$ ($0$) if $f$ is a quark (lepton).
\section{Amplitudes}
\label{app:Amplitudes}

In this Appendix, we report the amplitudes that we employed throughout the main text, expressed in terms of spinor-helicity variables.

\subsection{3-point tree amplitudes}
Here, the analytically continued 3-point tree amplitudes in the holomorphic ($\hol$) and antiholomorphic ($\ahol$) configurations belonging to the different sectors of the Lagrangian are displayed on the left and on the right, respectively.
They are completely constrained, up to an overall factor (the coupling constant), by locality, Poincaré invariance, and dimensional analysis~\cite{Benincasa:2007xk}. Indeed, in full generality they read as follows
\begin{align}
    \mathcal M_3^\hol(1^{h_1},2^{h_2},3^{h_3}) = g_\hol \agl{1}{2}^{a_3}\agl{2}{3}^{a_1}\agl{3}{1}^{a_2}\,, &&
    \mathcal M_3^\ahol(1^{h_1},2^{h_2},3^{h_3}) = g_\ahol \sqr{1}{2}^{\bar a_3}\sqr{2}{3}^{\bar a_1}\sqr{3}{1}^{\bar a_2}\,,
\end{align}
with $\bar a_i=-a_i$,
\begin{align}
    a_1= h_1-h_2-h_3\,,&&
    a_2= h_2-h_3-h_1\,, &&
    a_3= h_3-h_1-h_2\,,
\end{align}
and the mass dimensions of the coupling constants only depend on the helicities:
\begin{align}
    [g_\hol]=1+h_1+h_2+h_3\,, && 
    [g_\ahol]=1-h_1-h_2-h_3\,.
\end{align}
Locality implies $[g_\hol],[g_\ahol]<1$, therefore we can infer that the holomorphic (antiholomorphic) configuration is the consistent one if $h_1+h_2+h_3<0$ ($h_1+h_2+h_3>0$).
The case where $h_1+h_2+h_3=0$ is trivial, as it can only correspond to a cubic scalar interaction, where $h_1=h_2=h_3=0$.

\paragraph{$\mathcal L_{\text{LO}}$.}
The lowest order Lagrangian
\begin{equation}
    \mathcal L_{\text{LO}} = -\frac{1}{4}G_{\mu\nu}^a G^{a,\mu\nu}-\frac{1}{4}F_{\mu\nu}F^{\mu\nu}+i\bar f_i \gamma^\mu D_\mu f_i - y_i h \bar f_i f_i
\end{equation}
generates
\begin{align}
    \mathcal M(1^-_{f_i},2^+_{\bar f_j},3^-_\gamma) &= \sqrt 2 e Q_f \delta^{ij} \frac{\agl{1}{3}^2}{\agl{1}{2}}\,, &
     \mathcal M(1^-_{f_i},2^+_{\bar f_j},3^+_\gamma) &= \sqrt 2 e Q_f \delta^{ij}\frac{\sqr{3}{2}^2}{\sqr{1}{2}}\,, \\
     \mathcal M(1^-_{f_i^I},2^+_{\bar f_j^J},3^-_{g^a}) &= \sqrt 2 g_s c_f T^a_{IJ}\delta^{ij} \frac{\agl{1}{3}^2}{\agl{1}{2}}\,, &
     \mathcal M(1^-_{f_i^I},2^+_{\bar f_j^J},3^+_{g^a}) &= \sqrt 2 g_s c_f T^a_{IJ} \delta^{ij}\frac{\sqr{3}{2}^2}{\sqr{1}{2}}\,, \\
     \mathcal M(1^-_{g^a},2^-_{g^b},3^+_{g^c}) &= i\sqrt 2 g_s f^{abc}\frac{\agl{1}{2}^3}{\agl{1}{3}\agl{3}{2}}\,,  &
     \mathcal M(1^+_{g^a},2^+_{g^b},3^-_{g^c}) &=-i\sqrt 2 g_s f^{abc}\frac{\sqr{1}{2}^3}{\sqr{1}{3}\sqr{3}{2}}\,,  \\
     \mathcal M(1^-_{f_i},2^-_{\bar f_j},3_h) &= - y_i\delta^{ij}\agl{1}{2}\,, &
     \mathcal M(1^+_{f_i},2^+_{\bar f_j},3_h) &= - y_i \delta^{ij} \sqr{1}{2} \,.
\end{align}

\paragraph{$\mathcal L_{\phi}$.}
The ALP effective Lagrangian
\begin{equation}
    \mathcal{L}_\phi = 
    \frac{\tilde{\mathcal{C}}_\gamma}{\Lambda}\,\phi\, F\tilde F
    + \frac{\tilde{\mathcal{C}}_g}{\Lambda}\,\phi\, G\tilde G
    + \mathcal{Y}^{ij}_P\,\phi\, \bar{f}_i i\gamma_5 f_j 
     + \frac{\mathcal{C}_\gamma}{\Lambda}\,\phi\, FF
    + \frac{\mathcal{C}_g}{\Lambda}\,\phi\, GG
    + \mathcal{Y}^{ij}_S\,\phi\, \bar{f}_i f_j
\end{equation}
generates
\begin{align}
    \mathcal M(1^-_\gamma,2^-_\gamma,3_\phi) &= -\frac{2}{\Lambda}(\mathcal C_\gamma + i \tilde{\mathcal{C}}_\gamma)\agl{1}{2}^2\,, &
    \mathcal M(1^+_\gamma,2^+_\gamma,3_\phi) &= -\frac{2}{\Lambda}(\mathcal C_\gamma - i \tilde{\mathcal{C}}_\gamma)\sqr{1}{2}^2\,, \\
    \mathcal M(1^-_{g^a},2^-_{g^b},3_\phi) &= -\frac{2}{\Lambda}(\mathcal C_g + i \tilde{\mathcal{C}}_g)\delta^{ab}\agl{1}{2}^2\,, &
    \mathcal M(1^+_{g^a},2^+_{g^b},3_\phi) &= -\frac{2}{\Lambda}(\mathcal C_g - i \tilde{\mathcal{C}}_g)\delta^{ab}\sqr{1}{2}^2\,, \\
    \mathcal M(1^-_{f_i},2^-_{\bar f_j},3_\phi)&=(\mathcal Y_S^{ij}-i\mathcal Y_P^{ij})\agl{1}{2}\,, &
    \mathcal M(1^+_{f_i},2^+_{\bar f_j},3_\phi)&=(\mathcal Y_S^{ij}+i\mathcal Y_P^{ij})\sqr{1}{2} \,.
\end{align}

\paragraph{$\mathcal L^{(5)}$.}
The relevant dimension-$5$ Lagrangian invariant under $SU(3)_c\times U(1)_{\text{em}}$ and built of SM particles consists of dipole operators
\begin{equation}
    \mathcal L^{(5)} = \frac{c_{\text{M}}^{ij}}{\Lambda}\bar f_i \sigma^{\mu\nu} f_j F_{\mu\nu} + \frac{c_{\text{E}}^{ij}}{\Lambda}\bar f_i \sigma^{\mu\nu}i\gamma_5 f_j F_{\mu\nu} + \frac{c_{\text{CM}}^{ij}}{\Lambda}\bar f_i \sigma^{\mu\nu}T^a f_j G^a_{\mu\nu} + \frac{c_{\text{CE}}^{ij}}{\Lambda}\bar f_i \sigma^{\mu\nu}i\gamma_5 T^a f_j G^a_{\mu\nu}
\end{equation}
and generates
\begin{align}
    \mathcal M(1^-_{f_i},2^-_{\bar f_j},3^-_\gamma) &= \frac{2\sqrt 2}{\Lambda}C_\gamma^{ij}\agl{1}{3}\agl{2}{3}\,, &
    \mathcal M(1^+_{f_i},2^+_{\bar f_j},3^+_\gamma)&=- \frac{2\sqrt 2}{\Lambda}(C_\gamma^{ij})^*\sqr{1}{3}\sqr{2}{3}\,,  \\
    \mathcal M(1^-_{f_i^I},2^-_{\bar f_j^J},3^-_{g^a}) &= \frac{2\sqrt 2}{\Lambda}C_g^{ij}T^a_{IJ}\agl{1}{3}\agl{2}{3}\,, &
    \mathcal M(1^+_{f_i^I},2^+_{\bar f_j^J},3^+_{g^a})&= - \frac{2\sqrt 2}{\Lambda}(C_g^{ij})^*T^a_{IJ}\sqr{1}{3}\sqr{2}{3}\,,
\end{align}
with
\begin{align}
    C_\gamma^{ij}=c_{\text{M}}^{ij}-i\, c_{\text{E}}^{ij}\,, && C_g^{ij}=c_{\text{CM}}^{ij}-i\,c_{\text{CE}}^{ij}\,.
\end{align}

\paragraph{$\mathcal L^{(6)}$.}
The relevant dimension-$6$ Lagrangian invariant under $SU(3)_c\times U(1)_{\text{em}}$ and built of SM particles only consists of
\begin{equation}
    \mathcal L^{(6)} = \frac{D_G}{3\Lambda^2} f^{abc} G_\mu^{a, \nu} G_{\nu}^{b, \rho} G_{\rho}^{c, \mu} + \frac{d_G}{3\Lambda^2} f^{abc} G_\mu^{a, \nu} G_{\nu}^{b, \rho} \tilde G_{\rho}^{c, \mu}
\end{equation}
and generates
\begin{align}
    \mathcal M(1^-_{g^a},2^-_{g^b},3^-_{g^c}) = i\frac{\sqrt 2}{\Lambda^2}C_Gf^{abc}\agl{1}{2}\agl{2}{3}\agl{1}{3}\,, &&
     \mathcal M(1^+_{g^a},2^+_{g^b},3^+_{g^c}) = i\frac{\sqrt 2}{\Lambda^2}C_G^* f^{abc}\sqr{2}{1}\sqr{3}{2}\sqr{3}{1}\,,
\end{align}
with
\begin{equation}
    C_G= D_G + i\, d_G\,.
\end{equation}

\subsection{4-point tree amplitudes}
Here, the 4-point tree amplitudes needed for the calculations are displayed.
With the symbol $*$ we denote the region in the space of the couplings of the theory where only the gauge couplings $e$ and $g_s$ are different from zero.
\begin{align}
     \mathcal M|_*(1^-_{f_i},2^+_{\bar f_j},3^-_\gamma,4^+_\gamma) &= -2e^2Q_f^2\delta^{ij}\frac{\agl{1}{3}\sqr{4}{2}}{\agl{1}{4}\sqr{3}{1}}\,, \\
    \mathcal M|_*(1^+_{f_i},2^-_{\bar f_j},3^-_\gamma,4^+_\gamma) &= -2e^2Q_f^2\delta^{ij}\frac{\agl{2}{3}\sqr{4}{1}}{\agl{1}{4}\sqr{3}{1}}\,, \\
    \mathcal M|_*(1^-_{f_i^I},2^+_{\bar f_j^J},3^-_{g^a},4^+_{g^b}) &= -2g_s^2c_f^2T^a_{IK}T^b_{KJ}\delta^{ij}\frac{\agl{1}{3}\sqr{4}{2}}{\agl{1}{4}\sqr{3}{1}}\,, \\
    \mathcal M|_*(1^+_{f_i^I},2^-_{\bar f_j^J},3^-_{g^a},4^+_{g^b}) &= -2g_s^2c_f^2T^b_{IK}T^a_{KJ}\delta^{ij}\frac{\agl{2}{3}\sqr{4}{1}}{\agl{1}{4}\sqr{3}{1}}\,, \\
    \mathcal M|_*(1^-_{g^a},2^-_{g^b},3^+_{g^c},4^+_{g^d}) &= -2g_s^2\agl{1}{2}^4\bigg(\frac{f^{abe}f^{cde}}{\agl{1}{2}\agl{2}{3}\agl{3}{4}\agl{4}{1}}+\frac{f^{ace}f^{bde}}{\agl{1}{3}\agl{3}{2}\agl{2}{4}\agl{4}{1}}\bigg)\,,\\
    \mathcal M|_*(1^-_{f_i^I},2^-_{\bar f_j^J},3^+_{f_k^K},4^+_{\bar f_l^L}) &= -2 (e^2 Q_f^2 \delta_{IL}\delta_{KJ}+g_s^2c_f^2T^a_{IL}T^a_{KJ}) \delta^{il}\delta^{jk} \frac{\agl{1}{2}\sqr{4}{3}}{\agl{1}{4}\sqr{4}{1}}\,, \\
    \mathcal M|_*(1^-_{f_i^I},2^+_{\bar f_j^J},3^-_{f_k^K},4^+_{\bar f_l^L}) &= +2(e^2Q_f^2\delta_{IJ}\delta_{KL}+g_s^2c_f^2T^a_{IJ}T^a_{KL})\delta^{ij}\delta^{kl} \frac{\agl{1}{3}\sqr{4}{2}}{\agl{1}{2}\sqr{2}{1}}\nonumber \\
    &\quad - 2(e^2Q_f^2\delta_{IL}\delta_{KJ}+g_s^2c_f^2T^a_{IL}T^a_{KJ}) \delta^{il}\delta^{jk} \frac{\agl{1}{3}\sqr{4}{2}}{\agl{1}{4}\sqr{4}{1}}\,, \\
    \mathcal M|_*(1^-_{f_i^I},2^+_{\bar f_j^J},3^+_{f_k^K},4^-_{\bar f_l^L}) &= +2(e^2Q_f^2\delta_{IJ}\delta_{KL}+g_s^2c_f^2T^a_{IJ}T^a_{KL})\delta^{ij}\delta^{kl} \frac{\agl{1}{4}\sqr{3}{2}}{\agl{1}{2}\sqr{2}{1}}\,,\\
    \mathcal M(1^-_{f_i},2^+_\gamma,3^-_{\bar f_j},4_h) &= -\sqrt{2} y_i \delta^{ij} e Q_f \frac{\agl{1}{3}^2}{\agl{1}{2}\agl{2}{3}}\,, \\
    \mathcal M(1^-_{f_i^I},2^+_{g^a},3^-_{\bar f_j^J},4_h) &= -\sqrt{2} y_i \delta^{ij} g_s c_f T^a_{IJ} \frac{\agl{1}{3}^2}{\agl{1}{2}\agl{2}{3}} \,,\\
    \mathcal M(1^-_{f_i},2^-_\gamma,3^+_{\bar f_j},4_\phi) &= \frac{2\sqrt 2}{\Lambda} e Q_f \delta^{ij}(\mathcal C_\gamma + i\tilde{\mathcal C}_\gamma)\frac{\agl{1}{2}^2}{\agl{1}{3}}\,,\\
    \mathcal M(1^-_{f_i^I},2^-_{g^a},3^+_{\bar f_j^J},4_\phi) &= \frac{2\sqrt 2}{\Lambda} g_s c_f T^a_{IJ} \delta^{ij}(\mathcal C_g + i\tilde{\mathcal C}_g)\frac{\agl{1}{2}^2}{\agl{1}{3}}\,,\\
    \mathcal M(1^-_{f_i},2^-_{\bar f_j},3^+_\gamma,4_\phi) &= -\sqrt 2 e Q_f (\mathcal Y_S^{ij}-i\mathcal Y_P^{ij})\frac{\agl{1}{2}^2}{\agl{1}{3}\agl{2}{3}} \,,\\
    \mathcal M(1^-_{f_i^I},2^-_{\bar f_j^J},3^+_{g^a},4_\phi) &= -\sqrt 2 g_s c_f T^a_{IJ} (\mathcal Y_S^{ij}-i\mathcal Y_P^{ij})\frac{\agl{1}{2}^2}{\agl{1}{3}\agl{2}{3}}\,,\\
    \mathcal M(1^-_{g^a},2^-_{g^b},3^+_{g^c},4_\phi) &= -i\frac{2\sqrt 2}{\Lambda}g_sf^{abc}(\mathcal C_g+i\tilde{\mathcal C}_g)\frac{\agl{1}{2}^3}{\agl{1}{3}\agl{2}{3}}\,.
\end{align}

\section{Infrared anomalous dimensions}
\label{app:IRanomalousdimensions}

The on-shell method is not directly sensitive to the UV anomalous dimension associated with a certain operator $\gamma_{i\leftarrow j}$ but rather to the difference between it and the IR anomalous dimension matrix, $\delta_{ij} \gamma_{i,\text{IR}}$. 
Its knowledge is a necessary ingredient for the method, and therefore it is of paramount importance to understand how to treat it properly. 

There are two main approaches to IR divergences within the scope of the on-shell method. 
The first one consists in taking the IR anomalous dimensions to be external inputs from other computations. 
For instance, at one-loop level the IR anomalous dimension can be parametrized, in any gauge theory, as 
\begin{equation}
    \gamma_{\text{IR}}^{(1)} (\{p_i\}, \mu) = \frac{g^2}{4 \pi^2} \sum_{i<j} T^a_{ik} T^a_{kj} \log \frac{\mu}{-s_{ij}} + \sum_i \gamma_i^{\text{coll.}}
\end{equation}
where $T_{ik}^a$ are the gauge-group generators acting on the particle $i$ \cite{Becher:2009cu}. 
The first term of the IR anomalous dimension stems from soft wide-angle IR radiation, whereas the second one describes the effects arising from hard, collinear divergences. 

Alternatively, one can compute the IR anomalous dimensions via on-shell techniques by making use of the on-shell method \cite{Caron-Huot:2016cwu}.
Indeed,
IR divergences do not depend specifically on the gauge-invariant operator appearing within the definition of a form factor, but only on its external states. As a consequence, one can compute these quantities by simply considering a local, gauge-invariant operator with a vanishing UV anomalous dimension and allowing for two-particle interactions.
In this respect, a natural candidate is given by the energy-momentum tensor $T_{\mu\nu}$.
Since the energy-momentum tensor has to be conserved also at the quantum level, its UV anomalous dimension has to vanish, i.e.,~$\gamma_{T} = 0$, and we are left with 
\begin{equation}
\label{eq:Master_Formula_LO}
- \gamma_{\text{IR}}^{(1)} \, F_T = D \, F_T \qquad \implies \qquad \gamma_{\text{IR}}^{(1)} = -\frac{D \, F_T}{F_T} = \frac{1}{\pi} \frac{(\mathcal M F_T)^{(1)} }{F_T}\,.
\end{equation}

In this appendix, we are going to make use of the on-shell method to compute the IR collinear anomalous dimensions associated with the external particle states related to those operators we have considered within the main text.

\subsection{$\phi \bar f f$  and $\phi \bar f i  \gamma_5 f$ operators}\label{sec:IRphiffb}

\begin{figure}[htbp]
\centering
\includegraphics[width=\textwidth,page=13]{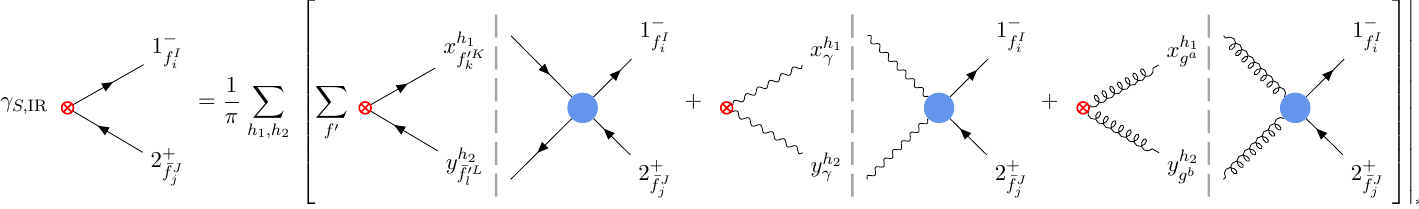}
\caption{Diagrammatic formula for computing $\gamma_{S,\text{IR}}$.}
\label{fig:phiffIR}
\end{figure}

The IR anomalous dimension $\gamma_{S,\text{IR}}$ associated with the operators $\phi \bar f_i f_j$  and $\phi \bar f_i i  \gamma_5 f_j$ can be
computed through the master formula
\begin{equation}
    \gamma_{S,\text{IR}} F_T^{\alpha\beta\dot\alpha\dot\beta}|_{*}(1^-_{f_i^I},2^+_{\bar f_j^J})=\frac{1}{\pi}(\mathcal M F_T^{\alpha\beta\dot\alpha\dot\beta})|_{*}(1^-_{f_i^I},2^+_{\bar f_j^J})\,,
\end{equation}
which diagrammatically reads as in Fig.~\ref{fig:phiffIR}.

On the left-hand side, we have the form factor of the fermion stress-energy tensor
\begin{equation}
    F_T^{\alpha\beta\dot\alpha\dot\beta}|_{*}(1_{f_i^I}^{-},2_{\bar f_j^J}^{+}) = \delta^{ij}\delta_{IJ}\mathcal T^{\alpha\beta\dot\alpha\dot\beta}_{12}
\end{equation}
where we have defined
\begin{equation}
    \mathcal T^{\alpha\beta\dot\alpha\dot\beta}_{12} = \frac{1}{2}\left( \lambda_1^\alpha \lambda_1^\beta \tilde\lambda_1^{\dot\alpha} \tilde\lambda_2^{\dot\beta} + \lambda_1^\alpha \lambda_1^\beta \tilde\lambda_2^{\dot\alpha} \tilde\lambda_1^{\dot\beta} - \lambda_1^\alpha \lambda_2^\beta \tilde\lambda_2^{\dot\alpha} \tilde\lambda_2^{\dot\beta} - \lambda_2^\alpha \lambda_1^\beta \tilde\lambda_2^{\dot\alpha} \tilde\lambda_2^{\dot\beta}  \right)\,,
\end{equation}
while, on the right-hand side, the convolution is expanded allowing for all possible intermediate states
\begin{align}
    (\mathcal M F_T^{\alpha\beta\dot\alpha\dot\beta})|_{*}(1_{f_i^I}^{-},2_{\bar f_j^J}^{+}) &= \sum_{h_1,h_2}\int d\text{LIPS}_2\,
    \bigg[\sum_{f'}
     \mathcal M|_{*} (1_{f_i^I}^{-},2_{\bar f_j^J}^{+}; x_{f_k'^K}^{h_1},y_{\bar f_l'^L}^{h_2})F_T^{\alpha\beta\dot\alpha\dot\beta}|_{*}(x_{f_k'^K}^{h_1},y_{\bar f_l'^L}^{h_2})\nonumber \\
    &\quad + \mathcal M|_{*} (1_{f_i^I}^{-},2_{\bar f_j^J}^{+}; x_\gamma^{h_1},y_\gamma^{h_2})F_T^{\alpha\beta\dot\alpha\dot\beta}|_{*}(x_\gamma^{h_1},y_\gamma^{h_2})\nonumber \\
    &\quad+ \mathcal M|_{*} (1_{f_i^I}^{-},2_{\bar f_j^J}^{+}; x_{g^c}^{h_1},y_{g^d}^{h_2})F_T^{\alpha\beta\dot\alpha\dot\beta}|_{*}(x_{g^c}^{h_1},y_{g^d}^{h_2})
    \bigg]\,.
\end{align}
The amplitudes that give a non-vanishing contribution are
\begin{align}
    \mathcal M|_{*} (1_{f_i^I}^{-},2_{\bar f_j^J}^{+}; x_{f_k'^K}^{-},y_{\bar f_l'^L}^{+})\delta_{KL} &= 
    -2\delta_{IJ}\bigg[\frac{e^2Q_f^2+C_Fg_s^2c_f^2}{\agl{1}{x}\sqr{x}{1}}\delta_{ff'}\delta^{ik}\delta^{jl} + N_{f'}\frac{e^2Q_f Q_{f'}}{\agl{1}{2}\sqr{2}{1}} \delta^{ij}\delta^{kl}\bigg]\nonumber \\&\quad\times \agl{1}{y}\sqr{x}{2}
    \,, \\
    \mathcal M|_{*} (1_{f_i^I}^{-},2_{\bar f_j^J}^{+}; x_{f_k'^K}^{+},y_{\bar f_l'^L}^{-})\delta_{KL} &= 
    -2\delta_{IJ} N_{f'}e^2Q_f Q_{f'}\delta^{ij}\delta^{kl} \frac{ \agl{1}{x}\sqr{y}{2}}{\agl{1}{2}\sqr{2}{1}}\,,  \\
    \mathcal M|_*(1^-_{f_i^I},2^+_{\bar f_j^J};x^-_\gamma,y^+_\gamma) &= -2e^2Q_f^2\delta^{ij}\delta_{IJ}\frac{\agl{1}{y}\sqr{x}{2}}{\agl{1}{x}\sqr{y}{1}}\,, \\
    \mathcal M|_*(1^-_{f_i^I},2^+_{\bar f_j^J};x^+_\gamma,y^-_\gamma) &= -2e^2Q_f^2\delta^{ij}\delta_{IJ}\frac{\agl{1}{x}\sqr{y}{2}}{\agl{1}{y}\sqr{x}{1}}\,,\\
    \mathcal M|_*(1^-_{f_i^I},2^+_{\bar f_j^J};x^-_{g^a},y^+_{g^b})\delta^{ab} &= -2C_Fg_s^2c_f^2\delta^{ij}\delta_{IJ}\frac{\agl{1}{y}\sqr{x}{2}}{\agl{1}{x}\sqr{y}{1}}\,, \\
    \mathcal M|_*(1^-_{f_i^I},2^+_{\bar f_j^J};x^+_{g^a},y^-_{g^b})\delta^{ab} &= -2C_Fg_s^2c_f^2\delta^{ij}\delta_{IJ}\frac{\agl{1}{x}\sqr{y}{2}}{\agl{1}{y}\sqr{x}{1}}\,,
\end{align}
(where $N_{f'}=c_{f'}N_c + (1 - c_{f'})$, namely $N_{f'}=N_c$ if $f'=q$ and $N_{f'}=1$ if $f'=\ell$) which are respectively multiplied by
\begin{align}
    F_T^{\alpha\beta\dot\alpha\dot\beta}|_{*}(x_{f_k'^K}^{-},y_{\bar f_l'^L}^{+}) &= \delta^{kl}\delta_{KL}\mathcal T^{\alpha\beta\dot\alpha\dot\beta}_{xy} \,,&
    F_T^{\alpha\beta\dot\alpha\dot\beta}|_{*}(x_{f_k'^K}^{+},y_{\bar f_l'^L}^{-}) &= -\delta^{kl}\delta_{KL}\mathcal T^{\alpha\beta\dot\alpha\dot\beta}_{yx}\,,\\
    F_T^{\alpha\beta\dot\alpha\dot\beta}|_{*}(x_{\gamma}^{-},y_{\gamma}^{+}) &= -2\lambda_x^\alpha \lambda_x^\beta \tilde\lambda_y^{\dot\alpha} \tilde\lambda_y^{\dot\beta}\,, &
    F_T^{\alpha\beta\dot\alpha\dot\beta}|_{*}(x_{\gamma}^{+},y_{\gamma}^{-}) &= -2\lambda_y^\alpha \lambda_y^\beta \tilde\lambda_x^{\dot\alpha} \tilde\lambda_x^{\dot\beta}\,, \\
    F_T^{\alpha\beta\dot\alpha\dot\beta}|_{*}(x_{g^a}^{-},y_{g^b}^{+}) &= -2\delta^{ab}\lambda_x^\alpha \lambda_x^\beta \tilde\lambda_y^{\dot\alpha} \tilde\lambda_y^{\dot\beta}\,, &
    F_T^{\alpha\beta\dot\alpha\dot\beta}|_{*}(x_{g^a}^{+},y_{g^b}^{-}) &= -2\delta^{ab}\lambda_y^\alpha \lambda_y^\beta \tilde\lambda_x^{\dot\alpha} \tilde\lambda_x^{\dot\beta}
    \,.
\end{align}

\subparagraph{Angular integration.}
The calculation of the phase-space integral with the angular parameterization is as follows.
The amplitudes read
\begin{align}
    \mathcal M|_{*} (1_{f_i^I}^{-},2_{\bar f_j^J}^{+}; x_{f_k'^K}^{-},y_{\bar f_l'^L}^{+})\delta_{KL} &= 
    2\delta_{IJ}\bigg[(e^2Q_f^2+C_Fg_s^2c_f^2)\delta_{ff'}\delta^{ik}\delta^{jl}\frac{\cos^2\theta}{\sin^2\theta}\nonumber\\ &\quad + N_{f'} e^2Q_f Q_{f'} \delta^{ij}\delta^{kl}\cos^2\theta \bigg]\,, \\
    \mathcal M|_{*} (1_{f_i^I}^{-},2_{\bar f_j^J}^{+}; x_{f_k'^K}^{+},y_{\bar f_l'^L}^{-})\delta_{KL} &= -2\delta_{IJ} N_{f'} e^2Q_f Q_{f'}\delta^{ij}\delta^{kl} \sin^2\theta e^{2i\phi}\,, \\
    \mathcal M|_*(1^-_{f_i^I},2^+_{\bar f_j^J};x^-_\gamma,y^+_\gamma) &= -2e^2Q_f^2\delta^{ij}\delta_{IJ}\frac{\cos\theta}{\sin\theta}e^{-i\phi}\,, \\
    \mathcal M|_*(1^-_{f_i^I},2^+_{\bar f_j^J};x^+_\gamma,y^-_\gamma) &= 2e^2Q_f^2\delta^{ij}\delta_{IJ}\frac{\sin\theta}{\cos\theta}e^{3i\phi}\,, \\
    \mathcal M|_*(1^-_{f_i^I},2^+_{\bar f_j^J};x^-_{g^a},y^+_{g^b})\delta^{ab} &= -2C_Fg_s^2c_f^2\delta^{ij}\delta_{IJ}\frac{\cos\theta}{\sin\theta}e^{-i\phi}\,, \\
    \mathcal M|_*(1^-_{f_i^I},2^+_{\bar f_j^J};x^+_{g^a},y^-_{g^b})\delta^{ab} &= 2C_Fg_s^2c_f^2\delta^{ij}\delta_{IJ}\frac{\sin\theta}{\cos\theta}e^{3i\phi}\,,
\end{align}
and the integration in the azimuthal angle $\phi$ yields
\begin{align}
    \int_0^{2\pi}\frac{d\phi}{2\pi}\,F_T^{\alpha\beta\dot\alpha\dot\beta}|_{*}(x_{f_k'^K}^{-},y_{\bar f_l'^L}^{+}) &= \cos^2\theta[-1+2\cos(2\theta)]\delta^{kl}\delta_{KL}\mathcal T^{\alpha\beta\dot\alpha\dot\beta}_{12}\,,\\
    \int_0^{2\pi}\frac{d\phi}{2\pi}\,F_T^{\alpha\beta\dot\alpha\dot\beta}|_{*}(x_{f_k'^K}^{+},y_{\bar f_l'^L}^{-})e^{2i\phi} &= -\sin^2\theta [1+2\cos(2\theta)]\delta^{kl}\delta_{KL}\mathcal T^{\alpha\beta\dot\alpha\dot\beta}_{12}\,, \\
    \int_0^{2\pi}\frac{d\phi}{2\pi}\,F_T^{\alpha\beta\dot\alpha\dot\beta}|_{*}(x_{\gamma}^{-},y_{\gamma}^{+})e^{-i\phi} &= -4\cos^3\theta\sin\theta \mathcal T^{\alpha\beta\dot\alpha\dot\beta}_{12}\,, \\
    \int_0^{2\pi}\frac{d\phi}{2\pi}\,F_T^{\alpha\beta\dot\alpha\dot\beta}|_{*}(x_{\gamma}^{+},y_{\gamma}^{-})e^{3i\phi} &= 4\cos\theta\sin^3\theta \mathcal T^{\alpha\beta\dot\alpha\dot\beta}_{12}\,,\\
    \int_0^{2\pi}\frac{d\phi}{2\pi}\,F_T^{\alpha\beta\dot\alpha\dot\beta}|_{*}(x_{g^a}^{-},y_{g^b}^{+})e^{-i\phi} &= -4\delta^{ab}\cos^3\theta\sin\theta \mathcal T^{\alpha\beta\dot\alpha\dot\beta}_{12}\,, \\
    \int_0^{2\pi}\frac{d\phi}{2\pi}\,F_T^{\alpha\beta\dot\alpha\dot\beta}|_{*}(x_{g^a}^{+},y_{g^b}^{-})e^{3i\phi} &= 4\delta^{ab}\cos\theta\sin^3\theta \mathcal T^{\alpha\beta\dot\alpha\dot\beta}_{12}\,.
\end{align}
Therefore, the remaining integral to compute is
\begin{align}
    (\mathcal M F_T^{\alpha\beta\dot\alpha\dot\beta})|_{*}(1^-_{f_i^I},2^+_{\bar f_j^J}) &= \frac{1}{16\pi}\int_0^{\pi/2}2\sin\theta\cos\theta \,d\theta\,\bigg[2\sum_{f'}2 N_{f'}e^2Q_f Q_{f'}\sin^4\theta [1+2\cos(2\theta)]\nonumber \\
    &\quad +2\sum_{f'}2 N_{f'}e^2Q_f Q_{f'}\cos^4\theta[-1+2\cos(2\theta)]\nonumber \\
    &\quad +4(e^2Q_f^2+C_Fg_s^2c_f^2)\frac{\cos^4\theta}{\sin^2\theta}[-1+2\cos(2\theta)]\nonumber \\
    &\quad  +8(e^2Q_f^2+C_Fg_s^2c_f^2)(\cos^4\theta + \sin^4\theta)
    \bigg]F_T^{\alpha\beta\dot\alpha\dot\beta}|_{*}(1^-_{f_i^I},2^+_{\bar f_j^J})\nonumber \\
    &= \frac{1}{4\pi}(e^2Q_f^2+C_Fg_s^2c_f^2)\int_0^{\pi/2}2\sin\theta\cos\theta \,d\theta\,\bigg[\frac{\cos^4\theta}{\sin^2\theta}[-1+2\cos(2\theta)]\nonumber\\
    &\quad +2(\cos^4\theta + \sin^4\theta) \bigg]
    F_T^{\alpha\beta\dot\alpha\dot\beta}|_{*}(1^-_{f_i^I},2^+_{\bar f_j^J})\,,
\end{align}
which implies
\begin{equation}\label{eq:gammaS_IR_angular}
    \gamma_{S,\text{IR}} = \frac{1}{4\pi^2}(e^2Q_f^2+C_Fg_s^2c_f^2)\int_0^{\pi/2}2\sin\theta\cos\theta \,d\theta\,\bigg[\frac{\cos^4\theta}{\sin^2\theta}[-1+2\cos(2\theta)]+2(\cos^4\theta + \sin^4\theta) \bigg]\,.
\end{equation}

\subparagraph{Stokes integration.}
The calculation of the phase-space integral with the Stokes parameterization is as follows.
The amplitudes read
\begin{align}
    \mathcal M|_{*} (1_{f_i^I}^{-},2_{\bar f_j^J}^{+}; x_{f_k'^K}^{-},y_{\bar f_l'^L}^{+})\delta_{KL} &= 
2\delta_{IJ}\bigg[(e^2Q_f^2+C_Fg_s^2c_f^2)\delta_{ff'}\delta^{ik}\delta^{jl}\frac{1}{z \zb}\nonumber\\ &\quad + N_{f'} e^2Q_f Q_{f'} \delta^{ij}\delta^{kl}\frac{1}{1+z\zb} \bigg]\,, \\
    \mathcal M|_{*} (1_{f_i^I}^{-},2_{\bar f_j^J}^{+}; x_{f_k'^K}^{+},y_{\bar f_l'^L}^{-})\delta_{KL} &= -2\delta_{IJ} N_{f'} e^2Q_f Q_{f'}\delta^{ij}\delta^{kl}\frac{\zb^2}{1+z\zb}\,, \\
    \mathcal M|_*(1^-_{f_i^I},2^+_{\bar f_j^J};x^-_\gamma,y^+_\gamma) &= 2e^2Q_f^2\delta^{ij}\delta_{IJ}\frac{1}{\zb}\,, \\
    \mathcal M|_*(1^-_{f_i^I},2^+_{\bar f_j^J};x^+_\gamma,y^-_\gamma) &= -2e^2Q_f^2\delta^{ij}\delta_{IJ}\frac{\zb^2}{z}\,, \\
    \mathcal M|_*(1^-_{f_i^I},2^+_{\bar f_j^J};x^-_{g^a},y^+_{g^b})\delta^{ab} &= 2C_Fg_s^2c_f^2\delta^{ij}\delta_{IJ}\frac{1}{\zb}\,, \\
    \mathcal M|_*(1^-_{f_i^I},2^+_{\bar f_j^J};x^+_{g^a},y^-_{g^b})\delta^{ab} &= -2C_Fg_s^2c_f^2\delta^{ij}\delta_{IJ}\frac{\zb^2}{z}\,,
\end{align}
which lead to
\begin{equation}
    (\mathcal M F_T^{\alpha\beta\dot\alpha\dot\beta})|_{*}(1^-_{f_i^I},2^+_{\bar f_j^J}) =  -\frac{3}{8\pi}(e^2Q_f^2+C_Fg_s^2c_f^2) \ F_T^{\alpha\beta\dot\alpha\dot\beta}|_{*}(1^-_{f_i^I},2^+_{\bar f_j^J})
\end{equation}
and thus
\begin{equation}\label{eq:gammaS_IR_stokes}
    \gamma_{S,\text{IR}} = -\frac{3}{8\pi^2}(e^2Q_f^2+C_Fg_s^2c_f^2)\,.
\end{equation}

\subsection{$\phi FF$  and $\phi F\tilde F$ operators}\label{sec:IRphigammagamma}

\begin{figure}[htbp]
\centering
\includegraphics[width=\textwidth,page=11]{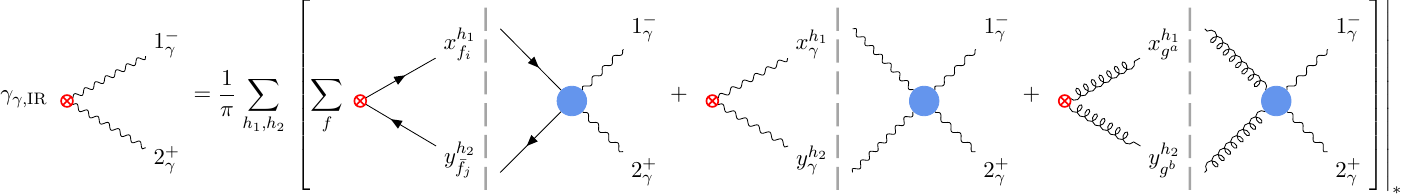}
\caption{Diagrammatic formula for computing $\gamma_{\gamma,\text{IR}}$.}
\label{fig:phigammagammaIR}
\end{figure}

The IR anomalous dimension $\gamma_{\gamma,\text{IR}}$ associated with the operators $\phi FF$  and $\phi F\tilde F$ can be computed through the master formula
\begin{equation}
    \gamma_{\gamma,\text{IR}} F_T^{\alpha\beta\dot\alpha\dot\beta}|_{*}(1^-_\gamma,2^+_\gamma)=\frac{1}{\pi}(\mathcal M F_T^{\alpha\beta\dot\alpha\dot\beta})|_{*}(1^-_\gamma,2^+_\gamma)\,,
\end{equation}
which diagrammatically reads as in Fig.~\ref{fig:phigammagammaIR}.

On the left-hand side, we have the form factor of the photon stress-energy tensor 
\begin{equation}
    F_T^{\alpha\beta\dot\alpha\dot\beta}|_{*}(1^-_\gamma,2^+_\gamma) = -2 \lambda_1^\alpha \lambda_1^\beta \tilde\lambda_2^{\dot\alpha} \tilde\lambda_2^{\dot\beta}\,,
\end{equation}
while, on the right-hand side, the convolution is expanded allowing for all possible intermediate states
\begin{align}
    (\mathcal M F_T^{\alpha\beta\dot\alpha\dot\beta})|_{*}(1^-_\gamma,2^+_\gamma) &= \sum_{h_1,h_2}\int d\text{LIPS}_2\,
    \bigg[
    \sum_f  \mathcal M|_{*} (1^-_\gamma,2^+_\gamma; x_{f_i}^{h_1},y_{\bar f_j}^{h_2})F_T^{\alpha\beta\dot\alpha\dot\beta}|_{*}(x_{f_i}^{h_1},y_{\bar f_j}^{h_2})\nonumber \\
    &\quad + \mathcal M|_{*} (1^-_\gamma,2^+_\gamma; x_\gamma^{h_1},y_\gamma^{h_2})F_T^{\alpha\beta\dot\alpha\dot\beta}|_{*}(x_\gamma^{h_1},y_\gamma^{h_2})\nonumber \\
    &\quad + \mathcal M|_{*} (1^-_\gamma,2^+_\gamma; x_{g^a}^{h_1},y_{g^b}^{h_2})F_T^{\alpha\beta\dot\alpha\dot\beta}|_{*}(x_{g^a}^{h_1},y_{g^b}^{h_2})
    \bigg]\,.
\end{align}
The only nonvanishing amplitudes are
\begin{align}
    \mathcal M|_{*} (1^-_\gamma,2^+_\gamma; x_{f_i}^{-},y_{\bar f_j}^{+}) &= 2e^2 Q_f^2 \delta^{ij}\frac{\agl{1}{y}\sqr{2}{x}}{\agl{2}{y}\sqr{1}{y}}\,, \\
    \mathcal M|_{*} (1^-_\gamma,2^+_\gamma; x_{f_i}^{+},y_{\bar f_j}^{-}) &= 2e^2 Q_f^2 \delta^{ij}\frac{\agl{1}{x}\sqr{2}{y}}{\agl{2}{y}\sqr{1}{y}}\,,
\end{align}
which are respectively multiplied by
\begin{align}
F_T^{\alpha\beta\dot\alpha\dot\beta}|_{*}(x_{f_i}^{-},y_{\bar f_j}^{+}) = \delta^{ij}\mathcal{T}_{xy}^{\alpha\beta\dot{\alpha}\dot{\beta}}\,, \qquad 
F_T^{\alpha\beta\dot\alpha\dot\beta}|_{*}(x_{f_i}^{+},y_{\bar f_j}^{-}) = -\delta^{ij}\mathcal{T}_{yx}^{\alpha\beta\dot{\alpha}\dot{\beta}}\,.
\end{align}

\subparagraph{Angular integration.}
The calculation of the phase-space integral with the angular parameterization is as follows.
The amplitudes read
\begin{align}
    \mathcal M|_{*} (1^-_\gamma,2^+_\gamma; x_{f_i}^{-},y_{\bar f_j}^{+}) &= 2e^2 Q_f^2 \delta^{ij}\frac{\cos\theta}{\sin\theta}e^{i\phi}\,, \\
    \mathcal M|_{*} (1^-_\gamma,2^+_\gamma; x_{f_i}^{+},y_{\bar f_j}^{-}) &= -2e^2 Q_f^2 \delta^{ij}\frac{\sin\theta}{\cos\theta}e^{3i\phi}\,,
\end{align}
and the integration in the azimuthal angle $\phi$ yields
\begin{align}
    \int_0^{2\pi}\frac{d\phi}{2\pi}\,F_T^{\alpha\beta\dot\alpha\dot\beta}|_{*}(x_{f_i}^{-},y_{\bar f_j}^{+})e^{i\phi} &= -2\delta^{ij}\lambda_1^\alpha \lambda_1^\beta \tilde\lambda_2^{\dot\alpha} \tilde\lambda_2^{\dot\beta} \cos^3\theta\sin\theta\,,  \\
    \int_0^{2\pi}\frac{d\phi}{2\pi}\,F_T^{\alpha\beta\dot\alpha\dot\beta}|_{*}(x_{f_i}^{+},y_{\bar f_j}^{-})e^{3i\phi} &=2\delta^{ij}\lambda_1^\alpha \lambda_1^\beta \tilde\lambda_2^{\dot\alpha} \tilde\lambda_2^{\dot\beta} \cos\theta\sin^3\theta \,.
\end{align}
Therefore, the remaining integral to compute is
\begin{align}
    (\mathcal M F_T^{\alpha\beta\dot\alpha\dot\beta})|_{*}(1^-_\gamma,2^+_\gamma) &= -4e^2 \lambda_1^\alpha \lambda_1^\beta \tilde\lambda_2^{\dot\alpha} \tilde\lambda_2^{\dot\beta}\frac{1}{8\pi}\int_0^{\pi/2}2\sin\theta\cos\theta \,d\theta\, (\cos^4\theta+\sin^4\theta)\sum_fQ_f^2 \nonumber \\
    &= -2\lambda_1^\alpha \lambda_1^\beta \tilde\lambda_2^{\dot\alpha} \tilde\lambda_2^{\dot\beta}\frac{e^2}{6\pi}\sum_fQ_f^2\,,
\end{align}
which implies
\begin{equation}\label{eq:gammaIRphoton}
    \gamma_{\gamma,\text{IR}} = \frac{e^2}{6\pi^2} \sum_f Q_f^2\,.
\end{equation}

\subparagraph{Stokes integration.}
By exploiting the Stokes parameterization, the amplitudes read
\begin{align}
    \mathcal M|_{*} (1^-_\gamma,2^+_\gamma; x_{f_i}^{-},y_{\bar f_j}^{+}) &= -2e^2 Q_f^2 \delta^{ij}\frac{1}{z}\,, \\
    \mathcal M|_{*} (1^-_\gamma,2^+_\gamma; x_{f_i}^{+},y_{\bar f_j}^{-}) &= 2e^2 Q_f^2 \delta^{ij}\frac{\zb^2}{z}\,,
\end{align}
and plugging everything together we obtain
\begin{align}
    (\mathcal M F_T^{\alpha\beta\dot\alpha\dot\beta})|_{*}(1^-_\gamma,2^+_\gamma) 
    &= -2\lambda_1^\alpha \lambda_1^\beta \tilde\lambda_2^{\dot\alpha} \tilde\lambda_2^{\dot\beta}\frac{e^2}{6\pi}\sum_fQ_f^2\,,
\end{align}
which implies Eq.~\eqref{eq:gammaIRphoton}.

\subsection{$\phi GG$  and $\phi G\tilde G$ operators}\label{sec:IRphigg}

\begin{figure}[htbp]
\centering
\includegraphics[width=\textwidth,page=12]{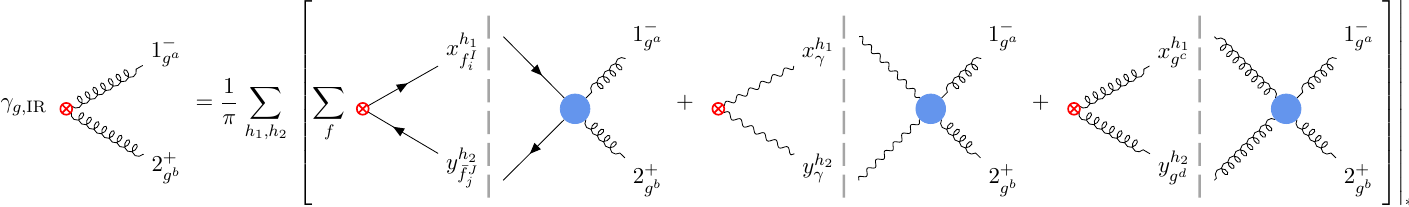}
\caption{Diagrammatic formula for computing $\gamma_{g,\text{IR}}$.}
\label{fig:phigluongluonIR}
\end{figure}

The IR anomalous dimension $\gamma_{g,\text{IR}}$ associated with the operators $\phi GG$  and $\phi G\tilde G$ can be computed through the master formula
\begin{equation}
    \gamma_{g,\text{IR}} F_T^{\alpha\beta\dot\alpha\dot\beta}|_{*}(1^-_{g^a},2^+_{g^b})=\frac{1}{\pi}(\mathcal M F_T^{\alpha\beta\dot\alpha\dot\beta})|_{*}(1^-_{g^a},2^+_{g^b})
\end{equation}
which diagrammatically reads as in Fig.~\ref{fig:phigluongluonIR}.

On the left-hand side, we have the form factor of the gluon stress-energy tensor 
\begin{equation}
    F_T^{\alpha\beta\dot\alpha\dot\beta}|_{*}(1^-_{g^a},2^+_{g^b}) = -2 \delta^{ab} \lambda_1^\alpha \lambda_1^\beta \tilde\lambda_2^{\dot\alpha} \tilde\lambda_2^{\dot\beta}\,,
\end{equation}
while, on the right-hand side, the convolution is expanded allowing for all possible intermediate states
\begin{align}
    (\mathcal M F_T^{\alpha\beta\dot\alpha\dot\beta})|_{*}(1^-_{g^a},2^+_{g^b}) &= \sum_{h_1,h_2}\int d\text{LIPS}_2\,
    \bigg[
    \sum_f  \mathcal M|_{*} (1^-_{g^a},2^+_{g^b}; x_{f_i^I}^{h_1},y_{\bar f_j^J}^{h_2})F_T^{\alpha\beta\dot\alpha\dot\beta}|_{*}(x_{f_i^I}^{h_1},y_{\bar f_j^J}^{h_2})\nonumber \\
    &\quad + \mathcal M|_{*} (1^-_{g^a},2^+_{g^b}; x_\gamma^{h_1},y_\gamma^{h_2})F_T^{\alpha\beta\dot\alpha\dot\beta}|_{*}(x_\gamma^{h_1},y_\gamma^{h_2})\nonumber \\
    &\quad + \mathcal M|_{*} (1^-_{g^a},2^+_{g^b}; x_{g^c}^{h_1},y_{g^d}^{h_2})F_T^{\alpha\beta\dot\alpha\dot\beta}|_{*}(x_{g^c}^{h_1},y_{g^d}^{h_2})
    \bigg]\,.
\end{align}
The amplitudes that give a nonvanishing contribution are
\begin{align}
    \mathcal M|_*(1^-_{g^a},2^+_{g^b};x_{f_i^I}^{-},y_{\bar f_j^J}^{+})\delta_{IJ} &= 2T_Fg_s^2c_f^2\delta^{ij}\delta^{ab}\frac{\agl{1}{y}\sqr{x}{2}}{\agl{2}{y}\sqr{y}{1}}\,, \\
    \mathcal M|_*(1^-_{g^a},2^+_{g^b};x_{f_i^I}^{+},y_{\bar f_j^J}^{-})\delta_{IJ} &= 2T_Fg_s^2c_f^2\delta^{ij}\delta^{ab}\frac{\agl{1}{x}\sqr{y}{2}}{\agl{2}{y}\sqr{y}{1}}\,,\\
    \mathcal M|_* (1^-_{g^a},2^+_{g^b};x_{g^c}^{-},y_{g^d}^{+}) \delta^{cd} &= -2C_A g_s^2\delta^{ab}\frac{\agl{1}{y}^4}{\agl{1}{x}\agl{x}{2}\agl{2}{y}\agl{y}{1}}\,, \\
    \mathcal M|_* (1^-_{g^a},2^+_{g^b};x_{g^c}^{+},y_{g^d}^{-}) \delta^{cd} &= -2C_A g_s^2\delta^{ab}\frac{\agl{1}{x}^4}{\agl{1}{x}\agl{x}{2}\agl{2}{y}\agl{y}{1}}\,,
\end{align}
which are respectively multiplied by
\begin{align}
& 
F_T^{\alpha\beta\dot\alpha\dot\beta}|_{*}(x_{f_i^I}^{-},y_{\bar f_j^J}^{+}) = \delta^{ij}\delta_{IJ}\mathcal{T}_{xy}^{\alpha\beta\dot{\alpha}\dot{\beta}}\,, 
\qquad & &
F_T^{\alpha\beta\dot\alpha\dot\beta}|_{*}(x_{f_i^I}^{+},y_{\bar f_j^J}^{-}) = -\delta^{ij}\delta_{IJ}\mathcal{T}_{yx}^{\alpha\beta\dot{\alpha}\dot{\beta}}\,, 
\\ 
&
F_T^{\alpha\beta\dot\alpha\dot\beta}|_{*}(x_{g^c}^{-},y_{g^d}^{+}) = -2\delta^{cd}\lambda_x^\alpha \lambda_x^\beta \tilde\lambda_y^{\dot\alpha} \tilde\lambda_y^{\dot\beta}\,, 
& & F_T^{\alpha\beta\dot\alpha\dot\beta}|_{*}(x_{g^c}^{+},y_{g^d}^{-}) = -2\delta^{cd}\lambda_y^\alpha \lambda_y^\beta \tilde\lambda_x^{\dot\alpha} \tilde\lambda_x^{\dot\beta}\,.
\end{align}

\subparagraph{Angular integration.}
The calculation of the phase-space integral with the angular parameterization is as follows.
The amplitudes read
\begin{align}
    \mathcal M|_*(1^-_{g^a},2^+_{g^b};x_{f_i^I}^{-},y_{\bar f_j^J}^{+})\delta_{IJ} &= 2T_Fg_s^2c_f^2\delta^{ij}\delta^{ab}\frac{\cos\theta}{\sin\theta}e^{i\phi}\,, \\
    \mathcal M|_*(1^-_{g^a},2^+_{g^b};x_{f_i^I}^{+},y_{\bar f_j^J}^{-})\delta_{IJ} &= -2T_Fg_s^2c_f^2\delta^{ij}\delta^{ab}\frac{\sin\theta}{\cos\theta}e^{3i\phi}\,,\\
    \mathcal M|_* (1^-_{g^a},2^+_{g^b};x_{g^c}^{-},y_{g^d}^{+}) \delta^{cd} &= 2C_A g_s^2\delta^{ab}\frac{\cos^2\theta}{\sin^2\theta}\,, \\
    \mathcal M|_* (1^-_{g^a},2^+_{g^b};x_{g^c}^{+},y_{g^d}^{-}) \delta^{cd} &= 2C_A g_s^2\delta^{ab}\frac{\sin^2\theta}{\cos^2\theta}e^{4i\phi}\,,
\end{align}
and the integration in the azimuthal angle $\phi$ yields
\begin{align}
    \int_0^{2\pi}\frac{d\phi}{2\pi}\,F_T^{\alpha\beta\dot\alpha\dot\beta}|_{*}(x_{f_i^I}^{-},y_{\bar f_j^J}^{+})e^{i\phi} &= -2\delta^{ij}\delta_{IJ}\lambda_1^\alpha \lambda_1^\beta \tilde\lambda_2^{\dot\alpha} \tilde\lambda_2^{\dot\beta}\cos^3\theta\sin\theta\,, \\
    \int_0^{2\pi}\frac{d\phi}{2\pi}\,F_T^{\alpha\beta\dot\alpha\dot\beta}|_{*}(x_{f_i^I}^{+},y_{\bar f_j^J}^{-})e^{3i\phi} &=2\delta^{ij}\delta_{IJ}\lambda_1^\alpha \lambda_1^\beta \tilde\lambda_2^{\dot\alpha} \tilde\lambda_2^{\dot\beta} \cos\theta\sin^3\theta\,, \\
    \int_0^{2\pi}\frac{d\phi}{2\pi}\,F_T^{\alpha\beta\dot\alpha\dot\beta}|_{*}(x_{g^c}^{-},y_{g^d}^{+}) &= -2\delta^{cd}\lambda_1^\alpha \lambda_1^\beta \tilde\lambda_2^{\dot\alpha} \tilde\lambda_2^{\dot\beta}\cos^4\theta\,, \\
    \int_0^{2\pi}\frac{d\phi}{2\pi}\,F_T^{\alpha\beta\dot\alpha\dot\beta}|_{*}(x_{g^c}^{+},y_{g^d}^{-})e^{4i\phi} &= -2\delta^{cd}\lambda_1^\alpha \lambda_1^\beta \tilde\lambda_2^{\dot\alpha} \tilde\lambda_2^{\dot\beta}\sin^4\theta\,.
\end{align}
Therefore, the remaining integral to compute is
\begin{multline}
    (\mathcal M F_T^{\alpha\beta\dot\alpha\dot\beta})|_{*}(1^-_{g^a},2^+_{g^b}) = -2 \delta^{ab} \lambda_1^\alpha \lambda_1^\beta \tilde\lambda_2^{\dot\alpha} \tilde\lambda_2^{\dot\beta}g_s^2\frac{1}{8\pi}\int_0^{\pi/2}2\sin\theta\cos\theta \,d\theta \,\\ \times \bigg[2T_F(\cos^4\theta + \sin^4\theta)\sum_f c_f^2 + C_A\frac{\cos^8\theta + \sin^8\theta}{\cos^2\theta \sin^2\theta}\bigg]\,,
\end{multline}
which implies
\begin{equation}\label{eq:gammaIRgluon_angular}
    \gamma_{g,\text{IR}} =T_F\frac{g_s^2}{6\pi^2}\sum_f c_f^2 +C_A\frac{g_s^2}{8\pi^2}\int_0^{\pi/2}2\sin\theta\cos\theta \,d\theta \, \frac{\cos^8\theta + \sin^8\theta}{\cos^2\theta \sin^2\theta} \,.
\end{equation}

\subparagraph{Stokes integration.}
Using the Stokes parameterization, the amplitudes read
\begin{align}
    \mathcal M|_*(1^-_{g^a},2^+_{g^b};x_{f_i^I}^{-},y_{\bar f_j^J}^{+})\delta_{IJ} &= -2T_Fg_s^2c_f^2\delta^{ij}\delta^{ab}\frac{1}{z}\,, \\
    \mathcal M|_*(1^-_{g^a},2^+_{g^b};x_{f_i^I}^{+},y_{\bar f_j^J}^{-})\delta_{IJ} &= 2T_Fg_s^2c_f^2\delta^{ij}\delta^{ab}\frac{\zb^2}{z}\,,\\
    \mathcal M|_* (1^-_{g^a},2^+_{g^b};x_{g^c}^{-},y_{g^d}^{+}) \delta^{cd} &= 2C_A g_s^2\delta^{ab}\frac{1}{z \zb}\,, \\
    \mathcal M|_* (1^-_{g^a},2^+_{g^b};x_{g^c}^{+},y_{g^d}^{-}) \delta^{cd} &=2C_A g_s^2\delta^{ab}\frac{ \zb^3}{z }\,,
\end{align}
and yield
\begin{equation}\label{eq:gammaIRgluon_stokes}
    \gamma_{g,\text{IR}} =-\frac{g_s^2}{8\pi^2}\bigg(\frac{11}{3}C_A -\frac{4}{3}T_F\sum_fc_f^2\bigg) \,.
\end{equation}

\bibliographystyle{JHEP}
\bibliography{references.bib}

\end{document}